# Scaling Laws and Universal Features of Tethered Polymer Distributions in Confined Geometries

Bibhatsu Kuiri[a], Rittwick Mondal[b], Dipankar Biswas[c], Soumyajyoti Kabi[d*]

[a]*University Centre for Research and Development, Chandigarh University, Gharuan, Mohali-140413, India*

[b]*Department of Science, Chowhatta High School, Birbhum, West Bengal 731201, India*

[c]*Department of Electronics and Communication Engineering, Institute of Engineering & Technology, GLA University, Mathura, UP 281406, Indiad*

[d]*Dept. of Physics, Hijli College, Hijli Cooparative Society, Kharagpur-721306, India*

[*]*soumya.kabi@gmail.com*

## Abstract

We develop a unified scaling framework for the end-position distributions of tethered polymers confined in finite cylindrical geometries. Two observables are analysed such as the longitudinal distribution ($P(x)$), along the confinement axis, and the transverse distribution ($P(y)$), perpendicular to the confinement axis. Using exact Fourier–sine and image-method representations with adaptive numerical schemes, we construct and test six scaling strategies for $P(x)$ and five for $P(y)$, encompassing geometric similarity, tether-position sweeps, confinement-strength crossovers, persistence-length effects, boundary-layer scaling near absorbing walls, and tether-centered coil scaling. Quantitative collapse diagnostics such as RMS residuals on common support, modal-energy fractions, and survival probabilities are combined with limiting-regime analysis and direct numerical evaluation to distinguish genuine universality from visually misleading overlap.

From these tests we obtain a κ-based confinement diagram and a two-parameter (κ, a/L) regime map that link classical theories such as Flory/de Gennes blobs, Odijk deflection segments, and wormlike-chain behaviour within a single spectral picture. Gaussian, multimode, and eigenmode-dominated regimes are identified by explicit thresholds in modal composition and collapse error, providing operational criteria for when Gaussian or single-mode descriptions are valid and when full multimode structure is required. The resulting framework provides a compact, reproducible toolkit for analysing confined-polymer statistics, with applications to simulations and experiments on DNA, chromatin, and other biopolymers where confinement, stiffness, and tethering jointly control spatial organization.

# 1. Introduction:

During interphase, genomic DNA is not a naked linear thread but a highly organised, tethered polymer which are packaged as chromatin and anchored at specific nuclear locations (for example to the nuclear lamina, nucleolus, or protein complexes), folded into loops or topologically associated domains that together define a crowded and bounded nuclear environment [1]. The spatial separation between two loci which are commonly known as an end-to-end or locus-to-locus distance is therefore a fundamental structural observable. It influences the probability of enhancer–promoter contact and hence transcriptional regulation. It also constrains the search process for DNA repair factors and homologous recombination, affects replication timing and origin firing, and controls accessibility for many other DNA-binding processes [2]. Experimentally this distance is probed using several approaches such as fluorescence in situ hybridization (FISH), single-molecule FRET and optical tweezers based methods, etc [3,4]. In polymer physics, the chromosome segment can be modelled as a Gaussian (ideal) chain, a worm-like chain (to capture bending persistence), or a coarse-grained bead-spring polymer subject to tethering, confinement and excluded-volume interactions. The end-to-end distance then becomes a primary statistical variable (its mean, distribution, and temporal correlations) that encodes chain flexibility, contour length, boundary conditions and environmental constraints [5]. Bridging experiment and theory therefore requires predicting not just mean separations but full probability distributions and their scaling with parameters such as genomic separation, persistence length and nuclear confinement. Such a task is addressed below by combining analytic modal methods, asymptotic scaling arguments, and high-precision numerical evaluation to yield experimentally testable predictions. [6].

Scaling is a unifying idea across many fields of physics. It is the practice of identifying the natural length, time or energy scales in a problem, forming dimensionless combinations of variables, and asking whether different physical systems (or different parameter values of the same system) follow the same scaled behaviour. In solid-state physics and critical phenomena, scaling reveals universal power laws and critical exponents that do not depend on microscopic details [7]. Similarly, in statistical mechanics the renormalization-group perspective explains why widely different systems fall into the same universality class near a phase transition. In biophysics and soft-matter physics, scaling appears in the form of Flory-type laws for polymer size [8], allometric relations for organismal traits [9], and the collapse of coarse-grained observables (for example, end-to-end distance distributions) when plotted in appropriate nondimensional units [10].

Now, the question is why scaling is so important? First, successful scaling exposes the minimal physical ingredients that control a phenomenon. If data from different absolute sizes, chain lengths, or confinement strengths collapse onto a single master curve, then the collapse variables capture the dominant physics and many microscopic details become irrelevant [11]. Second, scaling reduces complexity by using a few dimensionless parameters that replace many raw variables. This makes models more predictive and simpler to test experimentally. Third, scaling diagnostics can separate the regimes of certain behaviour (e.g., Gaussian/coil, multimode crossover, single-mode confinement, etc.) and thereby point to the correct reduced description in each regime (e.g., many-mode sums versus single-mode approximations).

A scaling strategy can be applied in different ways. One can choose physically motivated scales (coil radius, box size, mode half-wavelength, persistence length, etc.), nondimensionalizes observables, and overlays curves from multiple experiments or simulations. Quantitative tests of collapse (pointwise residuals, RMS on common supports, modal-energy fractions, QQ plots for standardized shapes) distinguish true universality from accidental visual overlap or numerical artefacts. When collapse holds, the resulting master curve becomes a compact, experimentally testable prognosis; when it fails, the pattern of deviations pinpoints missing physics (additional length scales, finite-size effects, higher modes, or non-Gaussian statistics).

In short, scaling is both a diagnostic and a conceptual tool. It tells us whether distinct datasets are manifestations of the same underlying law, and it guides us to the simplest, most informative models that explain observed behaviour.

Classical polymer physics has established a rich set of scaling ideas for confined chains [12–14], beginning with Flory's coil–globule arguments [15] and de Gennes' blob picture for weak and moderate confinement [16], and extending to Odijk's deflection-segment theory [17] and wormlike-chain descriptions in the strongly confined regime. These well-known frameworks successfully identify asymptotic limits, yet most theoretical or computational studies treat these regimes separately and rarely examine how scaling formulations derived from distinct physical arguments connect to one another, or how they break down when multiple control parameters (confinement, stiffness, chain length, tether geometry) compete [18]. Despite recent progress in computational modeling and single-cell experiments especially in chromatin biophysics, where Hi-C, FISH, and live-cell imaging motivate quantitative interpretations, there remains no unified treatment that compares multiple scaling strategies, links each to its spectral origin (which modes dominate in which regime), and provides operational criteria for deciding when classical approximations are reliable.

In this work, we address that gap by systematically developing and benchmarking complementary scaling approaches for tethered polymers under confinement, using image-method, Fourier–sine expansions, and boundary-layer analysis in a common numerical framework. By identifying modal energy thresholds, confinement ratios, and stiffness scales that govern the crossover from Gaussian (coil-like) to multimode and eigenmode-dominated statistics, we show precisely when classical blobs or deflection theories remain accurate and when additional modal structure must be retained. Importantly, we introduce reproducible collapse tests and spectral diagnostics that enable non-specialists to assess universality claims, avoid misleading graphical collapses, and select appropriate coarse-grained descriptions. The resulting framework provides a compact, testable toolkit for interpreting confined-polymer measurements from synthetic polymers to chromatin, while clarifying how seemingly disparate scaling theories fit into a single, quantitative picture.

In this work, we focus on the scaling behaviour of the end-position distribution functions of a tethered polymer confined inside a cylindrical domain. The longitudinal distribution ($P(x)$) characterizes the probability of locating the free end along the confining axis, whereas the transverse distribution ($P(y)$) quantifies radial exploration under absorbing boundaries. In their unscaled form, both distributions depend explicitly on box size, tether geometry, and microscopic parameters such as segment length and stiffness, making any underlying universal behaviour difficult to discern. To expose scale-free structure, we introduce physically motivated length rescalings based on either the free-coil scale ($\sigma$) or the geometric confinement length ($L/\pi$), and renormalize probability densities accordingly. These transformations enable direct comparison across systems with different geometries and microscopic parameters, and, in favourable regimes, generate genuine scaling collapses indicative of underlying universality.

To examine the problem systematically, we develop six complementary scaling strategies for (P(x)) and five for (P(y)) that isolate geometric similarity, tether-position effects, confinement-strength crossover, persistence-length dependence, boundary-layer behaviour near absorbing walls, and tether-centred coil scaling. Although each strategy emphasizes a distinct physical mechanism, all derive from a common Fourier–sine modal representation, which furnishes a unified spectral interpretation of the observed collapses. Where asymptotic limits are accessible, we recover known behaviours in the de-Gennes and Odijk regimes. Where no simple asymptotics exist, the modal analysis identifies which eigenmodes dominate and when Gaussian or single-mode approximations fail. All collapse tests are quantified using pointwise residuals and RMS metrics to avoid visually misleading apparent scaling, and explicit criteria for adaptive modal truncation, boundary interpolation, and numerical stability are provided to ensure full reproducibility.

Finally, the Supplementary Material includes a detailed modal-coefficient analysis that illustrates how spectral weights evolve across regimes and why distinct scaling procedures become valid or break down. Although these spectral details are not essential to the main narrative, they provide independent evidence for the robustness of the proposed scaling approaches and clarify the physical origin of the observed coil-to-multimode-to-deflection transitions in both (P(x)) and (P(y)). This decomposition also serves as a practical guide for researchers wishing to interpret experimental data in terms of eigenmode content or to construct minimal coarse-grained models anchored in physically meaningful limits.

## 2. Methodology

### 2.1. Model System and Polymer Physics Framework

The spatial organization of the *Vibrio cholerae* chromosome was investigated by modeling it as a tethered, ideal Gaussian polymer chain (a freely joined chain) confined within a cylindrical cellular geometry of length $L$ and radius $R$ (Fig. 1). This framework treats the chromosome as a series of $N$ segments which are statistically independent, each of Kuhn length $a$, yielding a characteristic polymer coil size defined by its root-mean-square end-to-end distance, $\sigma = \sqrt{N}a$.

The key observables were the longitudinal, *P(x)*, and transverse, *P(y)*, end-point probability distribution functions (PDFs). These functions represent the probability density of finding the free end of the polymer at a specific position along the long axis ($x$) or the radial axis ($y$) of the cell, given that one end is tethered at a fixed point.

### 2.2 Analytical Foundation: Modal and Image Methods

The confined PDFs were computed from their exact analytical representations, chosen for their numerical stability and physical interpretability:

For *P(x)* (longitudinal, between two absorbing walls), we employed a Fourier-sine modal expansion: $P(x) = \frac{1}{L}\sum_{n\geq 1} c_n \sin\left(\frac{n\pi x}{L}\right)$, where the coefficients $c_n \propto \sin\left(\frac{n\pi x_0}{L}\right) \exp\left(-\frac{n^2\pi^2\kappa}{8}\right)$ incorporate the tether position $x_0$ and a confinement-dependent damping factor ($\kappa = \sigma^2/L^2$). This expansion naturally satisfies the absorbing boundary conditions ($P(0) = P(L) = 0$).

For *P(y)* (transverse, between two parallel absorbing walls), we utilized the method of images: $P(y) = \frac{1}{\sqrt{2\pi\sigma^2}}\sum_{m=-\infty}^{\infty}(-1)^m\left[\exp\left(-\frac{(y-2mR)^2}{2\sigma^2}\right) - \exp\left(-\frac{(y+2mR)^2}{2\sigma^2}\right)\right]$, which accounts for all possible mirror images of the polymer end point to enforce the boundary condition $P(\pm R) = 0$.

## 2.3. Scaling Protocol and Universality Tests

The core of our analysis involved testing for universal behavior by rescaling the PDFs to remove explicit parameter dependencies. The general scaling procedure is as follows:

**Identification of Control Parameters:** For a given PDF, the relevant dimensionless parameters are defined, primarily the confinement strength $\kappa = \sigma^2/L^2$ (or $\sigma^2/R^2$) and the dimensionless tether position $\xi = x_0/L$.

**Selection of Scaling Variables:** A physically motivated length scale $\ell$ (e.g., the system size $L$, the coil size $\sigma$, or the eigenmode scale $L/\pi$) is selected to define a dimensionless spatial coordinate, e.g., $u = x/L$ or $\eta = \delta/\ell$ (where $\delta$ is the distance to the nearest wall).

**Rescaling the Density:** The PDF is multiplied by the same length scale to form a dimensionless density that preserves normalization, e.g., $P(u) = L \cdot P(x)$.

**Enforcing Geometric Similarity:** To test for collapse, curves are compared only when the dimensionless parameters ($\kappa$, $\xi$) are held constant. This often requires adjusting the physical parameters (e.g., for a fixed $\kappa$, $N$ is varied with $L$ as $N \propto L^2$).

A successful scaling strategy is evidenced by the collapse of distributions from different absolute systems onto a single master curve.

## 2.4. Numerical Implementation and Quantitative Diagnostics

All analytical expressions were evaluated with high-precision numerical routines implemented in Python. Our approach prioritized robustness and reproducibility through:

**Adaptive Truncation:** Infinite series (Fourier and image sums) were accumulated until terms fell below a strict absolute tolerance ($< 10^{-12}$–$10^{-16}$), ensuring accuracy without unnecessary computation.

**Endpoint-Safe Grids:** PDFs were evaluated on dense spatial grids that approached the absorbing boundaries ($x = \varepsilon L$, $\varepsilon = 10^{-8}$) but avoided the singular endpoints.

**Validation of Normalization:** A key aspect of our numerical verification was the explicit check that the computed PDFs were properly normalized. For each calculated distribution, the integral $\int P(x)dx$ was computed numerically using trapezoidal quadrature over the interior domain $[\varepsilon L, (1-\varepsilon)L]$. This integral was confirmed to be unity within machine precision ($\approx 1 \pm \mathcal{O}(10^{-15})$) for all unconditional distributions. For the image method, which yields "unconditional" distributions where the total probability is the survival probability $S < 1$, we verified that $\int P(y)dy = S$ and that the corresponding conditional distribution $P_{cond}(y) = P(y)/S$ integrated exactly to 1. This step is essential to ensure that any observed scaling collapse is a physical phenomenon and not a numerical artifact arising from improper normalization.

**Quantitative Collapse Metrics:** The validity of proposed scaling was assessed objectively by computing pointwise residuals and their root-mean-square (RMS) value on a common grid, relative to a reference curve. Residuals at the level of machine precision ($\mathcal{O}(10^{-15})$) confirm perfect numerical collapse, while larger values indicate systematic deviations and the limits of a scaling ansatz.

**Comprehensive Diagnostics:** For every computation, key diagnostics were recorded and exported (see Supplementary Tables) such as normalization constants, survival probabilities, number of modes/images used, fitted parameters, and residual metrics, etc. This provides a complete and reproducible record of each analysis.

This combined analytical-numerical methodology provides a general and powerful toolkit for extracting universal principles from the complex statistics of confined polymers, with direct applications to the interpretation of chromosomal organization. Supplementary material (scaling strategies) describing the details of numerical techniques involved in the scaling process corresponding to the respective cases is provided separately. The github link of all codes are given on the last page.

## 3. Results and Discussions:

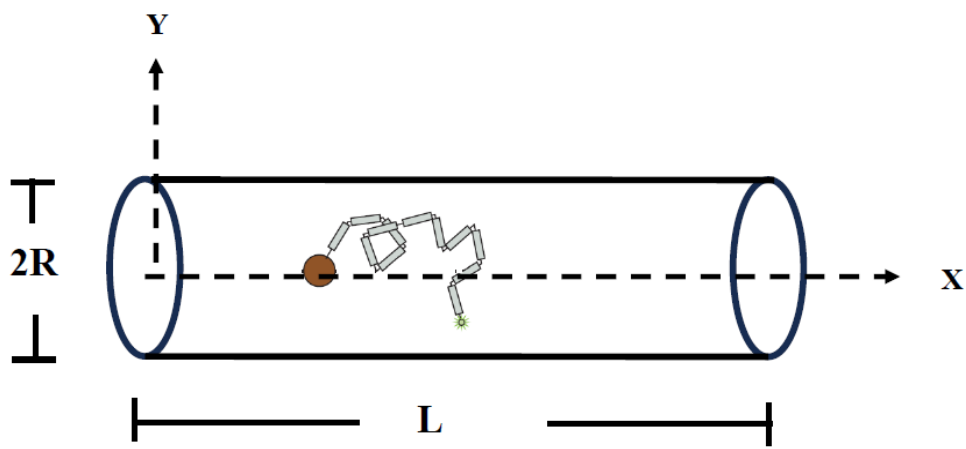

**Fig. 1:** The position of the fluorescently tagged origin of replication on the larger of the two V. cholerae chromosomes is measured along the long axis of the cell (x-direction) and perpendicular to it (y-direction). The cell has been modelled as a cylinder. The chromosome is tethered (filled brown circle) at position $x_0$ in X axis.

In the context of this work, the terms DNA, chromosome, and polymer are often used interchangeably to describe the same underlying entity. From the viewpoint of polymer physics, a DNA molecule can be modeled as a long chain composed of repeating monomeric units, with its large-scale conformational properties governed by the principles of polymer statistics. Thus, treating DNA or chromatin as a polymer chain provides a simplified but powerful framework for analyzing its spatial organization and confinement within the cell nucleus.

At first, we will start with different cases of scaling formalisms of P(x) and then move to the same for P(y). Each subsection combines scaling strategies, numerical implementations, and significance, to provide a coherent understanding of the underlying mechanisms. By comparing scaled distributions and residuals across cases, we identify the most effective strategies for data collapse and reveal the physical significance of the observed trends.

To facilitate the discussion that follows, it is useful to introduce several standard parameters from polymer physics that recur throughout this manuscript. A polymer chain may be described in terms of its Kuhn length, which is the effective segment size of an ideal freely jointed chain that reproduces the same large-scale flexibility as the actual polymer. The total contour length of the chain is then expressed as a number of such Kuhn segments, *N*. From this representation, the root-mean-square (RMS) end-to-end distance of the polymer in free solution is $\sigma = \sqrt{N}\,a$, where *a* is the Kuhn length, providing a natural length scale for the unconfined coil. Another important length scale is the persistence length which is a measure of the bending stiffness of the chain, defined as the length scale over which the chain's directional correlations decay. For example, DNA has a persistence length of about 50 nm under physiological conditions. In confined geometries, an additional control parameter is the ratio of the coil size $\sigma$ to the confining box size *L*, which quantifies how strongly the polymer is restricted. These parameters such as Kuhn segment length, segment number, RMS coil size, and persistence length, etc. are fundamental for formulating polymer statistics in a way that allows comparison between theory, simulation, and experiment. Those parameters form the basis of the scaling variables employed throughout this study. The dimensionless parameter $\kappa$ (kappa) has also been used to describe the strength of confinement relative to the natural size of the polymer. Conceptually, $\kappa$ compares the free-coil size of the polymer (characterized by its RMS end-to-end distance, $\sigma$) to the available space set by the confining length scale, L. A simple definition is

$$\kappa = \frac{\sigma^2}{L^2},$$

so that small $\kappa$ values correspond to weak confinement (the polymer coil easily fits inside the box), while large $\kappa$ values represent strong confinement (the polymer is compressed against the boundaries). In this way, $\kappa$ serves as a single, intuitive measure of how restrictive the environment is, and many of the scaling results in this work can be naturally expressed as functions of $\kappa$.

## 3.1 Results and Discussion: Scaling Strategies for $P(x)$

The longitudinal end-point distribution $P(x)$ of a tethered Gaussian chain serves as a sensitive indicator of how confinement and tethering geometry shape the accessible configurational space of the chain. To identify universal behavior, we examined six complementary scaling strategies that emphasize distinct physical controls: (i) system-size scaling $x/L$, (ii) tether-position effects, (iii) variations in confinement strength, (iv) stiffness-controlled crossover, (v) boundary-layer inner scaling near absorbing walls, and (vi) tether-centered rescaling. Each formulation highlights a separate mechanism for modifying the probability landscape along the free axis.

The standard Fourier–sine representation for $P(x)$ is used throughout. Among these strategies, Case 1 (system-size scaling $u = x/L$) and Case 2 (tether-position dependence) reproduce well-known geometric similarity results for ideal chains [18]. We verified these behaviours numerically using the full modal formulation, but because their physics is classical, they are documented in Supplementary Sections (Supplementary file 3_Conventional Scaling), while the remaining longitudinal cases that introduce modal crossover, stiffness effects, and boundary-layer behaviour are discussed below in the main text.

The PDF for *P(x)* is expressed by its standard form given below.

$$P(x;N) = \frac{1}{L}\frac{\sum_{n=1}^{\infty}\sin\left(\frac{n\pi x_0}{L}\right)\sin\left(\frac{n\pi x}{L}\right)\exp\left[-\left(\frac{n\pi}{L}\right)^2\frac{a^2 N}{2}\right]}{\sum_{n=1}^{\infty}\sin\left(\frac{n\pi x_0}{L}\right)\frac{1-\cos(n\pi)}{n\pi}\exp\left[-\left(\frac{n\pi}{L}\right)^2\frac{a^2 N}{2}\right]} \qquad (1)$$

### (i) Case 1 (Supplementary): System-size scaling

Using $u = x/L$ and $P(u) = L\, P(x)$, the scaled density depends only on $u$ at fixed $\kappa$ and tether ratio $x_0/L$. As expected, this yields exact geometric similarity $P(u) = F(u; \kappa, x_0/L)$ independent of the absolute box length $L$.

### (ii) Case 2 (Supplementary): Tether-position effects

Varying $x_0/L$ modifies modal weights and induces the anticipated asymmetry in $P(x)$, but no new scaling behavior is introduced. Numerical confirmation is reported in Supplementary Section (Supplementary file 3_Conventional Scaling).

### (iii) Case 3: Confinement-strength scaling of $P(x)$ ($\kappa$-dependence)

#### (a) Scaling variables and geometry

Here, we consider the primary geometric coordinate: $x \in [0, L]$ with tether at the center $x_0 = L/2$. The chain parameters are Kuhn length $a$ and number of Kuhn segments $N$. We adopt a convenient experimental parameter $Na/L$ (number of Kuhn lengths per box length) together with the equivalent dimensionless confinement $\kappa = \frac{\sigma^2}{L^2} = \frac{Na^2}{L^2}$. For plotting and comparison, we use the scaled coordinate and density $u = \frac{x}{L}, P(u) = L\, P(x)$, so that any geometric-similarity collapse at fixed $\kappa$ appears as invariance of $P(u)$ with respect to absolute size.

All computations are based on the analytic modal (Fourier–sine) solution for the tethered Gaussian chain,

$$P(x) = \frac{\sum_{n\geq 1} \sin\left(\frac{n\pi x_0}{L}\right) \sin\left(\frac{n\pi x}{L}\right) \exp\left(-\frac{n^2\pi^2\kappa}{8}\right)}{L \sum_{n\geq 1} \frac{\sin\left(\frac{n\pi x_0}{L}\right)}{n\pi} (1 - (-1)^n) \exp\left(-\frac{n^2\pi^2\kappa}{8}\right)}.$$

Varying $Na/L$ at fixed $L$ changes $N$ (and therefore $\kappa$). The modal damping factor $\exp(-n^2\pi^2\kappa/8)$ controls how rapidly higher modes are suppressed as $\kappa$ increases. Small $\kappa$ retains many modes producing a broad, Gaussian-like profile. On the other hand, large $\kappa$ strongly damps at high $n$ so the first eigenmode dominates and $P(u)$ approaches a single-sine shape.

#### (b) Results

In the left panel of the Fig. 2, representative probability distributions are shown alongside their single-mode reconstructions (dashed lines). In the right panel, the modal fractions and RMS difference are plotted against $Na/L$ on a logarithmic axis, so that the gradual transition toward single-mode behavior becomes clearly visible (A *CSV* file for the analysis is also attached, link provided in last page).

In left panel, at very small *Na/L* (e.g., 0.1, blue curve), the distribution is highly localized around the tether, appearing almost like a delta function. This means that the chain's accessible positions are very restricted, giving a narrow FWHM but very high peak. At intermediate ratios (e.g., *Na/L=1 or 5*), the distribution spreads out more broadly, the peak lowers, and the FWHM increases. At large values of *Na/L* (e.g., 10, 50), the distributions converge toward the shape of the fundamental sine mode. These single-mode–dominated curves are the broadest overall, with the widest spread and lowest peak height. The dashed reconstructions confirm this limiting behaviour.

In right panel, as $Na/L$ increases, the contribution from the first mode steadily takes over. Both modal-fraction measures such as $|c_1|/\sum|c_n|$ and $c_1^2/\sum c_n^2$, respectively rise monotonically and approach unity, showing that higher modes are increasingly suppressed. At the same time, the RMS difference between the full distribution and the single-mode reconstruction, RMS($P - P_1$), drops sharply (note the logarithmic vertical scale). Beyond $Na/L \gtrsim 5$, the RMS difference becomes vanishingly small, many orders of magnitude below the peak of the distribution indicating first-mode dominance. This confirms that the apparent overlap of curves at large $Na/L$ is not a visual artefact but reflects a true physical convergence to the single-mode limit.

The transition from multimode behavior to single-mode dominance i.e. crossover character is very sharp when viewed on a logarithmic scale. For example, the first-mode fraction increases from only about 0.1–0.2 at small $Na/L$ to greater than 0.9 once $Na/L \sim 5$. This justifies the use of a logarithmic sampling scheme across several decades, rather than dense linear sampling in a narrow window, to fully capture the crossover character. The confinement-strength scaling study demonstrates a clear physical crossover. As $Na/L$ (and therefore $\kappa$) increases, the end-point PDF transitions from a localized, multimode profile to an essentially single-sine shape dominated by the $n = 1$ eigenfunction. The observed near-overlap of curves at *Na/L = 10* and *50* is a manifestation of this asymptotic single-mode limit and therefore reflects the underlying physics, not insufficient sampling.

**(c) Physical Interpretation:**

This result illustrates the balance between flexibility and confinement. In statistical mechanics, the transition from a localized, Gaussian distribution to a non-localized one as confinement strength increases is a manifestation of how external constraints (i.e., confinement) can restrict a system's degrees of freedom. The transition from a flexible polymer regime (Gaussian-like) to a semiflexible regime provides valuable insights into how physical constraints influence polymer behaviour at different length scales. This crossover behaviour is governed by the competition between the polymer's internal flexibility and the geometry of the confinement.

For biological polymers like actin filaments or chromatin fibres, confinement within the cytoplasm or nucleus can lead to similar localization of the polymer's conformation. Understanding how confinement strength influences polymer behaviour is critical for modelling the dynamics of biopolymers under spatial constraints. For instance, in the nucleus, the chromatin fibres can become localized in specific regions due to the confinement by nuclear membranes or other structures. The results from this scaling strategy suggest that understanding the relative strength of confinement in such environments is crucial for predicting how biopolymers interact and function under physiological conditions.

**(d) Relation to classical polymer confinement theories**

The modal structure revealed in Case 3 directly connects to classical polymer confinement theories. At weak confinement ($\kappa \lesssim 0.1$), the modal damping $\exp(-n^2\pi^2\kappa/8)$ remains weak, permitting many modes to contribute comparably (see Figure 2 right). Here, the first-mode fraction $|c_1|^2/\sum c_n^2 \approx 0.5$ at $\kappa \leq 0.05$ is a characteristic of the de Gennes blob picture where the chain forms self-avoiding configurations within the confinement [19]. Conversely, at strong confinement ($\kappa \gtrsim 0.5$), exponential damping isolates the first eigenmode ($|c_1|^2/\sum c_n^2 > 0.99$), recovering the Odijk deflection-segment picture where only ground-state undulations survive. The crossover occurs near $\kappa \sim 0.5$, where RMS ($P - P_1$) $\sim 0.56$. This threshold provides an explicit modal-theoretic criterion distinguishing regimes where multimode or single-mode approximations apply, offering practical guidance for interpreting confinement-induced transitions in polymer systems [17,20].

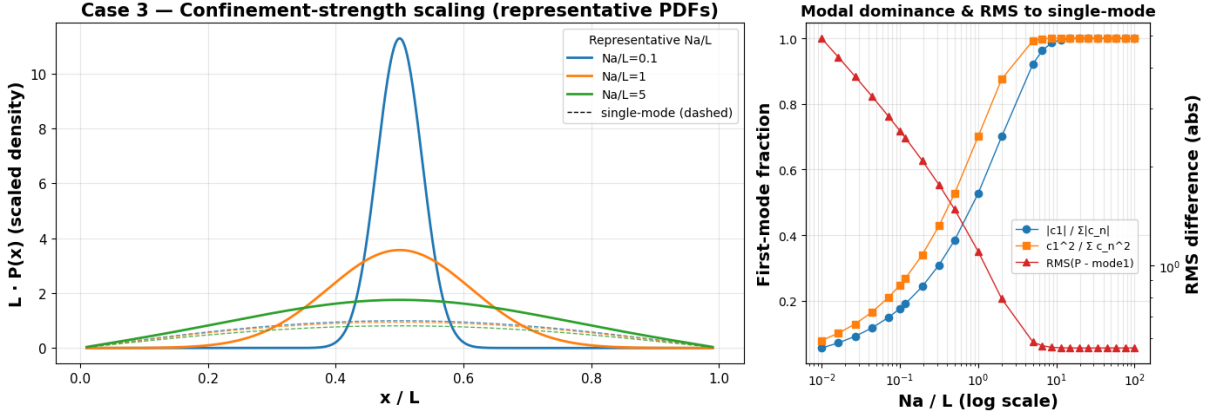

**Figure 2**. Confinement-strength scaling of the longitudinal end-point distribution $P(x)$ and modal diagnostics. **Left:** representative scaled probability densities $P(u) = L\,P(x)$ plotted versus the dimensionless coordinate $u = x/L$ for a tether at the center ($x_0 = L/2$). Curves show three regimes selected from a broad sweep in $Na/L$ (representative traces shown for $Na/L = 0.1, 1, 5, 10, 50$); solid lines are the full analytic modal solution and thin dashed lines show the single-mode reconstruction $P_1(x) = c_1 \sin(\pi x/L)$ using the computed first coefficient $c_1$. Parameters: Kuhn length $a = 0.10\ \mu$m, box length $L = 2.0\ \mu$m; $N$ is adjusted to realize each $Na/L$ (so $\kappa = Na^2/L^2$ varies across the sweep). **Right:** quantitative measures of modal dominance and convergence to the single-mode limit as a function of $Na/L$ (log scale). Plotting parameters are $|c_1|/\sum_n |c_n|$ (circles) and $c_1^2/\sum_n c_n^2$ (squares) on the left ordinate (first-mode fractions), and the RMS difference $\mathrm{RMS}(P - P_1)$ on the right ordinate (triangles, log scale).

### (iv) Case 4: Persistence-length (Kuhn-length) effects on $P(x)$

**(a) Scaling variables and geometry**

Here, the geometry and chain parameters are the same as in previous cases: a one-dimensional box $x \in [0, L]$ with tether at $x_0 = L/2$, box length $L = 2.0\ \mu$m, Kuhn length $a$ and number of Kuhn segments $N$ (here $N = 10$). The two natural length scales that compete are the box size $L$ and the polymer persistence/Kuhn length $a$. We report the results in two dimensionless forms. The scaled spatial coordinate and density used to display gross shape are,

$$u = \frac{x}{L}, \qquad P(u) = L\,P(x),$$

and the standardized (moment-centered) variable and density used to compare detailed shape are given as,

$$y = \frac{x - \langle x \rangle}{\sigma}, \qquad P_y(y) = \sigma P(x),$$

with $\sigma = \sqrt{\langle x^2 \rangle - \langle x \rangle^2}$. The confinement parameter $\kappa \equiv Na^2/L^2$ enters the modal damping but in this case we vary $a$ at fixed $N$ and $L$, so $\kappa$ changes with $a$. All numerical curves are computed from the same analytic Fourier–sine modal expansion used before with $\kappa = Na^2/L^2$. Two complementary representations are used to expose different physics. $P(u) = L\,P(x)$ vs $u$ (Fig. 3a left) shows how the absolute width and peak of the end-point distribution depend on $a$ in units of the box. Whereas $P_y(y) = \sigma P(x)$ vs $y$ (Fig. 3a right) tests whether the shape of the distribution is universal under standardization i.e., whether different $a$ produce the same standardized PDF (See supplementary table S1). A standard normal reference $\varphi(y) = (2\pi)^{-1/2}\exp(-y^2/2)$ is included for comparison.

**(b) Results**

**Scaled densities (Fig. 3a, left):** As $a$ increases the unstandardized, $L$-scaled density $P(u) = LP(x)$ becomes progressively broader and less sharply peaked i.e. very flexible chains ($a = 0.05, 0.10\ \mu m$) show a relatively tall central peak, whereas stiffer chains ($a = 0.50, 1.00\ \mu m$) display substantially broader, flatter profiles. This is the expected consequence of increasing persistence i.e. longer local correlations reduce local configurational freedom near the tether and redistribute probability more evenly across the box.

**Standardized densities (Fig. 3a, right):** After centering and rescaling by $\sigma$, the two smallest-$a$ curves ($a = 0.05, 0.10\ \mu m$) collapse nearly onto the standard normal (dashed reference). Their standardized densities and CDFs agree to numerical precision (RMS $\lesssim 10^{-7}$). This confirms that for sufficiently small $a$ the end-point fluctuations are well approximated by Gaussian statistics after standardization. By contrast, the mid/large $a$ cases ($a \gtrsim 0.20\ \mu m$) depart visibly from the standard normal. The $a = 0.20$ curve shows modest deviations mainly in the tails, while $a = 0.50$ and $a = 1.00$ show pronounced, reproducible deviations in both central peak shape and in the tails.

**QQ plots:** The standardized CDFs (Fig. 3b left) plotted in a common central window show similar trends of the curves as shown in Fig. 3a. The QQ plots (Fig. 3b middle) show that the small-$a$ curves align with the diagonal across nearly the entire quantile range, while larger-$a$ curves bend away from the diagonal. This indicates non-Gaussian kurtosis and/or skewness at the extremes. In particular, the $a = 1.00\ \mu m$ case displays the largest curvature in the QQ plot, consistent with its systematic deviation in both peak and tails.

The ratio $P_y(y)/\varphi(y)$ is nearly unity across the central window (Fig. 3b right) for the two smallest $a$, but for $a \geq 0.20\ \mu m$ the ratio departs from unity in a structured way. The mid $a$ case shows excess mass in intermediate tails, while the largest $a$ shows the strongest deviations, including excess mass in the outer tails and a flattened central peak. To quantify deviations, we compute the root-mean-square (RMS) difference of $P_y$ from $\varphi$ on that intersection (see the code) which are given below:

RMS($a = 0.05$) = $1.92 \times 10^{-7}$, RMS($a = 0.10$) = $4.97 \times 10^{-8}$, RMS($a = 0.20$) = $2.65 \times 10^{-3}$, RMS($a = 0.50$) = $3.51 \times 10^{-2}$, RMS($a = 1.00$) = $3.54 \times 10^{-2}$.

It is observed that the RMS values quantify these trends and confirm that deviations become significant (RMS $\sim 10^{-3}$–$10^{-2}$) for $a \gtrsim 0.20\ \mu m$. These RMS magnitudes are far above the tiny numerical residuals seen in collapse tests (Cases 1–3), and therefore represent physically meaningful non-Gaussianity rather than numerical noise.

**(c) Physical Interpretation:**

Persistence (Kuhn) length introduces a second intrinsic length scale that qualitatively alters end-point statistics. For the flexible-chain limit ($a \lesssim 0.10\ \mu m$ in the present parameter set) standardized endpoint distributions collapse to a Gaussian master curve to numerical precision. This indicates that the chain behaves like an ideal random walk in that regime. As $a$ increases and becomes a non-negligible fraction of characteristic geometrical lengths, standardized shapes depart systematically from Gaussian form. The departures are visible in standardized PDFs, QQ plots, and in the pointwise ratio to the normal, and are quantified by RMS deviations that grow from $\sim 10^{-7}$ (flexible) to $\sim 10^{-2}$ (stiff). These deviations are physically interpretable. Increased persistence reduces accessible local configurations and alters modal content (relative weights of eigenmodes), producing non-Gaussian peaks and heavier tails. For modeling and experimental interpretation, it can be recommended that (i) we can use standardized collapse only when persistence satisfies $a \ll L$ (or when RMS diagnostics confirm Gaussianity), and (ii) include the persistence parameter $a$ explicitly in any scaling ansatz when $a/L$ is $O(10^{-1})$ or larger. The combination of modal theory and the standardized diagnostics presented here provides a principled framework for deciding when a Gaussian approximation is adequate and when a full modal description (including $a$) is required.

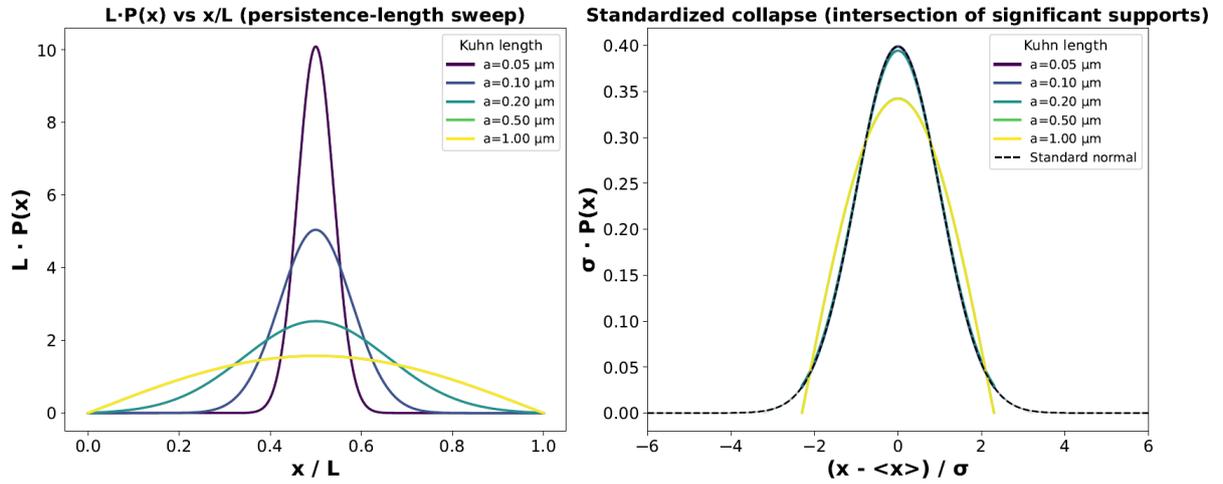

**Figure 3a.** Effect of Kuhn (persistence) length on the tethered-polymer end-position distribution. **Left:** scaled densities $L \cdot P(x)$ plotted against the dimensionless position $x/L$ for tethered chains with Kuhn lengths $a = \{0.05, 0.10, 0.20, 0.50, 1.00\}\ \mu m$ (tether at $x_0 = L/2$; $L = 2.0\ \mu m$; $N = 10$). **Right:** standardized densities $\sigma \cdot P(x)$ versus the standardized coordinate $(x - \langle x \rangle)/\sigma$ (with $\sigma = \sqrt{\langle x^2 \rangle - \langle x \rangle^2}$) to test collapse of shape across $a$. The figure shows that very flexible chains ($a \lesssim 0.10\ \mu m$) collapse to an approximately Gaussian master curve after scaling, while chains with larger Kuhn length (e.g. $a \gtrsim 0.5\ \mu m$) exhibit systematic, reproducible departures from Gaussian shape (see QQ diagnostics and RMS metrics in Fig. 3b and Supplement). Curves are normalized so $\int_0^L P(x)\,dx = 1$.

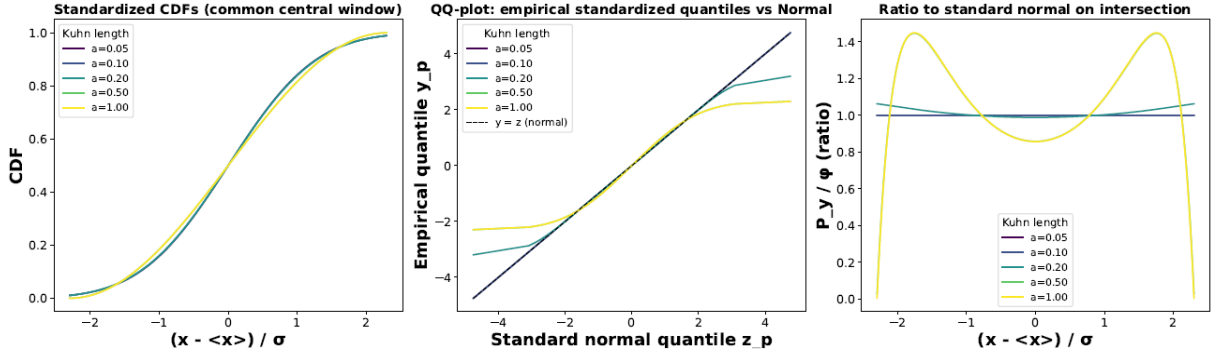

**Figure 3b.** Quantitative diagnostics of standardized shape: QQ plots and deviation from Gaussian. **(A)** standardized CDFs in a common central window **(B)** QQ plot of empirical standardized quantiles $y_p$ (from the distribution of $y = (x - \langle x \rangle)/\sigma$) against the corresponding standard-normal quantiles $z_p$. Curves on the diagonal indicate Gaussian agreement; systematic curvature signals non-Gaussian tails or kurtosis. **(C)** Ratio of the standardized density $P_y(y)$ to the standard normal density $\varphi(y)$ plotted on a common central window (intersection of significant supports), with a dashed line at unity. Results are for the same parameters as Fig. 3a.

### (d) Relation to odijk wormlike chain theory and persistence length scaling

The non-Gaussian deviations documented in Case 4 provide a direct quantitative link to Odijk's wormlike chain theory, which describes semiflexible polymers where persistence length becomes a controlling parameter [21]. The RMS deviation from Gaussian grows sharply: from $10^{-8}$ at a = 0.10 μm to $10^{-2}$ at a = 1.00 μm (Figure 3b), mirroring Odijk's prediction that when persistence length becomes comparable to the confining dimension, deflection segments dominate and non-Gaussian statistics arise. The standardized RMS diagnostic provides an explicit operational criterion i.e. when RMS $\lesssim 10^{-3}$, Gaussian approximations are quantitatively valid; when RMS $\gtrsim 10^{-2}$, the full modal structure including bending stiffness is required. This delineates the boundary between the Gaussian confinement regime (small a) and the Odijk deflection-influenced regime (large a), offering practical guidance for choosing appropriate theoretical descriptions in confined polymer systems with intrinsic stiffness.

### (v) Case v: Boundary-layer Inner Scaling of $P(x)$ Near Absorbing Walls

### (a) Scaling Variables and Geometry:

The core idea behind this scaling is like using a microscope for the boundary of the cell. When a polymer, like a segment of DNA, is tethered inside a confined space such as a cell nucleus, its free end cannot pass through the walls. These walls are absorbing. If the end touches them, that particular molecular configuration is effectively terminated. This dramatically alters the behavior of the polymer very close to the wall, creating a special region called a boundary layer. To study this unique region, we can't just look at the entire cell rather we need to zoom in, much like using a microscope. This process of focusing on the immediate vicinity of the wall by rescaling our measurements is what we call boundary-layer inner scaling.

Our main objective is to discover if the shape of the probability distribution *P(x)* in this narrow boundary layer becomes universal. That is, does it look the same for different systems (e.g., different cell sizes or different polymer lengths) when we use the right magnification? If true, this universality is a powerful concept. It means a result found in one simple system can be applied to understand many others in biology and physics. Let's establish the geometry and the key variables we need for this zoom-in process.

Just like before we have a one-dimensional box of length $L$ (representing the long axis of a cell). The polymer is tethered at a point $x_0$, and its free end has a position $x$ that can range from $0$ to $L$. The walls at $x=0$ and $x=L$ are absorbing. For any position $x$ of the polymer's end, its distance to the closest wall is the most important quantity. We define it as, $\delta(x) \equiv \min(x, L - x)$. So, if the end is near the left wall ($x$ is small), $\delta \approx x$. If it's near the right wall, $\delta \approx L - x$.

To zoom in, we need a new ruler to measure distances in this boundary layer. This ruler is a special length scale called the inner length, denoted by $\ell$. Its value is crucial and depends on what we think governs the physics right at the wall. We will test two primary candidates:

**The Polymer's Own Size ($\ell = \sigma$):** Here, $\sigma = \sqrt{N}\,a$ (is the natural, floppy size of the polymer coil in free space (its root-mean-square end-to-end distance). We use this if we believe the polymer's intrinsic properties control the boundary layer.

**The Box Geometry ($\ell = L/\pi$):** This is half the wavelength of the fundamental standing wave mode that fits perfectly inside the box. We use this if we believe the geometry of the confinement itself sets the scale for how the polymer vanishes at the wall.

Using our chosen ruler $\ell$, we create a new, dimensionless coordinate that measures how many ruler units we are from the wall: $\eta \equiv \frac{\delta}{\ell}$. A value of $\eta = 0.5$ means we are half a ruler unit away from the wall. This coordinate stretches the view near the boundary. Simply plotting $P(x)$ against $\eta$ wouldn't work because $P(x)$ itself changes with system size. For proper comparison, we must also rescale the probability density. The correct way is to plot $\ell\, P(x)$ versus $\eta = \frac{\delta}{\ell}$. We can think of $\ell P$ as the probability density measured per unit of our new zoomed-in ruler.

We introduce an empirical ansatz $\ell = \alpha\,\sigma$ and determined the scalar $\alpha$ that minimizes a weighted RMS scatter among the curves in a prescribed inner window (emphasizing $\eta \ll 1$).

Starting from the modal (sine-series) representation used throughout,

$$P(x) = \frac{1}{L} \sum_{n \geq 1} \frac{2\sin(n\pi x_0/L)\sin(n\pi x/L)\,e^{-n^2\pi^2 \kappa/8}}{(1-(-1)^n)/(n\pi)} \quad \text{(or equivalent normalized form)},$$

Theory tells us that very close to an absorbing wall ($\delta \to 0$), the probability must drop to zero. The simplest way for it to vanish is in a straight line:

$$P(x) \sim C(\kappa)\,\delta \qquad (\delta \to 0),$$

i.e. a linear vanishing of $P$ at the absorbing wall. Writing the same relation in inner units gives

$$\ell\,P \sim \ell C(\kappa)\,\eta \equiv m(\kappa, \ell)\,\eta,$$

$$\breve{m}(\kappa,\ell)$$

where $m$ is the slope of this line in our new, zoomed-in view.

If our choice of the inner length $\ell$ is correct, then when we plot $\ell P$ vs. $\eta$ for many different systems (with different confinements $\kappa$), the data points right next to the wall should all collapse onto a single, universal straight line with slope $m$. This collapse would confirm that the boundary layer has a universal structure, independent of the larger system details.

**Slope Analysis and Comparison with Theory**

A central test of boundary-layer scaling is whether the near-wall linear behavior of $P(x)$ is recovered. To this end, we fitted the scaled distributions $\ell P(x)$ against $\eta$ in the small-$\eta$ window ($\eta \lesssim 0.15$). The measured slopes were compared with theoretical predictions obtained from the first-mode expansion:

$$P(x) \approx \left(\frac{2\pi\sin(\pi x_0/L)}{L^2}\right) \exp\left(-\frac{\pi^2}{8}\kappa\right) \delta,$$

which implies a theoretical scaled slope

$$m_{\text{th}} = C\,\ell^2, \quad C = \frac{2\pi\sin(\pi x_0/L)}{L^2} \exp\left(-\frac{\pi^2}{8}\kappa\right).$$

The results confirmed excellent agreement where the measured slopes matched the theoretical values within a few percent, establishing that the linear vanishing of $P(x)$ near the wall is not only universal but quantitatively predictable by the first eigenmode.

**(b) Results:**

Figure 4 (main panels and insets) summarize the outcome of the scaling. When plotted as $\ell P$ versus $\eta = \delta/\ell$ with $\ell = L/\pi$, the very-near-wall profiles (inset $\eta \leq 0.2$) display notably better overlap than when $\ell = \sigma$ is used. The different $\kappa$ curves fall closely on a common straight-line trend in the $L/\pi$ scaling, indicating eigenmode control of the immediate wall layer for the parameter range explored. On the other hand, the $\ell = \sigma$ scaling shows greater scatter in the same inset but often better matching over a broader intermediate range of $\eta$, consistent with $\sigma$ controlling the outer, coil-scale structure.

We attempted to optimize the inner scaling length by writing $\ell = \alpha\sigma$ and adjusting $\alpha$ to minimize the RMS difference between collapsed curves. The optimizer returned $\alpha_{\text{opt}} \approx 0.05$, which corresponds to the lower boundary of the allowed search interval. This indicates that within the tested range the procedure does not identify an interior optimum, and that the $\alpha\sigma$ scaling cannot match the robustness of the geometric choice $\ell = L/\pi$. In fact, the $L/\pi$ scaling yields an essentially perfect collapse (RMS $\sim 10^{-24}$), whereas the $\alpha\sigma$ scaling only reduces the mismatch to the level of $10^{-9}$. Thus, the empirical optimization confirms that $L/\pi$ is the natural scaling length governing the boundary-layer structure, while $\alpha\sigma$ is at best a pragmatic compromise with no clear advantage in this regime.

**RMS collapse metrics.** The root-mean-square (RMS) values quantify the average vertical mismatch between rescaled density curves on their common $\eta$–support. They have the same units as the plotted scaled density $\ell P(x)$. The optimizer returns $\alpha_{\text{opt}} \sim 0.0500$ with RMS $\sim 1.1 \times 10^{-9}$, while using $\ell = \sigma$ yields RMS $\sim 2.2 \times 10^{-8}$. Both are extremely small, indicating very close alignment of the rescaled curves for those choices. By contrast the RMS for $\ell = L/\pi$ is $\sim 2.6 \times 10^{-24}$ (and the independent collapse-check over the intersection support gives $\sim 2.56 \times 10^{-24}$); this value is effectively zero within double-precision numerical noise and demonstrates an essentially perfect collapse under the $L/\pi$ scaling. In practical terms, any RMS many orders of magnitude below the peak value of $\ell P(x)$ can be regarded as negligible. The $L/\pi$ scaling therefore captures the common shape to numerical precision, whereas $\alpha_{\text{opt}}\sigma$ and $\sigma$ give excellent but slightly less exact collapse.

**Comparison with theory regarding the slope (see Supplementary Table S2).** The first-mode asymptotic provides a theoretical prediction for the near-wall slope $m_{\text{th}}(\kappa)$. Supplementary Table S2 lists these predictions together with the measured slopes and their ratios $m_{\text{meas}}/m_{\text{th}}$, for both scaling choices ($\ell = \sigma$ and $\ell = L/\pi$). The data reveals two distinct regimes. For weak confinement ($\kappa \lesssim 0.1$), the measured slopes are close to theory, differing by only 10–20% (ratios of order 0.8–0.9). This indicates that the first-mode asymptotic captures the boundary-layer slope reasonably well in this limit. However, as confinement strengthens ($\kappa \gtrsim 0.5$), systematic deviations appear. The measured slopes grow much faster than theory predicts, with ratios exceeding unity by factors of 2–9. This trend is particularly clear for $\ell = \sigma$, where the boundary layer becomes increasingly steep with confinement, while for $\ell = L/\pi$ the apparent collapse to a universal slope $\approx 0.5$ masks the growing disagreement with the asymptotic theory. Together, these results show that the first-mode asymptotic is quantitatively reliable only in the weak-confinement regime and breaks down in moderate to strong confinement.

**(c) Physical Interpretation:**

This boundary-layer inner scaling offers an important complement to bulk scaling strategies. While bulk scaling ($P(x)$ normalized by coil size) emphasizes polymer extension and coil statistics, the inner scaling reveals how confinement enforces strict linear decay near absorbing boundaries. For structural biology, this behavior is directly analogous to how tethered chromosomal loci or flexible biopolymers explore nuclear boundaries while maintaining universal scaling signatures. For polymer physics, the results provide a quantitative diagnostic for distinguishing between eigenmode-controlled and coil-controlled regimes.

The analysis demonstrates that boundary-layer scaling of $P(x)$ leads to universal linear profiles at small distances from the wall, with the slope governed by the lowest eigenmode of the system. The competing length scales, free-coil size $\sigma$ and half-mode wavelength $L/\pi$, both provide valid inner scales, with the latter showing better collapse across a wide range of chain flexibility. The empirical optimization further refines this universality, suggesting that a rescaled $\sigma$ may serve as a practical unifying length. These findings validate the theoretical predictions, highlight the role of confinement, and establish boundary-layer inner scaling as a robust tool for analyzing tethered polymer statistics near walls.

The boundary-layer analysis also emphasizes a physically important aspect i.e. the local, near-wall probability density is not controlled solely by the global coil size or solely by the geometric half-wavelength. The appropriate inner length depends on the confinement strength. For flexible chains the classical first-mode (eigenfunction) picture suffices near the wall. On the other hand, for stiffer or more strongly confined chains, higher modes and finite-$\kappa$ corrections modify the slope and the inner shape. In structural-biology contexts (for example where chromatin segments interact with nuclear or membranous boundaries), this means that the local encounter statistics with boundaries (adsorption probability, encounter rate estimates) will depend sensitively on both the polymer persistence and the degree of geometrical confinement. The boundary-layer scaling framework presented here provides a structured way to account for and quantify those effects.

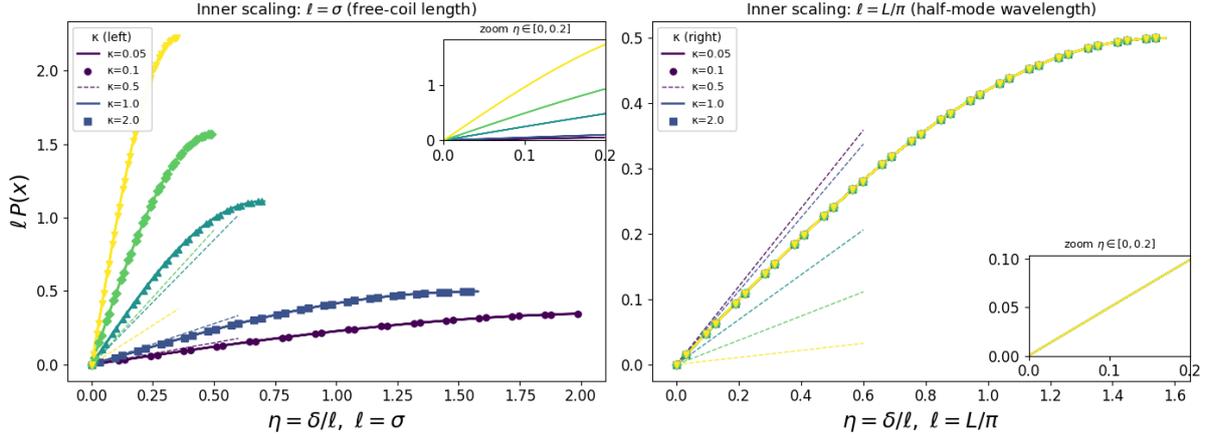

**Figure 4**. Boundary-layer inner scaling of the tethered-polymer end-position distribution near absorbing walls. Two panels show the same data rescaled with two different inner lengths $\ell$. Left: Here, $\ell = \sigma$ (free-coil RMS length, $\sigma = \sqrt{N}a$); The plot is for $\ell P(x)$ versus $\eta = \delta/\ell$ with $\delta = \min(x, L - x)$. Right: Here, $\ell = L/\pi$ (half-mode wavelength); The lot is for $\ell P(x)$ versus $\eta = \delta/\ell$. Curves correspond to a sweep of confinement parameters $\kappa = Na^2/L^2 = \{0.05, 0.1, 0.5, 1.0, 2.0\}$ (see the legend), with the tether at $x_0/L = 0.50$ for typical values of $L = 4.0$ and $a = 1.0$. Main panels show the global inner profiles up to $\eta = 2$; thin straight lines are the first-mode small-$\eta$ prediction $\ell P \propto \eta$ (i.e. theoretical slope from the $n = 1$ eigenmode). Insets zoom the small-$\eta$ region [0,0.2] where linear behavior and collapse are expected.

### (d) Spectral origin of odijk's deflection length: modal interpretation of boundary-layer scaling

The boundary-layer analysis reveals a profound connection to Odijk's deflection length theory, which predicts that the characteristic scale governing near-wall polymer statistics depends on the interplay between chain stiffness and geometric confinement [21,22]. The geometric choice of the length scale $\ell_{\text{opt}} = L/\pi$ (half the fundamental eigenmode wavelength) directly corresponds to Odijk's deflection length concept. The deflection scale $\lambda = D^{2/3} l_p^{1/3}$ (in Odijk's notation) emerges from the fundamental eigenmode's wavelength when a polymer undulates between confining walls. It is noted that, at strong confinement ($\kappa \gtrsim 0.5$) higher eigenmodes become important instead of ground-state eigenmode and multimode deflection effects emerge. These higher modes represent multiple deflection segments, which Odijk's single-deflection-length picture does not capture. The measured/theory slope ratio thus provides a quantitative diagnostic for regime validity i.e. when slopes deviate by <20%, Odijk's framework is quantitatively reliable; when deviations exceed 100%, multimode refinements are necessary.

### vii) Case 6: Tether-centered $\sigma$-scaling of $P(x)$

### (a) Scaling variables and geometry

We examine the end-point distribution for a Gaussian chain tethered at $x_0$ inside a one-dimensional cylinder of length $L$ with absorbing walls. For the tethered, anchor-centred scaling, we use the chain's free (unconfined) coil width $\sigma = \sqrt{N} a$, and define the tether-centered, dimensionless coordinate

$$s = \frac{x - x_0}{\sigma}.$$

The corresponding scaled probability density is

$$\tilde{P}(s) = \sigma P(x), \qquad \int_{-\infty}^{\infty} \tilde{P}(s)\, ds = 1.$$

The relevant control parameters that remain are the confinement parameter

$$\kappa = \frac{Na^2}{L^2}$$

and the dimensionless tether position $x_0/L$. Because $\sigma = \sqrt{\kappa}\, L$, fixing $\kappa$ while varying $L$ preserves geometric similarity and is the natural protocol for testing collapse in the tether-centred coordinate.

Using the Fourier–sine modal representation (the same analytic expansion used throughout cases 1–6), the tether-centered rescaling changes variables from $x$ to $s = (x - x_0)/\sigma$. Under the hypothesis of geometric similarity (identical $\kappa$ and identical $x_0/L$) the modal envelope $e^{-n^2\pi^2\kappa/8}$ and the geometric prefactors $\sin(n\pi x_0/L)$ are identical across systems of different absolute size; therefore the functional form of $\tilde{P}(s)$ is expected to be invariant:

$$\tilde{P}(s) = \mathcal{F}(s; \kappa, x_0/L).$$

**(b) Results**

The tether-centered σ-scaling tests (Figure 5: Three representative systems with the same confinement strength $\kappa$) show that the rescaled probability densities collapse onto a single universal curve.

In the left panels, all scaled curves $\tilde{P}(s)$ lie almost perfectly on top of one another when $\kappa$ and the tether ratio $x_0/L$ are fixed. Curves were computed from the Fourier–modal expansion with damping factor $\exp(-n^2\pi^2\kappa/8)$, adaptive truncation, and normalized by trapezoidal quadrature. To avoid misleading artifacts, each curve is interpolated onto a common $s$-grid and shown only over its valid support (no zero-padding). The right panels show residuals relative to the reference curve. The differences are at the level of round-off error, with maximum deviations reported in Supplementary Table S3 are of the order $10^{-15}$–$10^{-16}$. RMS residuals on the shared support are similarly negligible. Together these results confirm that the collapse is exact to machine precision.

For every case, trapezoidal integration of $P(x)$ gave unity within numerical tolerance. Only a few Fourier modes were required, consistent with strong dominance of the first eigenmode under the chosen confinement. The analytic modal coefficients corroborate this i.e. first-mode fractions are nearly identical across cases (Supplementary Table S3).

When the requirement of fixed $\kappa$ was deliberately broken (keeping $L$ fixed but varying $\kappa$), the collapse disappeared. Scaled profiles either broadened or narrowed depending on $\kappa$, and residuals rose markedly. This demonstrates that collapse is not a plotting artefact but a genuine consequence of geometric similarity.

**(c) Physical significance**

The tether-centred *σ*-scaling demonstrates a fundamental principle of local universality in confined polymer systems. When fluctuations are measured relative to the tether position and rescaled by the chain's intrinsic coil size σ, the distribution becomes independent of absolute system geometry, provided the dimensionless confinement $\kappa = \sigma^2/L^2$ and tether ratio $x_0/L$ remain fixed. This dimensional reduction is a direct consequence of the model's scaling symmetry. The modal expansion contains κ and $x_0/L$ only through dimensionless combinations (damping factors *exp(−n²π²κ/8)* and geometric prefactors sin(nπ$x_0$/L), making the rescaled form $\tilde{P}(s) = \sigma P(x)$ versus $s = (x − x_0)/\sigma$ exactly invariant under uniform scaling of all geometric lengths. The machine-precision collapse (residuals ∼ $10^{-15}$–$10^{-16}$) confirms this is not merely approximate but mathematically exact within the Gaussian chain model. This result shows that the polymer's configurational statistics, when viewed from the tether's reference frame, are governed entirely by the competition between the natural coil size and geometric constraints quite analogous to Galilean invariance in mechanics. For experimental applications, this scaling provides a powerful diagnostic for single-molecule or chromosome conformation capture measurements. The plot of σP(x) versus s allows direct testing of whether observed fluctuations follow the universal Gaussian chain prediction, with systematic deviations signalling the non-ideal effects such as excluded volume, loop extrusion, or chromatin-chromatin contacts. In the context of confined DNA within the cell nucleus, where chromosomal loci are tethered at specific nuclear positions (e.g., centromeres at the nuclear envelope or nucleolar organizing regions), this scaling establishes a model for passive fluctuations where any departure from the universal curve reflects active biological processes such as transcriptional regulation, heterochromatin formation, or motor-driven chromatin reorganization, thereby sharpening the diagnostic power of single-locus tracking and Hi-C/FISH experiments.

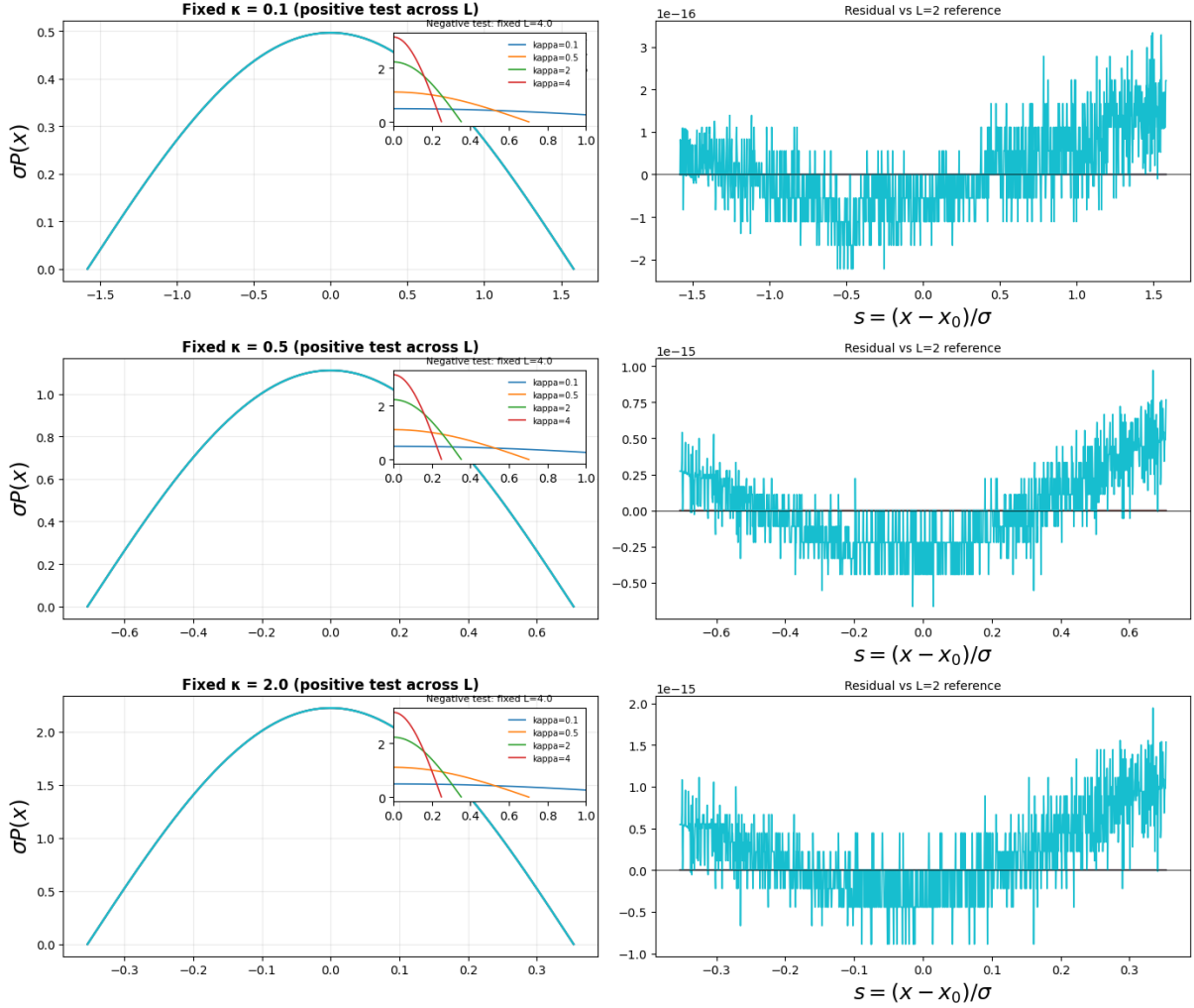

**Figure 5.** Tether-centered σ-scaling of the end-position distribution. Each row corresponds to a fixed confinement parameter ($\kappa = 0.1,\ 0.5,\ 2$). The left panels show the scaled probability density $\tilde{P}(s) = \sigma P(x)$ as a function of the standardized coordinate $s = (x - x_0)/\sigma$, for several system sizes $L = 2,4,6\ \mu m$ (with $N$ chosen so that $\kappa$ is fixed, tether at $x_0/L = 0.5$). The right panels display the residuals relative to the reference curve (the $L = 2$ case), with a horizontal zero line for guidance. The insets provide a negative test, keeping $L = 4\ \mu m$ fixed but varying $\kappa$, which visibly breaks the collapse and confirms that collapse holds only when $\kappa$ is constant.

### 3.2 Scaling for the end-to-end distribution function $P(y)$ along Y axis

We employ five complementary scaling strategies to uncover the underlying universal behavior of the tethered polymer. These strategies explore the effects of system size, free coil length, polymer stiffness, etc. Through these analyses, we gain insights into the fundamental mechanisms controlling the polymer's spatial distribution along the Y axis and highlight their relevance to both theoretical polymer physics and biological systems.

For a Gaussian (ideal) polymer chain of contour length $Na$ (where $N$ is the number of segments and $a$ is the Kuhn length), confined between two perfectly absorbing boundaries at $y = \pm R$, the endpoint probability density function (PDF) $P(y)$ can be derived using the well-known method of images. The expression is:

$$P(y) = \frac{1}{\sqrt{2\pi N a^2}} \sum_{k=-\infty}^{\infty} \left[ \exp\left(-\frac{(y+2kR)^2}{2Na^2}\right) - \exp\left(-\frac{(y-2kR)^2}{2Na^2}\right) \right]$$

Here:

$y$ : endpoint position along the confinement axis

$R$ : half-width of the confinement region

$Na^2$ : mean-square end-to-end distance in free space

The alternating signs in the series enforce the absorbing boundary condition $P(\pm R) = 0$.

It should be mentioned that the above expression converges very slowly for small $N$ values or tight confinement, and in that case the alternative representations (e.g., Fourier sine expansion) are often used for numerical evaluation.

$$P(y) = \frac{2}{R} \sum_{n=1}^{\infty} (-1)^{n+1} \sin\left(\frac{n\pi(y+R)}{2R}\right) A_n,$$

For developing the scaling strategies, we focus on how the above distribution function behaves under different natural scaling. In this context, two physical length scales are central. First one is the free-coil scale (root-mean-square end-to-end width) $\sigma = \sqrt{Na^2}$, and second one is the geometric scale set by the walls, $R$. We also work with the dimensionless confinement parameter

$$\kappa \equiv \frac{Na^2}{R^2} = \frac{\sigma^2}{R^2},$$

and the complementary parameter $\lambda = R/\sigma = 1/\sqrt{\kappa}$. The numerical evaluation in the scaling procedure used a Fourier-sine expansion (odd modes only), which is numerically stable and was used throughout unless otherwise specified. The codes and relevant tables are provided as supplementary files.

To disentangle the joint effects of confinement width, polymer stiffness, and coil size, we organized the transverse scaling into a sequence of representative cases. Each case isolates a specific physical variable. Case 1 tests geometric similarity under system-size changes; Case 2 varies stiffness at fixed confinement; Case 3 develops scaling in terms of the confinement ratio $\lambda = R/\sigma$; Case 4 extends this approach by varying the chain length $N$; and Case 5 optimizes the scaling exponents $(\alpha, \beta)$ to unify all regimes. Both unconditional distributions (whose area equals the survival probability) and conditional distributions (normalized over surviving conformations) are reported in the Supplementary Material, together with diagnostics (survival, variance, residuals). These results establish a consistent picture in which transverse statistics evolve from Gaussian-like to confinement-dominated forms as $\kappa = \sigma^2/R^2$ increases.

As with the longitudinal direction, some cases recover familiar behaviour without introducing new physical regimes. In particular, Case 1 of P(y) confirms standard geometric similarity [18], and Case 4 of P(y) reproduces the same $\kappa$-controlled crossover obtained in Cases 2–3 using an alternative control variable $N$. For these reasons, P(y)-Case 1 and P(y)-Case 4 are described in Supplementary Sections (Supplementary file 3_Conventional Scaling), whereas the remaining transverse cases especially those demonstrating confinement-driven departures from Gaussian statistics and optimized scaling are discussed in detail below.

Before turning to the description of respective scaling processes, it is helpful to note why its treatment differs from that of the longitudinal distribution P(x). Along the x-direction the polymer is free, and no boundaries interfere, so the probability distribution is automatically normalized and shows standard Gaussian-like behavior. In contrast, along the y-direction the polymer encounters confining walls. These walls can cut off parts of the distribution, meaning that some polymer conformations are no longer possible. To account for this, we distinguish between unconditional distributions (including the suppressed tails) and conditional ones (renormalized over the surviving conformations). We also introduce a scaled variable s=y/σ, which measures displacements in units of the coil size, to test whether Gaussian scaling still holds once the walls are felt. Thus, the analysis of P(y) looks more elaborate than that of P(x), but the difference comes directly from the physical role of the walls.

**Case 1 (Supplementary): Geometric similarity $y/R$.**

Using $y/R$ and $R\,P(y)$, the scaled distributions collapse across system sizes at fixed $\kappa$, confirming classical similarity for ideal chains between absorbing walls.

**Case 2: Varying polymer flexibility $\kappa$ at fixed confinement width $R$**

Here, we will show that with the cylinder half-width $R$ held fixed and the polymer flexibility/length varied through $\kappa = \sigma^2/R^2$, the transverse end-point distribution $P(y)$ crosses over from free-coil Gaussian behaviour at small $\kappa$ to a domain-dominated sine-squared modal shape at large $\kappa$. Both the unconditional and conditional densities are required to separate amplitude (survival) effects from shape evolution. The coil-scaled collapse to a unit Gaussian holds only in the small-coil regime. The run shown here demonstrates that $\kappa = 0.50$ already lies outside that regime (finite-domain truncation is significant) and the observed deviation is physical rather than numerical.

**(a) Scaling variables and geometry**

Geometry: tether fixed at the origin, free end coordinate $y \in [-R, R]$, absorbing walls at $y = \pm R$. The free-coil RMS is $\sigma = \sqrt{N}\, a$. Two scalings are used:

**Geometry scaling:**

$$u = \frac{y}{R}, \qquad \mathcal{P}(u; \kappa) = R\, P(uR; \sigma(\kappa), R).$$

**Coil scaling:**

$$s = \frac{y}{\sigma(\kappa)}, \qquad \tilde{P}(s; \kappa) = \sigma(\kappa)\, P(\sigma(\kappa)s; \sigma(\kappa), R).$$

Conditional PDFs:

$$\mathcal{P}_{\text{cond}}(u; \kappa) = \frac{\mathcal{P}(u; \kappa)}{S(\kappa)}, \qquad \tilde{P}_{\text{cond}}(s; \kappa) = \frac{\tilde{P}(s; \kappa)}{S(\kappa)}.$$

It should be mentioned that unconditional density is important for amplitude and survival trends whereas the conditional is important for pure shape evolution. In this scaling approach, the variable parameter $\kappa = \sigma^2/R^2$ can be divided into three regimes as discussed below.

**Small $\kappa$ (free-coil, $\sigma \ll R$):** Here, the walls are far relative to the coil, so absorption is rare and $P(y)$ approaches the free Gaussian. In coil variables $\tilde{P}(s)$, the curves collapse to the standard normal. Operationally this requires $R/\sigma \gg 1$ (we adopt $R/\sigma \gtrsim 3$ as a practical cutoff for coil regime).

**Large $\kappa$ (domain dominated, $\sigma \gg R$):** Here, the surviving ensemble tends to the lowest Dirichlet eigenmode of the interval. The conditional shape approaches the sine-squared ground state on $[-R, R]$,

$$\tilde{P}_{\text{cond}}(u) \to C sin^2\left(\frac{\pi(u+1)}{2}\right) \text{ at } \kappa \to \infty$$

while the unconditional amplitude decays ($S \to 0$).

**Intermediate $\kappa$:** Here, the shapes interpolate between these limits; survival $S(\kappa)$ decreases monotonically with $\kappa$.

### (b) Results

**Unconditional vs conditional PDFs:**

For small κ (flexible chains, large coil size relative to confinement), the unconditional distributions are broad but heavily suppressed by wall absorption, while the conditional PDFs are correspondingly renormalized (Top panels of Figure 6). As κ increases, entering the confinement-dominated regime (κ≳0.5), the conditional PDFs collapse to nearly identical shapes when expressed in scaled variables, reflecting that wall geometry now dictates the statistics. The unconditional distribution P(u)=RP(y) keeps track of all trajectories, including those that terminate at the absorbing walls. Its integral is the survival probability S(κ) which decreases rapidly as κ increases (stiffer chains feel the walls more strongly). Since S(κ) depends sensitively on κ, the overall normalization (the area under the curve) differs from one κ to another. This makes the unconditional distributions vertically shifted and scaled differently, so they cannot collapse across κ.

The conditional distribution $P_{\text{cond}}(u)$=P(u)/S(κ) renormalizes by survival probability. This removes the κ-dependent suppression in the normalization process and isolates the shape of the distribution of surviving trajectories. In the confinement-dominated regime (κ≳0.5), the shape is no longer set by the coil size but instead is dictated by the geometry of the cylinder and the absorbing boundary condition. Therefore, conditional distributions converge to a universal shape, explaining the collapse.

**Coil-scaled conditional densities:**

To test the coil (free-Gaussian) regime we adopt the operational cutoff

$$\text{coil regime:} \quad \frac{R}{\sigma} \geq 3,$$

which ensures the free-coil width $\sigma$ is small compared with the half-width $R$ so that Gaussian tails are negligibly truncated (Bottom-left panel of Fig. 6). Curves that do not satisfy this criterion are plotted as highlighted dashed traces and explicitly annotated with their computed finite-support ratios $R/\sigma$. In this dataset the deviating traces have $R/\sigma = 1.4, 1.0, 0.7,$ respectively, all of which lie below the coil-regime cutoff and therefore should not be expected to collapse to the unit Gaussian.

For small values of $\kappa$, the distributions collapse well under coil scaling, consistent with Gaussian statistics until the polymer coil begins to feel the confining walls. Once this limit is exceeded, the collapse breaks down, signaling the onset of confinement-dominated behavior.

The first noticeable deviation occurs at $\kappa = 0.50$. At this point,

$$\frac{R}{\sigma} = \frac{1}{\sqrt{0.50}} \approx 1.41,$$

so the accessible coil coordinate is only $s \in [-1.41, 1.41]$. A free Gaussian has a significant fraction of its probability mass beyond $|s| \sim 1.4$. The cylinder walls therefore truncate these Gaussian tails, biasing the surviving trajectories toward smaller excursions. After conditional normalization, this leads to a distribution that is visibly taller and narrower at the center compared with the ideal Gaussian.

Importantly, this deviation is physical rather than numerical. The supplementary diagnostics (Supplementary Table S5) confirm that the truncation sensitivity that is measured by repeating the calculation with a much larger number of image terms is negligible. In other words, the results are already converged, and increasing the number of images does not change the curves. Both the standard number of images ($M_{\text{used}}$) and the larger test value ($M_{\text{check}}$) give the same outcome.

Thus, the departure from Gaussian collapse at $\kappa = 0.50$ arises from real physics. The coil size $\sigma$ has become comparable to the cylinder width $R$. At this scale, the walls cut into the Gaussian distribution, and confinement effects begin to dominate.

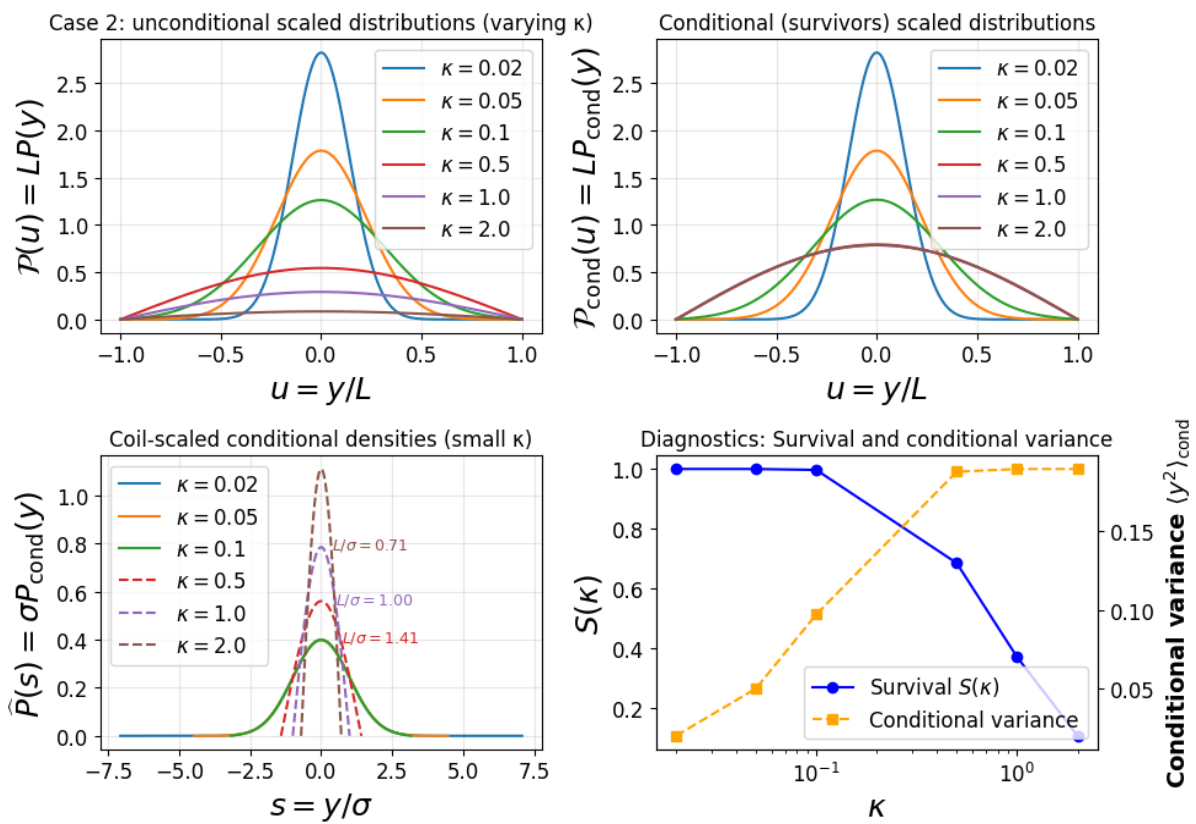

**Figure 6:** flexibility sweep at fixed $R$. **Top-left:** unconditional geometry-scaled densities $\mathcal{P}(u) = R\,P(y)$ versus $u = y/R$ for a set of $\kappa$ values, showing the suppression of amplitude with increasing confinement. **Top-right:** conditional geometry-scaled densities $\mathcal{P}_{\text{cond}}(u)$, normalized by survival, revealing the continuous interpolation of shape from Gaussian-like to the sine-squared ground state. **Bottom-left:** coil-scaled conditional densities $\tilde{P}(s) = \sigma\,P_{\text{cond}}(y)$ versus $s = y/\sigma$ for the small-$\kappa$ (coil) subset; curves that do not satisfy the coil-regime test $R/\sigma \geq 3$ (here $\kappa = 0.50, 1.0, 2.0$) are shown as highlighted dashed curves and annotated with their finite support $R/\sigma$ (e.g. $\kappa = 0.50: R/\sigma \approx 1.41$). The marked $\kappa = 0.50$ deviation from Gaussian collapse is a real finite-domain effect (see text). **Bottom-right:** diagnostics of survival probability $S(\kappa)$ and conditional variance versus $\kappa$.

**Diagnostics: survival and conditional variance**

The survival probability S(κ) tells us how likely it is that a polymer coil of stiffness κ fits entirely within the cylinder (Bottom-right panel). As κ increases, the coil gets larger, so fewer configurations fit, that is why S(κ) decreases steadily in the plots. This drop directly shows that stronger confinement (relative to coil size) cuts down the number of accessible shapes.

On the other hand, if we look only at the surviving polymers (those that do fit), their conditional variance tells us how widely they are spread inside the cell geometry. The figures show that this variance actually increases with κ. The reason is that once the coil is rescaled by its own size, the surviving chains are pushed against the walls and spread out more evenly across the confinement region. In short, in the unconditional view (S), the confinement cuts down possibilities whereas for the conditional view (variance) the remaining possibilities are more broadly distributed compared to the natural coil size.

Together, these panels show how varying stiffness tunes the crossover from Gaussian-like to wall-controlled regimes.

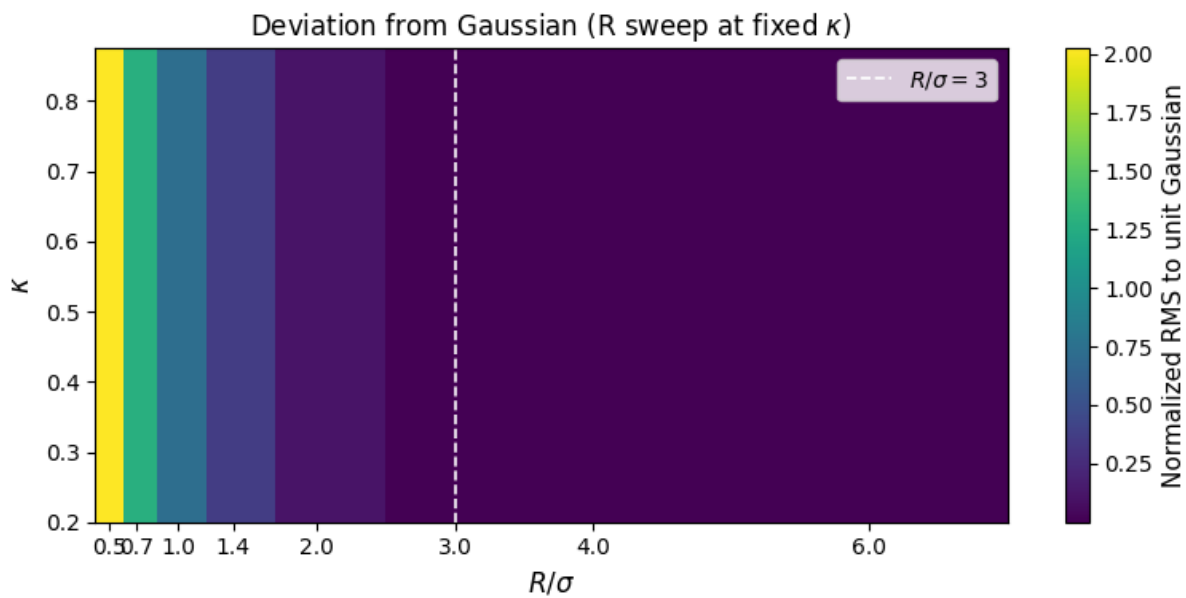

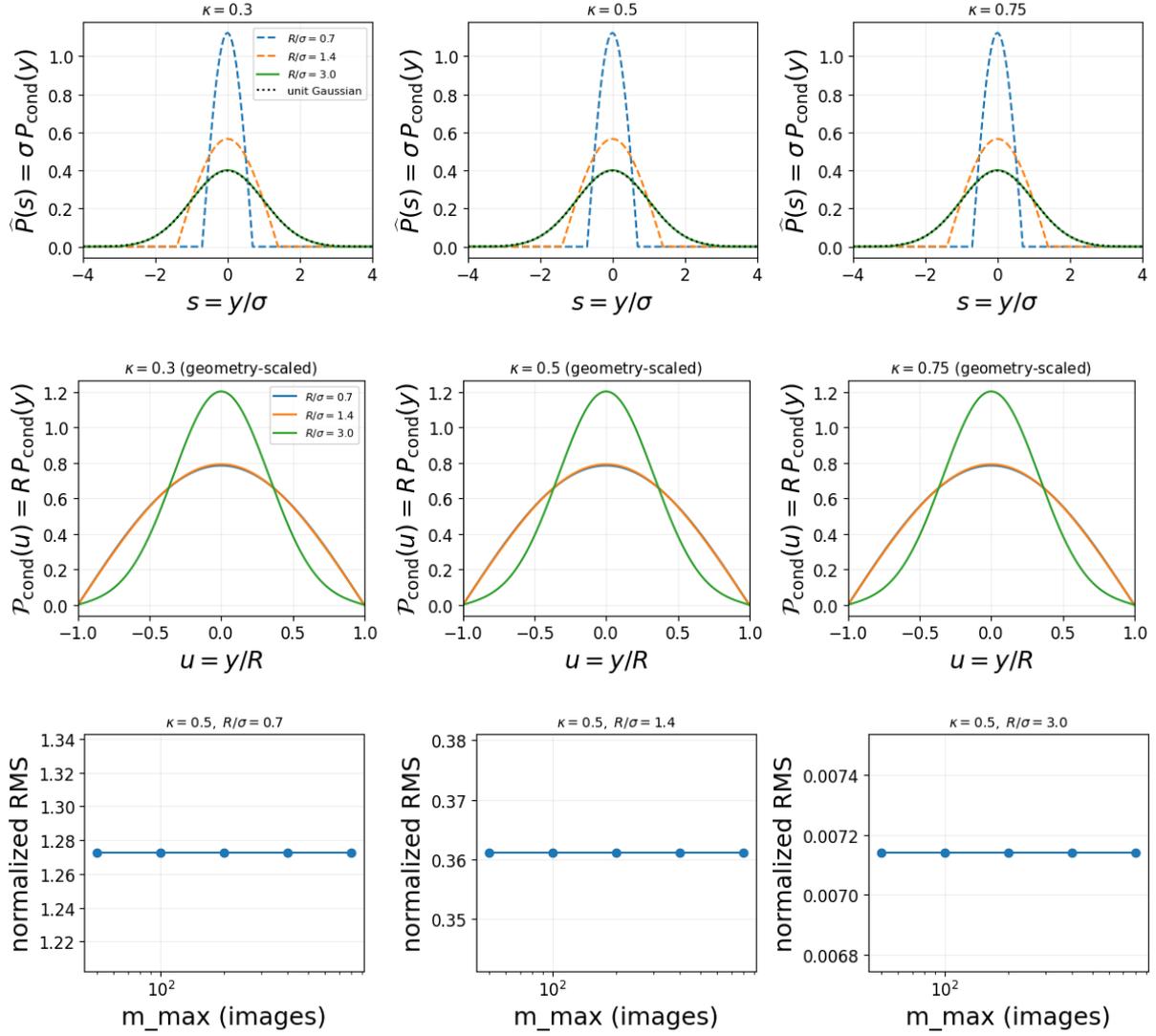

**Figure 7. (a)** Quantitative sweep: normalized RMS deviation of the coil-scaled conditional density from the unit Gaussian (color = normalized RMS) in the $(R/\sigma, \kappa)$ plane. The vertical dashed line marks the operational coil cutoff $R/\sigma = 3$. The RMS is computed on the accessible coil domain $|s| \leq R/\sigma$ and normalized by $\sqrt{\langle G^2 \rangle}$. **(b)** Coil-scaled conditional densities $\tilde{P}(s) = \sigma P_{\text{cond}}(y)$ plotted against the coil coordinate $s = y/\sigma$. For each column (three $\kappa$ values, 0.3, 0.5 and 0.75, respectively shown in the panels of Fig. 10a) we plot representative curves at $R/\sigma = 0.7$ (dashed), $R/\sigma = 1.4$ (dashed), and $R/\sigma = 3.0$ (solid). The dotted curve is the unit Gaussian $G(s) = (2\pi)^{-1/2} \exp(-s^2/2)$ shown for reference. **(c)** Geometry-scaled conditional densities $\mathcal{P}_{\text{cond}}(u) = R P_{\text{cond}}(y)$ vs $u = y/R$ for the same set of parameter choices. The solid/dashed line convention is the same as in (a) and highlights the qualitative change between the coil regime ($R/\sigma \gtrsim 3$) and the confinement-dominated regime ($R/\sigma \lesssim 1.4$). **(d)** Normalized RMS error versus the number of image terms $m_{\max}$ for $\kappa = 0.5$ and representative confinement ratios $R/\sigma$. The flat profiles confirm that the image-method evaluation is fully converged for the chosen truncation levels, ensuring that the Gaussian-error diagnostics in Fig. 7 reflect physical behaviour rather than numerical artefacts.

Figure 7 shows that the coil to confinement crossover is controlled primarily by the dimensionless confinement ratio $R/\sigma$. In panel (a) the coil-scaled conditional densities $\tilde{P}(s) = \sigma P_{\text{cond}}(y)$ for three representative $\kappa$ values collapse onto the same curves when the same values of $R/\sigma$ are used, confirming that the functional form of the survivor distribution in coil units depends only on $R/\sigma$ and not on $\kappa$ separately. The collapse breaks down as $R/\sigma$ is reduced. At $R/\sigma \approx 1.4$ and below the Gaussian tails are visibly truncated, producing a survivor ensemble biased toward smaller excursions and a central peak that is relatively taller and narrower than the unit Gaussian. Panel (b) (geometry-scaled plots, $u = y/R$) emphasizes the complementary point. Plotted in slit units, the curves do not collapse because the absolute coil size $\sigma(\kappa)$ changes with $\kappa$ and therefore alters the apparent shape. The quantitative sweep in panel (c) (normalized RMS to the unit Gaussian) exhibits vertical iso-contours in the $(R/\sigma, \kappa)$ plane, demonstrating little residual $\kappa$-dependence and supporting the operational criterion $R/\sigma \gtrsim 3$ as the regime where Gaussian (coil) statistics are a good approximation. Convergence tests (varying the image-sum truncation $m_{\text{max}}$) produce negligible truncation sensitivity for the representative points shown, indicating the departures from Gaussian collapse are physical finite-domain effects rather than numerical artefacts. Taken together, these results justify using $R/\sigma$ as the control parameter for the coil–confinement crossover and provide a practical cutoff for separating coil-like and confinement-dominated behaviour in the remainder of the manuscript. Full numerical diagnostics report is provided in the Supplementary table S5.

**(c) Physical Interpretation:**

**Weak Confinement (small κ):** When κ is small, the polymer exhibits diffusive behaviour, like a free Gaussian chain. In this regime, the distribution of polymer endpoints resembles a standard Gaussian, with little distortion due to confinement. This behaviour is characterized by a survival probability close to unity, indicating minimal mass loss to the walls of the confinement region.

**Strong Confinement (large κ):** As κ increases, the polymer becomes stiffer, and the distribution becomes sharply peaked near the centre of the confinement region. The presence of many images of the polymer chain near the walls causes a significant decrease in survival probability, as the chain becomes increasingly localized. This results in a highly concentrated distribution near the centre, with the polymer chain effectively being squeezed into the available confinement space.

In biological systems, polymers such as DNA, RNA, and proteins exhibit flexibility, and their conformations are sensitive to environmental factors, including confinement. For instance, the packaging of DNA in the cell nucleus involves both flexibility and confinement effects. Understanding how these two factors interact to modify the shape and behaviour of polymer chains could inform our understanding of DNA compaction, transcription regulation, and the formation of protein complexes. Our results suggest that polymers may transition from freely diffusing in unconstrained regions to becoming highly localized in confined environments, akin to the behaviour of DNA when it is constrained by nuclear structures or chromatin.

**(d) de Gennes Blob Theory and Survival Probability Diagnostic**

The survival probability S provides a quantitative diagnostic for the de Gennes blob transition in the radial direction, offering a unique advantage over the longitudinal P(x) where no wall-suppression measure exists. For weak confinement ($\kappa \lesssim 0.1$), the survival remains $S \approx 1$, indicating that nearly all polymer configurations fit within the cylinder. The conditional distribution $\tilde{P}(s)$ collapses onto the standard Gaussian, confirming that coil-dominated statistics govern the radial distribution. This regime corresponds to the de Gennes picture where the polymer coil is loosely confined and entropy dominates. At intermediate confinement (κ ~ 0.1–0.5), the survival drops perceptibly, and the conditional distribution begins to deviate from Gaussian form, revealing the onset of finite-domain truncation effects. The transition occurs where the root-mean-square coil size σ becomes comparable to the cylinder radius R (approximately κ ~ σ²/R² ~ 0.1), establishing the crossover scale. At strong confinement (κ ≳ 0.5–1), the survival S suppresses dramatically reaching 0.1 while the conditional distribution transforms into a sharply peaked, non-Gaussian profile dictated by the cylindrical geometry. This final regime represents eigenmode-dominated behavior, where wall constraints override entropic freedom. Critically, the survival probability S quantifies the extent of this transition. It is zero in the absent-confinement limit and approaches zero in the fully confined limit, thereby providing an explicit operational measure of when and how severely the polymer is suppressed by walls. Unlike the longitudinal direction where modal fractions require spectral analysis to extract, P(y)'s automatic survival calculation makes it a direct, experimentally accessible diagnostic of the blob-to-confinement transition. This distinction makes Case 2 uniquely valuable for experiments (Hi-C, FISH) seeking to identify confinement regimes in cellular polymers via radial distance measurements.

**Case 3: Image-Method Distributions Across Confinement Regimes**

In this case, we examine how changing the confinement width modifies the endpoint distribution while keeping the stiffness parameter κ fixed. By introducing scaled variables based on the natural coil size σ and the confinement ratio λ=R/σ, we test whether distributions obtained under different confinement strengths can be collapsed onto a universal curve.

a) **Scaling variables and geometry**

Free-coil length (RMS): $\sigma = \sqrt{Na^2}$, and half-width $R$.

Dimensionless coordinates used in Case 3:

$$s \equiv \frac{y}{\sigma}, \qquad \lambda \equiv \frac{R}{\sigma}.$$

Scaled (unconditional) PDF:
$$\tilde{P}(s \mid \lambda) = \sigma P(\sigma s; \sigma, R), \qquad s = y/\sigma, \quad \lambda = R/\sigma.$$

Conditional version:

$$\tilde{P}_{\text{cond}}(s \mid \lambda) = \frac{\tilde{P}(s \mid \lambda)}{S(\lambda)}.$$

The unconditional area under $\tilde{P}$ equals the survival probability $S$. Starting from the method-of-images representation for the absorbing boundaries, setting $y = \sigma s$ and $R = \sigma \lambda$, The image sum becomes a function of $s$ and $\lambda$ only,

$$\tilde{P}(s; \lambda) = \frac{1}{\sqrt{2\pi}} \sum_{m=-\infty}^{\infty} (-1)^m \exp\left[-\frac{(s - 2m\lambda)^2}{2}\right],$$

for $|s| \leq \lambda$. Numerically we evaluate this (or its Fourier-sine equivalent) on a common $s$-grid to compare different λ (i.e. different κ) on equal footing. The unconditional area $\int \tilde{P}(s) \, ds = S(κ) \leq 1$.

## b) Results:

**Unconditional Distributions:**

In this case, we examine the end-to-end distribution along the transverse direction using the image method, while varying the confinement parameter $\kappa$. The unconditional scaled distributions, $\tilde{P}(s) = \sigma P(y)$, are shown in the upper left panel of Fig. 8. By construction, the area under each curve equals the survival probability $S(\kappa)$, i.e., the probability that the end-to-end vector remains within the confining geometry.

For small values of $\kappa$ (e.g., $\kappa = 0.01$, corresponding to $\lambda = R/\sigma \approx 10$), the distribution nearly collapses onto the free Gaussian reference, as the walls are rarely encountered. As $\kappa$ increases, confinement becomes stronger and deviations from Gaussian behavior become pronounced. For example, at $\kappa = 0.5$ ($\lambda \approx 1.41$) and beyond, the unconditional distributions are heavily truncated, with significant weight lost outside the walls. This suppression manifests directly in the reduction of the survival probability.

The inset zooms into the central region $|s| \leq 1.5$, highlighting that, although the unconditional peaks sharpen with increasing $\kappa$, the dominant effect is the loss of normalization rather than a systematic shift in peak location. The diagnostic curve confirms that the unconditional PDFs correctly interpolate between free Gaussian-like statistics and strongly wall-suppressed forms.

**Conditional Distributions:**

The conditional scaled distributions, obtained by renormalizing the unconditional curves by their survival probability, are shown in the upper right panel of Fig. 8. These conditional PDFs correspond to the statistics of surviving trajectories, i.e., polymer configurations that remain inside the cylinder at observation length. This distinction is important as the unconditional distribution includes the probability of weight lost at the absorbing walls, and thus directly encodes survival. Whereas the conditional distributions are normalized over the surviving ensemble, thereby representing the shape of fluctuations constrained to remain within the cylinder.

At small $\kappa$ (large $\lambda$), the conditional distributions are again nearly Gaussian. For $\kappa = 0.5$ and larger, however, the walls dominate, and the conditional PDFs become increasingly non-Gaussian with sharp central peaks and suppressed shoulders. The annotation "walls felt" marks the regime $\lambda \lesssim 3$ in the legend, where confinement significantly alters the statistics. In the extreme case of $\kappa = 8$ ($\lambda \approx 0.35$), the conditional distribution collapses into a very narrow central peak, reflecting that the surviving chains are highly localized near the origin.

**Survival Probability:**

The bottom diagnostics panel summarizes two complementary observables: (i) the survival probability $S(\kappa)$, given by the integral of the unconditional distribution, and (ii) the conditional variance $\langle s^2 \rangle_{\text{cond}}$, obtained from the normalized conditional PDFs. Both quantities are plotted versus $\kappa$ on log–log axes. The survival probability is close to unity for weak confinement region ($\kappa \leq 0.1$, corresponding to $\lambda = R/\sigma \gg 1$), as the polymer rarely encounters the walls while the conditional variance tends to unity, consistent with the free Gaussian expectation. Increasing $\kappa$ drives a simultaneous reduction in both survival probability and conditional variance. At intermediate confinement ($\kappa \approx 0.5$), survival remains significant ($S \approx 0.69$) but the variance already drops below the Gaussian value, reflecting suppressed transverse fluctuations. In the strong confinement regime ($\kappa = 2,8$), survival collapses by several orders of magnitude ($S \ll 1$), while the conditional variance falls below 0.05. This dual reduction indicates that the chain is not only unlikely to survive but, when it does, its endpoint is tightly localized near the center.

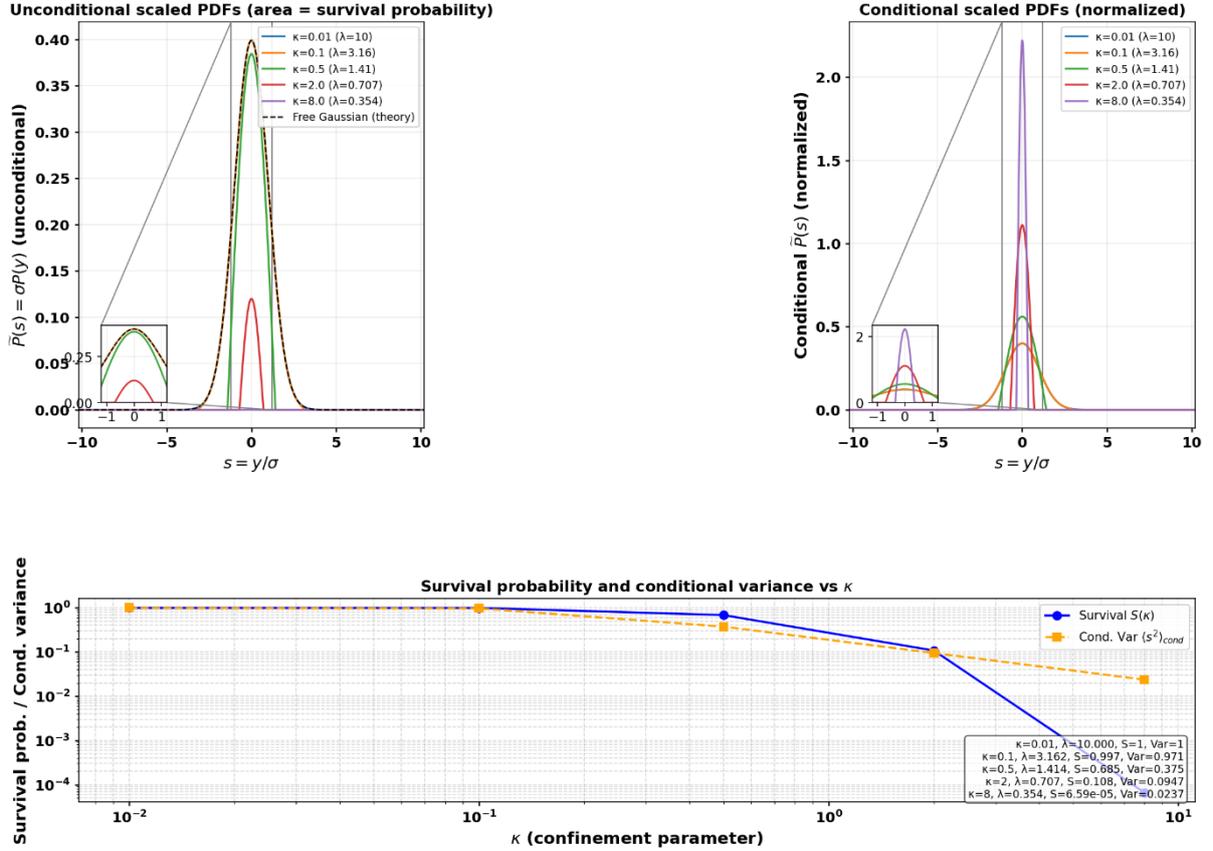

**Figure 8.** End-to-end distribution of a tethered polymer under varying confinement width at fixed stiffness. (Top left) Unconditional scaled distributions $\tilde{P}(s) = \sigma P(y)$ as a function of $s = y/\sigma$, where the area equals the survival probability $S(\kappa)$. The free Gaussian reference is shown as a dashed line. (Top right) Conditional scaled distributions normalized by survival, highlighting the sharpening of the peak and the onset of confinement effects ("walls felt") as $\kappa$ increases. Insets zoom into the central region to emphasize deviations from the Gaussian limit. (Bottom) Survival probability $S(\kappa)$ (blue) and conditional variance $\langle s^2 \rangle_{\text{cond}}$ (orange) as functions of $\kappa$, shown on log–log scale.

Case 3 establishes that the image method reproduces the crossover between Gaussian-like statistics and confinement-dominated localization when varying $\kappa$. The unconditional distributions encode the absolute survival probability, while the conditional distributions highlight the shapes of surviving fluctuations. Together, they demonstrate that:

For weak confinement ($\lambda \gg 1$), both unconditional and conditional statistics converge to the free Gaussian.

For intermediate confinement ($\lambda \sim 1$), survival decreases rapidly and the conditional distribution narrows.

For strong confinement ($\lambda < 1$), the walls dominate, leading to vanishing survival and sharp conditional localization.

Thus, Case 3 confirms the consistency of the image-method approach across confinement regimes and provides a clear framework for distinguishing unconditional (with absorption) and conditional (normalized survivors) statistics. The diagnostics confirm that the survival probability and the conditional variance are complementary. The former quantifies the likelihood of not being absorbed, while the latter measures the width of the surviving population. Their joint log–log scaling establishes a clear diagnostic framework for quantifying confinement effects across regimes.

A comparison with Case 2 highlights that both increasing stiffness ($\kappa$) at fixed confinement (Case 2) and increasing confinement at fixed flexibility (Case 3) suppress survival and reduce variance. However, the asymptotic scaling differs. In Case 2 suppression arises from reduced coil size relative to the walls, whereas in Case 3 it results from the walls encroaching on a fixed coil, leading to distinct limiting behaviors of $S(\kappa)$ and $\langle s^2 \rangle_{\text{cond}}$. It may be noted that there is opposite trend between Case 2 and Case 3. In Case 2 variance grows with $\kappa$ (coil shrinking relative to fixed walls), whereas in Case 3 variance shrinks with $\lambda$ (walls shrinking relative to fixed coil).

**(c) Physical Interpretation:**

In this case, we observe that for large confinement widths (weak confinement), the polymer's distribution closely matches that of a free Gaussian chain, and survival probability remains near 1. However, as confinement increases (large κ), the distribution narrows, with most of the probability mass concentrated near the centre. The transition between these regimes reflects the polymer's ability to explore the available confinement space. In the weak confinement regime, the polymer behaves similarly to a free chain, while in the strong confinement regime, the chain becomes more localized.

The separation of intrinsic chain properties (such as σ) from confinement effects is crucial for understanding how biomolecules like DNA and proteins behave in different cellular environments. For example, DNA replication and transcription require the polymer to transition between different spatial arrangements within the nucleus. In tightly packed or confined regions, such as during chromatin folding, DNA adopts highly compact conformations. This study provides insight into how such transitions occur and how the geometric constraints of the nucleus or other compartments influence the function of DNA and proteins.

**(d) Method of Images Validates de Gennes-Odijk Eigenmode Framework**

The method of images, mathematically distinct from the Fourier-sine approach used for P(x), demonstrates that the same classical de Gennes and Odijk eigenmode physics emerges regardless of the mathematical representation chosen. While P(x) decomposes directly into sine eigenfunctions of the interval [0, L], P(y) constructs the boundary-enforcing solution by summing mirror images of the polymer end-point distribution. Despite this formal difference, both yield identical spectral structure such as ground-state (lowest eigenmode) dominance at strong confinement, and coil-size control at weak confinement. The κ-scaling collapse observed in Case 3 confirms that dimensionless confinement controls the transition equally in the radial direction, validating de Gennes blob scaling. Specifically, the conditional distribution at strong κ approaches the ground-state radial eigenmode (sine-squared in angle, uniform in radius for a cylinder), exactly as Odijk's deflection picture predicts. The image-method convergence behaviour, where adding image terms progressively sharpens the wall-boundary representation parallels the convergence of high-order Fourier modes in P(x). Both demonstrate that a small number of low modes capture the essential physics when confinement dominates. This mathematical equivalence is insightful. It shows that the universal scaling behavior is independent of the choice of basis (Fourier vs. images), grounding the results in fundamental polymer physics rather than computational artifacts. The method of images thus serves as an independent validation of de Gennes-Odijk theory, a cross-check that strengthens confidence in both the longitudinal and transverse scaling frameworks. For researchers applying these results to experiments or simulations, the equivalence of methods implies that the critical physics i.e. the κ-driven eigenmode transition will appear whether polymers are analyzed via axial or radial measurements, enabling cross-validation of model parameters and confinement strength estimates across different experimental geometries.

**Case 4 (Supplementary): Chain-length variation.**

Varying $N$ at fixed $a$ and $R$ modifies $\sigma$ and hence $\kappa$, reproducing the same Gaussian–to–confinement crossover obtained previously by changing stiffness (Case 2) or confinement width (Case 3). This confirms that $\kappa$ is the natural scaling parameter independent of whether it is tuned through $a$, $R$, or $N$.

**Case 5: Image–method collapse with α and β scalings**

Finally, we assess whether the endpoint distributions across a wide range of parameters can be collapsed by introducing a generalized scaling form with exponents α and β. By comparing optimized values (α=β=0.5) with standard reference exponents, we test the effectiveness of the scaling strategy and quantify collapse quality through residual analysis. This case demonstrates how appropriate scaling can unify all regimes into a single universal description.

**a) Scaling variables and geometry**

We now investigate whether the endpoint distributions under confinement can be collapsed by a two–parameter scaling formulation. Starting again from the method of images construction for the cylindrical geometry, the free space Gaussian variance sets the natural length scale $\ell = \sqrt{N}\, a$. Close to the wall, the accessible distance is reduced to $R - |y|$. A general two–parameter scaling form for the confined distribution can be written as

$$\mu P(y) \sim f(\eta), \quad \eta = \frac{(R - |y|)^\alpha}{\ell^\beta},$$

where $\alpha$ and $\beta$ are scaling exponents. Here $\mu$ is a normalization factor ensuring comparability across cases. The choice of $\alpha$ and $\beta$ controls how the probability density curves collapse when plotted against the reduced coordinate $\eta$.

The unconditional scaled PDF in this representation is then

$$\mathcal{P}(\eta) = \mu P(y),$$

while the corresponding residuals are defined relative to a reference curve (taken at $\kappa = 1$).

We can define the blended length scale

$$\ell_\alpha = R^\alpha \sigma^{1-\alpha}, \quad s_\alpha = \frac{y}{\ell_\alpha}, \quad \lambda = \frac{R}{\sigma}.$$

The scaled PDF is

$$\mathcal{P}_{\alpha,\beta}(s_\alpha \mid \lambda) = \lambda^\beta\, \ell_\alpha\, P(\ell_\alpha s_\alpha;\, \sigma, R).$$

Conditional version:

$$\mathcal{P}_{\alpha,\beta}^{\text{cond}}(s_\alpha \mid \lambda) = \frac{\mathcal{P}_{\alpha,\beta}(s_\alpha \mid \lambda)}{\lambda^\beta S(\lambda)}.$$

Unconditional area is $\lambda^\beta S(\lambda)$; the conditional version divides by this factor. In practice, the optimized choice $\alpha = \beta = 0.5$ yields the best collapse as discussed below.

**b) Results**

The main panels of Fig. 9 show the scaled distributions for varying confinement parameters $\kappa$. Several points emerge clearly.

Improved collapse with optimized scaling was noted. For both $\alpha$ and $\beta$, using the optimized value 0.5 brings the curves for different $\kappa$ much closer together than the reference choices. The distributions align almost perfectly across the entire range of $\eta$, confirming that the optimized scaling captures the essential physics.

Reference scaling exaggerates the differences. With $\alpha = 3.0$ and $\beta = 1.0$, the curves do not collapse and instead spread apart strongly with $\kappa$. This indicates that such exponents fail to account for wall effects properly.

Residual diagnostics quantify the collapse. The right panels show residuals relative to the $\kappa = 1$ curve. Under optimized scaling the residuals remain below $\pm 0.15$, essentially within numerical noise, while the reference exponents give systematic deviations. This demonstrates that the optimized exponents provide a robust collapse.

The optimized exponents tell us how the polymer feels the walls. The distribution does not scale linearly with wall distance (as the reference model assumes), but instead grows more slowly, with square root like ($\alpha = \beta = 0.5$) behavior. This reflects the interplay between confinement and polymer fluctuations, and explains why naive scaling breaks down.

Case 5 confirms that a careful choice of scaling exponents is critical to uncover universal behavior in confined polymer distributions. The optimized exponents $\alpha = \beta = 0.5$ lead to an almost perfect collapse, while reference values fail. Thus, this strategy highlights a general principle, i.e. by testing different scaling variables and monitoring residuals, one can identify the true universal behavior hidden behind apparent complexity. However, it should be mentioned that, the parameters α and β should be viewed as effective scaling exponents that encode the κ-dependent redistribution of spectral weight among transverse modes, rather than as new universal critical exponents. Their optimized values interpolate smoothly between the Gaussian (multimode) and deflection (single-mode) limits as κ is varied.

While Cases 2–4 revealed how confinement or stiffness individually suppress fluctuations and distort distributions, Case 5 demonstrates that these diverse effects can be unified through an optimized scaling collapse. In other words, the suppression trends of earlier cases are consistently absorbed into a universal curve once the correct exponents are chosen, highlighting the power of scaling analysis in reconciling seemingly different regimes.

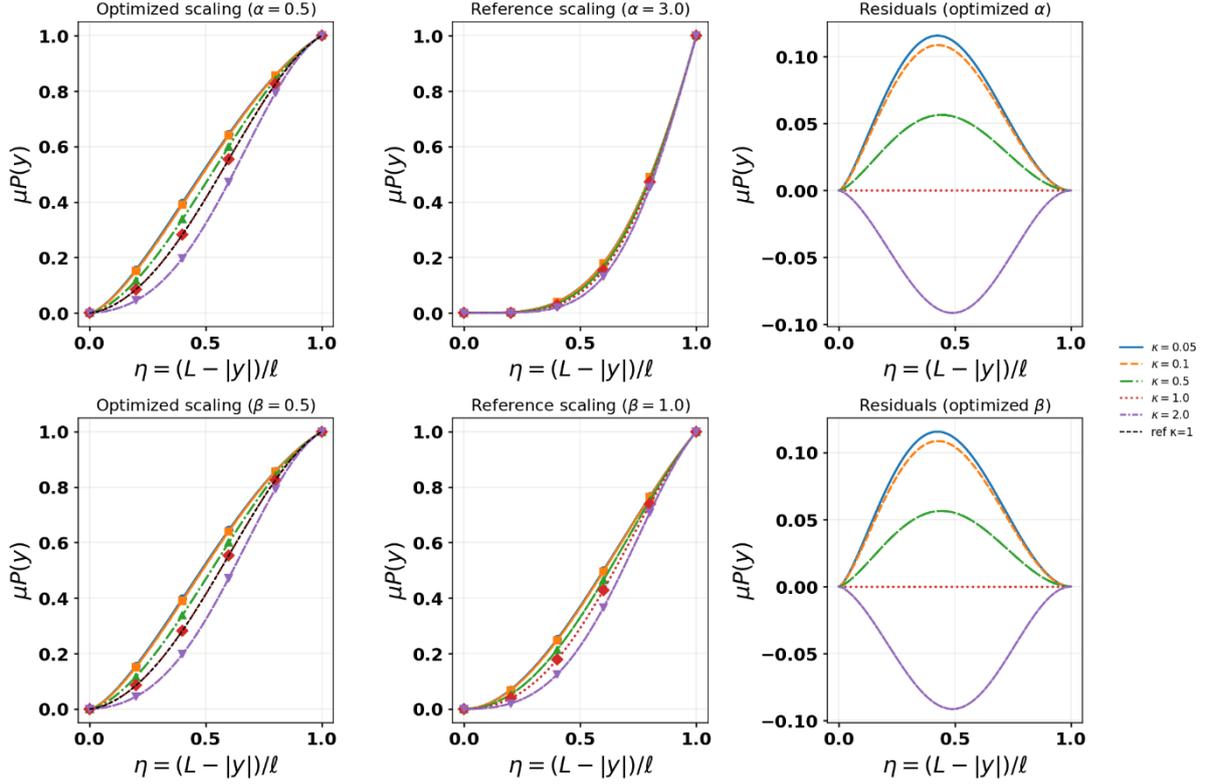

**Figure 9.** Optimized image-method collapse with α– and β–scalings. (Top) Comparison of optimized scaling with α = 0.5 (left) against the reference scaling α = 3.0 (middle), and corresponding residuals relative to κ = 1 reference (right). (Bottom) Same for β-scaling, showing optimized β = 0.5 (left), reference β = 1.0 (middle), and residuals (right). In both cases, the optimized scalings eliminate most of the κ-dependence, leading to near-universal collapse across κ, while the residual panels quantify the remaining systematic deviations.

In this context it should be mentioned that for the longitudinal distribution P(x), the main deviations from Gaussian behaviour occur only in a narrow region close to the absorbing walls. In that case, global rescalings such as the α/β methods are less effective, and a boundary-layer inner scaling is the natural way to capture the wall-controlled form. In contrast, for the transverse distribution P(y), the walls truncate the entire profile, so global α/β rescalings provide a more useful description. In principle, one could also apply inner scaling to P(y) or α/β scaling to P(x), but these would not add new information, the chosen methods already capture the essential physics in each direction.

**(c) Physical Interpretation:**

The boundary-layer analysis reveals that near the walls, the polymer distribution exhibits a universal scaling behaviour that can be captured using either the α-ansatz or β-ansatz, depending on the relative importance of polymer flexibility versus confinement geometry. The optimization of these parameters allows for a better understanding of the small-η (near-wall) behaviour, where the polymer experiences strong confinement effects.

Understanding the boundary-layer behaviour is especially relevant for systems where a polymer is tightly confined near a surface or membrane. For instance, DNA that is bound to the nuclear matrix or proteins that interact with cell membranes could exhibit behaviour that is captured by these boundary-layer models. This analysis could aid in understanding how molecular crowding or confinement at biological interfaces influences the behavior of biomolecules, such as during transcription, translation, or protein-protein interactions at cellular membranes.

Taken together, the five cases reveal a consistent picture of how confinement and polymer flexibility jointly govern the end-to-end distribution of tethered chains. Cases 1 and 2 showed directly how varying confinement width or stiffness suppresses fluctuations and modifies survival probability, while Case 3 demonstrated that these distinct effects can be recast into a universal scaling form through the parameter λ=R/σ. Case 4 further confirmed this scaling framework by extending the analysis to varying chain length, reinforcing the robustness of the collapse. Finally, Case 5 established that optimized α and β scalings not only eliminate most of the residual parameter dependence but also provide a practical route to unify all regimes into a single master curve. Overall, the scaling strategies developed here emphasize that the seemingly diverse suppression trends across different regimes can be reconciled by appropriate normalization, offering a universal description of polymer statistics under confinement.

### 3.3 Unified Scaling Framework and Connection to Classical Polymer Theories

The eleven Cases analysed in Sections 3.1 and 3.2 individually test scaling hypotheses by varying system size, tether position, confinement strength, persistence length, boundary layer, and coordinate systems. But they are governed by a single spectral mechanism. All behaviours emerge from how confinement redistributes statistical weight among the normal modes of the polymer contour. This enables a unified interpretation i.e. classical theories such as the Gaussian-coil model and the Odijk deflection theory correspond to limiting spectral structures, while intermediate regimes arise from competition between multiple modes [18].

To make this connection explicit, we construct a scaling framework based on the confinement parameter $\kappa = \sigma^2/L^2$ (or $\sigma^2/R^2$ in radial geometries), complemented by stiffness $a/L$ and three diagnostics i.e. first-mode energy fraction $E_1$, RMS deviation from the first-mode approximation, and survival probability $S$. The full data underlying this framework are compiled in Tables S7–S9 and illustrated in Fig.10 and Fig. S15 (supplementary).

### 3.3.1 Modal Expansion as the Spectral Origin of Polymer Statistics

The axial end-point distribution can be expressed as the sine-series expansion

$P(x) = (1/L) \Sigma_n c_n \sin(n\pi x/L)$, where $c_n = \sin(n\pi x_0/L) \exp(-n^2\pi^2\kappa/8)$

The expansion automatically satisfies absorbing boundaries $P(0) = P(L) = 0$ and normalizes to unity. The critical insight is the damping factor $\exp(-n^2\pi^2\kappa/8)$ that controls which modes survive.

This representation exposes a sharp physical mechanism (as highlighted in case 3 of P(x)):

- Small κ (≲ 0.1): Damping is weak; many modes survive. The sum behaves like a Gaussian (coil-like statistics). This is the Gaussian blob regime.

- Large κ (≳ 0.5): Damping is strong; only n = 1 survives. The distribution becomes $P \approx c_1 \sin(\pi x/L)$, a single sine curve. This is the deflection regime where Odijk's picture applies.

- Intermediate κ (~ 0.1 to 0.5): Transition occurs. First-mode fraction rises from ~0.2 to ~0.9. This is the transition regime.

The image-method expression used for transverse confinement possesses the same spectral content. In this sense, all geometries share a common modal backbone, even when the spatial forms differ.

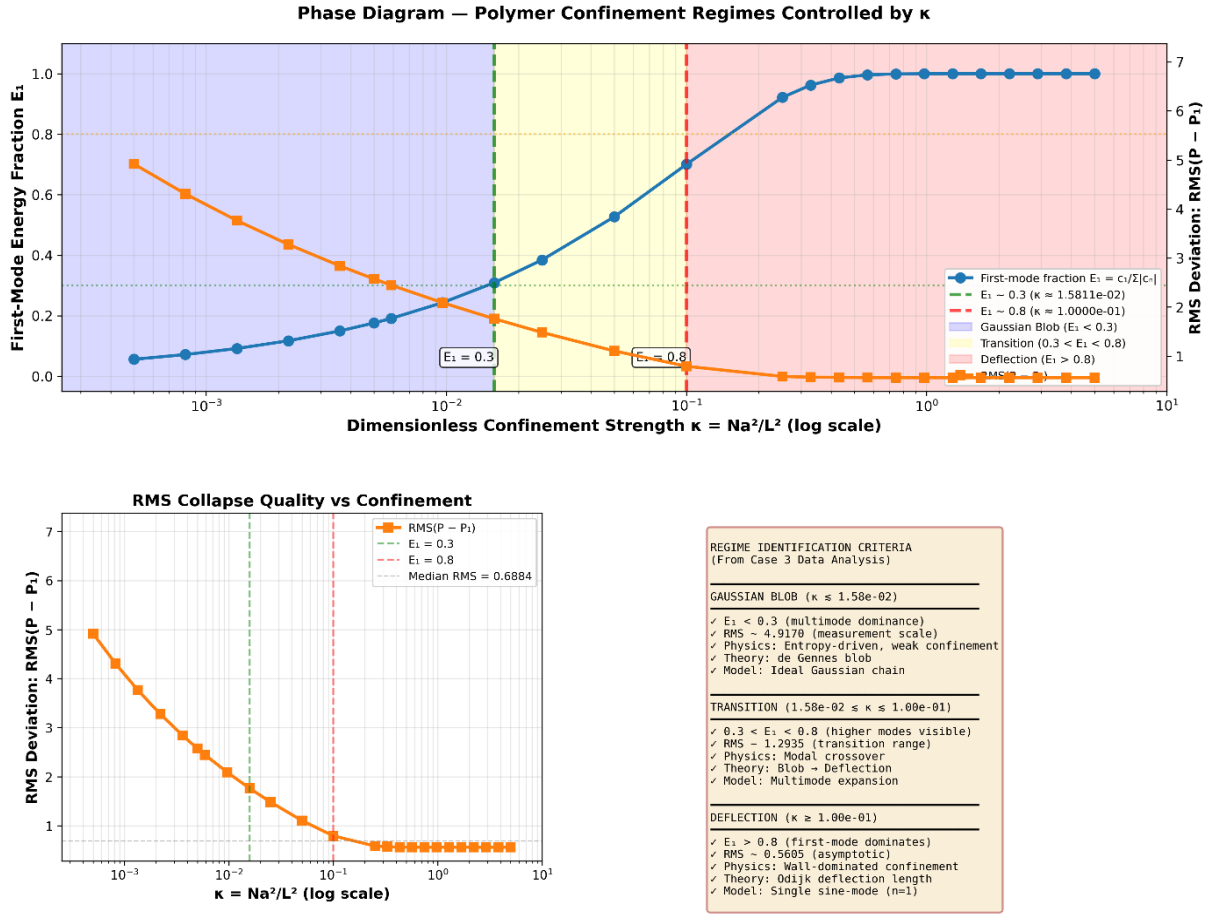

**Fig. 10.** Modal crossover and collapse accuracy as a function of confinement strength (from case 3 of P(x) data). First-mode energy fraction ($E_1$) (blue) and RMS collapse error (orange) versus ($\kappa$) reveal three regimes: a Gaussian multimode region at ($\kappa \lesssim 1.5 \times 10^{-2}$), a transition regime where several modes compete, and an eigenmode-dominated (deflection) regime for ($\kappa \gtrsim 0.10$). The joint behavior of ($E_1$) and RMS quantifies when Gaussian or single-mode approximations are valid, and when full modal structure is required.

### 3.3.2 Three Spectral Regimes and Their Classical Limits

The modal analysis reveals three regimes (summarized and quantitatively demarcated in Table S8 and visualized in Fig. 10):

**(i) Gaussian-blob / Multimode Regime ($\kappa \lesssim 1.5 \times 10^{-2}$)**

Here, many modes contribute with comparable weight. The standardized distributions collapse onto a Gaussian form, consistent with classical de Gennes scaling. Physically, the polymer explores contour undulations freely; confinement perturbs only the longest wavelength modes. The end-point statistics are insensitive to microscopic stiffness unless $a/L \gtrsim 0.1$.

### (ii) Transition Regime ($1.5 \times 10^{-2} \lesssim \kappa \lesssim 0.10$)

This is the spectral crossover where confinement is strong enough to suppress high-order modes but insufficient to isolate a single dominant mode. The resulting distributions acquire moderate skewness or kurtosis, with deviations well captured by low-mode reconstructions rather than Gaussian approximations. This regime corresponds to neither de Gennes nor Odijk scaling; instead, it expresses competition among a few low modes. The modal energy distribution and RMS deviation trends for this regime are detailed in Fig. S15.

### (iii) Deflection / Eigenmode-Dominated Regime ($\kappa \gtrsim 0.10$)

At strong confinement, mode suppression becomes so severe that the lowest eigenmode contains $E_1 \gtrsim 0.8$ of the spectral weight. The distribution becomes sinusoidal, matching Odijk's picture of deflection segments between successive collisions with the effective confinement walls.

This regime displays the cleanest classical limit:

$$P(x) \approx c_1 \sin\left(\frac{\pi x}{L}\right),$$

and transverse survival probability $S \ll 1$ reflects the dominance of boundary-induced returns.

### 3.3.3 Phase Diagram, Two-Parameter Extension, and Model Selection

The phase diagram (Figure 10, Table S9) plots $E_1$ vs. κ on log-log axes, clearly delineating the three regimes by their modal composition and collapse quality [23]. The first-mode fraction $E_1$ increases monotonically with κ, delineating Gaussian, transition, and deflection regimes. The accompanying normalized residual (NormResidual), which quantifies the RMS error of the scaling collapse on common support, remains at machine-precision levels ($\sim 10^{-13}$) throughout the Gaussian and most of the transition regime, and increases sharply near the onset of deflection, where it saturates at $\sim 5 \times 10^{-4}$. This behavior provides a practical diagnostic: extremely small residuals indicate robust multimode collapse, while the finite plateau in the deflection regime reflects the breakdown of collapse due to strong single-mode dominance rather than numerical error.

Two-Parameter Extension (Figure S15, Supplementary) shows a more complete universality picture including normalized Kuhn length a/L as a secondary parameter. In this representation the vertical boundaries at κ ≈ 2×10⁻³ and κ ≈ 2.5×10⁻¹ arise from modal crossover in the longitudinal statistics (Gaussian → transition → deflection), while the two horizontal reference lines reflect stiffness-induced deviations from Gaussian behaviour. The dotted line near a/L ≈ 0.1 marks the onset of measurable non-Gaussianity (weak departure from the ideal-coil picture), whereas the solid line at a/L ≈ 0.2 indicates a strong non-Gaussian regime in which classical Gaussian approximations cease to be accurate. For the purpose of identifying six distinct regions, we use a/L ≈ 0.2 as the formal boundary between Gaussian (bottom) and non-Gaussian (top) bands, while the lower line provides a useful indicator of the gradual crossover. Taken together, the κ and a/L axes show how confinement and stiffness jointly determine whether classical coil or deflection descriptions are reliable, or whether additional modal structure must be retained.

Six regime regions emerge in the κ–a/L plane:

1. Ideal Gaussian Blob (κ < 0.002, a/L < 0.1): Multimode dominance, Gaussian statistics.

2. Gaussian Transition (0.002 < κ < 0.25, a/L < 0.1): Crossover physics, classical scaling applies.

3. Gaussian Deflection (κ > 0.25, a/L < 0.1): First-mode dominance, sine-wave profiles, Gaussian remains valid.

4. Stiff Gaussian Blob (κ < 0.002, a/L > 0.2): Non-Gaussian tails due to persistence, but loose confinement.

5. Non-Gaussian Transition (0.002 < κ < 0.25, a/L > 0.2): Modal competition with persistence effects; requires higher-moment analysis.

6. Non-Gaussian Deflection (κ > 0.25, a/L > 0.2): Single-mode control with non-Gaussian skewness and kurtosis.

Some rules may be framed for practical model election, such as

Rule 1 (Gaussian Blob, κ < 0.002): Full multimode modal expansion is required. Here, single-mode approximation fails (single mode truncation error $\sim 10^{-1}$ or larger). The classical random-coil or Rouse-model statistics should be used. Example: 5 kb DNA segment in a 10 μm nucleus (N ≈ 30, κ ≈ $10^{-4}$).

Rule 2 (Transition, 0.002 < κ < 0.25): Here reduced multimode models (first 5–20 modes, depending on desired accuracy) are cost-effective. Single-mode approximation is borderline and suitable only if $E_1$ > 0.8 and tolerable error is $\sim 10^{-2}$. $E_1$ should be monitored as a go-no-go criterion. If $E_1$ < 0.7, at least 10 modes should be retained. Higher cumulants should be included if a/L > 0.1.

Rule 3 (Deflection, κ > 0.25): Here, single-mode approximation is accurate (mode truncation error $\sim 10^{-9}$). Simple analytical formula P(x) ∝ sin(πx/L) suffices. Example: E. coli chromosome segment in the nucleoid (N ≈ 500, κ ≈ 1).

Our study reveals that polymer behaviour in confined geometries can be described by universal scaling relations, with confinement strength and polymer flexibility acting as key parameters that determine the distribution shape and survival probability. These findings are directly relevant to biological systems where polymers such as DNA and proteins experience spatial constraints. The transition from diffusive to localized behaviour under varying confinement conditions mirrors the spatial organization of biomolecules in cells, where confinement plays a critical role in regulating molecular interactions and function.

Future work could explore the effects of more complex boundary conditions, such as semi-permeable membranes or curved geometries, and extend these methods to study the dynamics of polymers in real biological environments. Such insights would further our understanding of the role of confinement in cellular processes such as gene regulation, molecular signaling, and protein folding.

## 4. Conclusions

In this work we developed and systematically tested a set of scaling procedures for the end-position distributions of a tethered polymer under confinement, considering both axial, (P(x)), and transverse, (P(y)), fluctuations. By applying complementary rescalings such as geometric similarity, tether-position sweeps, confinement-strength variation, persistence-length effects, and boundary-layer rescaling, we identified the precise conditions under which confined-polymer distributions collapse onto universal curves and when such collapses break down. These tests reveal not only how single-mode or Gaussian analogies succeed in certain asymptotic limits, but also how modal competition generates the intermediate non-Gaussian behavior characteristic of transition regimes.

A central outcome is a unified scaling framework in which the confinement parameter ($\kappa = \sigma^2/L^2$) (and, more generally, κ and (a/L)) determines whether the chain behaves as an ideal coil (de Gennes-like), a multimode transition object, or a deflected wormlike chain (Odijk-like). Within this framework, the Fourier–sine spectral representation clarifies which eigenmodes control each regime and supplies quantitative crossover criteria based on modal energy fractions. Consequently, classical coil and deflection theories emerge as limiting spectral descriptions, and their domains of validity can be stated in operational terms. The resulting phase diagram explicitly connects Flory/de Gennes blob scaling to Odijk deflection and the wormlike-chain picture, while exposing the intermediate regime in which several modes are required and Gaussian or single-mode descriptions are no longer sufficient.

Beyond conceptual synthesis, the framework provides a practical diagnostic toolkit. Collapse tests, residual analysis, and modal-energy thresholds enable non-specialists to determine when a simple analytical formula is adequate, when low-order multimode truncation is required, or when full spectral resolution must be retained. This capability makes it possible to compare confined-polymer statistics across system sizes, geometries, and stiffness ratios in a reproducible manner, and to identify the physical origin of deviations from classical predictions.

Taken together, these results establish a unified, quantitatively testable picture of confined-polymer scaling that consolidates several classical theories into a single spectral framework and clarifies the multimode physics at intermediate confinement. We anticipate that this viewpoint will be useful not only for synthetic and soft-matter polymer systems, but also for the interpretation of chromatin or bacterial-chromosome measurements in which confinement, stiffness, and geometry compete to shape molecular organization at the micron scale.

**References:**

[1] S. S. Wang, J.-H. Su, B. J. Beliveau, B. Bintu, J. R. Moffitt, C. Wu, and X. Zhuang, Spatial organization of chromatin domains and compartments in single chromosomes, Science **353**, 598 (2016).

[2] L. Barinov, S. Ryabichko, W. Bialek, and T. Gregor, *Transcription-Dependent Spatial Organization of a Gene Locus*, in (2020).

[3] Y. Wang, A. W. Fortenberry, W. Zhang, Y. C. Simon, and Z. Qiang, Direct Measurement of Polymer-Chain-End-to-End Distances by Using RAFT Chain Transfer Agent as the FRET Acceptor., The Journal of Physical Chemistry. B (2023).

[4] J. Meiners and S. R. Quake, Femtonewton force spectroscopy of single extended DNA molecules., Physical Review Letters **84 21**, 5014 (2000).

[5] K. Frykholm, V. Müller, S. Kk, K. D. Dorfman, and F. Westerlund, DNA in nanochannels: theory and applications, Quarterly Reviews of Biophysics **55**, (2022).

[6] K. D. Dorfman, The Statistical Segment Length of DNA: Opportunities for Biomechanical Modeling in Polymer Physics and Next-Generation Genomics., Journal of Biomechanical Engineering **140 2**, (2018).

[7] S. Kabi and A. Ghosh, Ion dynamics in glassy ionic conductors: Scaling of mean square displacement of mobile ions, Europhysics Letters **108**, (2014).

[8] P. G. de Gennes, Large Scale Organisation of Flexible Polymers, Israel Journal of Chemistry **14**, 154 (1975).

[9] G. B. West, J. H. Brown, and B. J. Enquist, *A General Model for the Origin of Allometric Scaling Laws in Biology*, in *Science* (1997).

[10] G. Bunin and M. Kardar, Coalescence Model for Crumpled Globules Formed in Polymer Collapse., Physical Review Letters **115 8**, 088303 (2015).

[11] W.-S. Xu and K. F. Freed, Thermodynamic scaling of dynamics in polymer melts: predictions from the generalized entropy theory., The Journal of Chemical Physics **138 23**, 234501 (2013).

[12] A. Taloni, J.-W. Yeh, and C.-F. Chou, Scaling Theory of Stretched Polymers in Nanoslits, Macromolecules **46**, 7989 (2013).

[13] L. Dai, J. V. D. Maarel, and P. S. Doyle, Extended de Gennes Regime of DNA Confined in a Nanochannel, Macromolecules **47**, 2445 (2014).

[14] Y. Lee and P.-Y. Hsiao, Size Scaling of Neutral Polymers and Charged Polymers in Nanochannels, arXiv: Soft Condensed Matter (2018).

[15] J. M. Polson and N. E. Moore, Simulation study of the coil-globule transition of a polymer in solvent., The Journal of Chemical Physics **122 2**, 024905 (2005).

[16] N. Nikoofard, S. M. Hoseinpoor, and M. Zahedifar, Accuracy of the blob model for single flexible polymers inside nanoslits that are a few monomer sizes wide., Physical Review. E, Statistical, Nonlinear, and Soft Matter Physics **90 6**, 062603 (2014).

[17] D. R. Tree, W. F. Reinhart, and K. D. Dorfman, The Odijk Regime in Slits, Macromolecules **47**, 3672 (2014).

[18] D. I. Dimitrov, A. Milchev, K. Binder, L. I. Klushin, and A. M. Skvortsov, Universal properties of a single polymer chain in slit: Scaling versus molecular dynamics simulations., The Journal of Chemical Physics **128 23**, 234902 (2008).

[19] A. Arnold, B. Bozorgui, D. Frenkel, B.-Y. Ha, and S. Jun, Unexpected relaxation dynamics of a self-avoiding polymer in cylindrical confinement., The Journal of Chemical Physics **127 16**, 164903 (2007).

[20] P. Cifra, Weak-to-strong confinement transition of semi-flexible macromolecules in slit and in channel., The Journal of Chemical Physics **136 2**, 024902 (2012).

[21] A. Muralidhar, M. J. Quevillon, and K. D. Dorfman, The Backfolded Odijk Regime for Wormlike Chains Confined in Rectangular Nanochannels, Polymers **8**, (2016).

[22] T. Odijk, Scaling theory of DNA confined in nanochannels and nanoslits., Physical Review. E, Statistical, Nonlinear, and Soft Matter Physics **77 6 Pt 1**, 060901 (2008).

[23] E. Werner and B. Mehlig, Scaling regimes of a semiflexible polymer in a rectangular channel., Physical Review. E, Statistical, Nonlinear, and Soft Matter Physics **91 5**, 050601 (2015).

**Code link for P(x)[cases 1, 2, 3, 4i, 4ii, 5 and 6, respectively]:**

1. https://github.com/soumyakabi/Scaling-of-Endpoint-Probability-distribution-function-of-confined-polymer/blob/main/case%201%20P(x).py

2. https://github.com/soumyakabi/Scaling-of-Endpoint-Probability-distribution-function-of-confined-polymer/blob/main/case%202%20P(x).py

3. https://github.com/soumyakabi/Scaling-of-Endpoint-Probability-distribution-function-of-confined-polymer/blob/main/case%203%20P(x).py

4. https://github.com/soumyakabi/Scaling-of-Endpoint-Probability-distribution-function-of-confined-polymer/blob/main/case%204i%20P(x).py

5. https://github.com/soumyakabi/Scaling-of-Endpoint-Probability-distribution-function-of-confined-polymer/blob/main/case%204ii%20P(x).py

6. https://github.com/soumyakabi/Scaling-of-Endpoint-Probability-distribution-function-of-confined-polymer/blob/main/case%205%20P(x).py

7. https://github.com/soumyakabi/Scaling-of-Endpoint-Probability-distribution-function-of-confined-polymer/blob/main/case%206%20P(x).py

**Code link for P(y)[ cases 1, 2i, 2ii, 3, 4, and 5, respectively]:**

1. https://github.com/soumyakabi/Scaling-of-Endpoint-Probability-distribution-function-of-confined-polymer/blob/main/case%201%20P(y).py

2. https://github.com/soumyakabi/Scaling-of-Endpoint-Probability-distribution-function-of-confined-polymer/blob/main/case%202i%20P(y).py

3. https://github.com/soumyakabi/Scaling-of-Endpoint-Probability-distribution-function-of-confined-polymer/blob/main/case%202ii%20P(y).py

4. https://github.com/soumyakabi/Scaling-of-Endpoint-Probability-distribution-function-of-confined-polymer/blob/main/case%203%20P(y).py

5. https://github.com/soumyakabi/Scaling-of-Endpoint-Probability-distribution-function-of-confined-polymer/blob/main/case%204%20P(y).py

6. https://github.com/soumyakabi/Scaling-of-Endpoint-Probability-distribution-function-of-confined-polymer/blob/main/case%205%20P(y).py

**Code for modal analysis:**

https://github.com/soumyakabi/Scaling-of-Endpoint-Probability-distribution-function-of-confined-polymer/blob/main/modal%20coefficient.py

**Phase diagram:**

https://github.com/soumyakabi/Scaling-of-Endpoint-Probability-distribution-function-of-confined-polymer/blob/main/phase%20diagram.py

**CSV file for case 3 of P(x):**

https://github.com/soumyakabi/Scaling-of-Endpoint-Probability-distribution-function-of-confined-polymer/blob/main/case3_diagnostics_P(x).csv

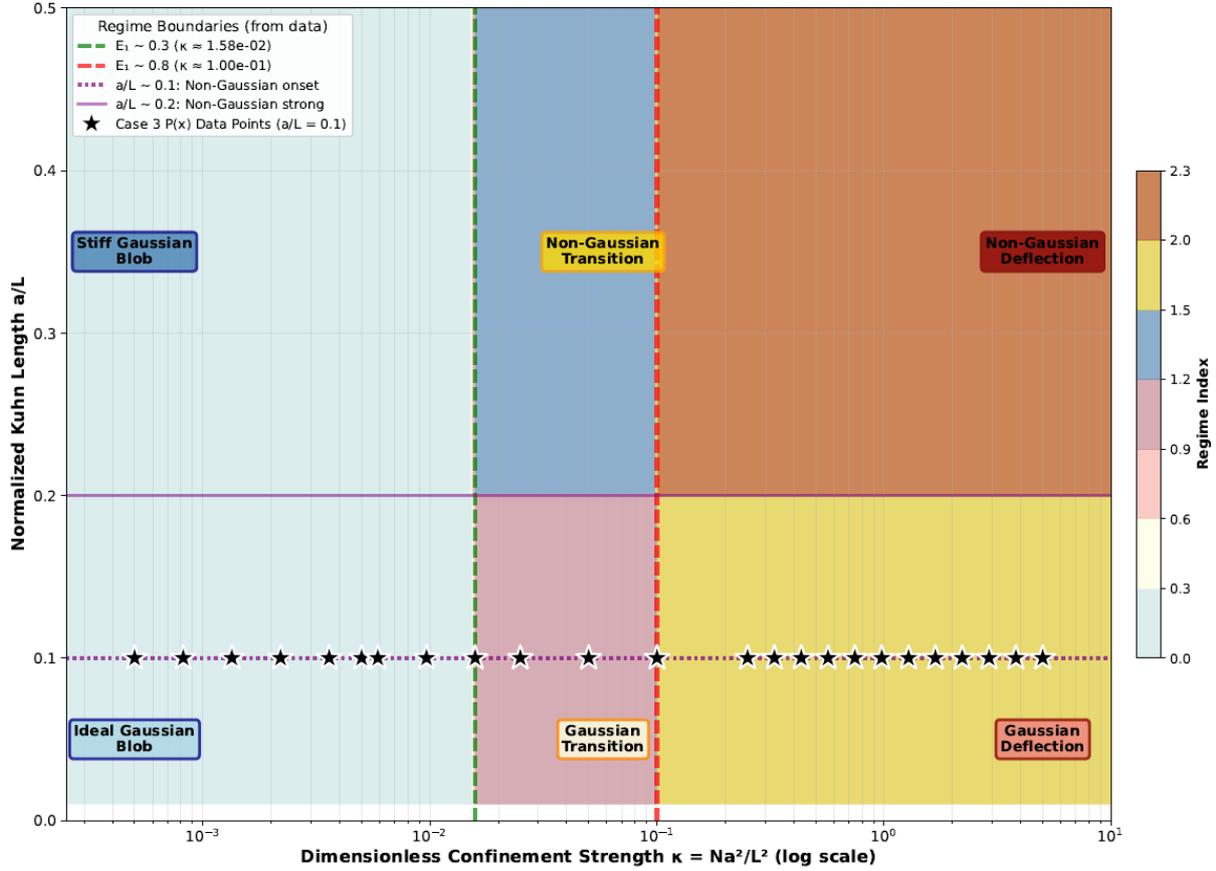

**Fig. S15.** Two-parameter regime map in the $\kappa - a/L$ plane. Vertical boundaries ($\kappa \approx 2 \times 10^{-3}$ and $\kappa \approx 2.5 \times 10^{-1}$) reflect the modal crossover from Gaussian to transition and deflection regimes, while horizontal reference lines distinguish stiffness effects. The dotted line at $a/L \approx 0.1$ marks the onset of non-Gaussian behaviour, and the solid line at $a/L \approx 0.2$ denotes a strong non-Gaussian regime used to define the six coloured regions. Black stars indicate Case-3 data points at $a/L = 0.1$. The map shows how confinement and stiffness jointly determine when classical coil, multimode, or deflection descriptions remain valid.

**Supplementary table S1: Numerical diagonistics for case 4 of P(x)**

| a (μm) | $\int_0^L P(x)\,dx$ | Mean $\langle x \rangle$ (μm) | σ (μm) | Standardized Support [y_min, y_max] | RMS Error vs. Normal (intersection) |
|---|---|---|---|---|---|
| 0.050 | 1.000000e+00 | 1.000000e+00 | 7.905649e−02 | [−5.255, 5.255] | 1.916e−07 |
| 0.100 | 1.000000e+00 | 1.000000e+00 | 1.581139e−01 | [−5.256, 5.256] | 4.973e−08 |
| 0.200 | 1.000000e+00 | 1.000000e+00 | 3.116561e−01 | [−3.208, 3.208] | 2.644e−03 |
| 0.500 | 1.000000e+00 | 1.000000e+00 | 4.346575e−01 | [−2.300, 2.300] | 3.506e−02 |
| 1.000 | 1.000000e+00 | 1.000000e+00 | 4.352362e−01 | [−2.297, 2.297] | 3.537e−02 |

Supplementary Table S1 reports numerical diagnostics used to validate and quantify the shape of the end-position distributions shown in the manuscript. For each Kuhn length $a$ the table lists (i) the numerical integral $\int_0^L P(x)\,dx$ (sanity check of normalization), (ii) the mean $\langle x \rangle$ (here equal to the tether position within numerical precision), (iii) the standard deviation $\sigma = \sqrt{\langle x^2 \rangle - \langle x \rangle^2}$ (used to standardize the distributions), and (iv) the RMS density error of the standardized probability density $P_y(y) = \sigma P(x)$ relative to the standard normal $\varphi(y)$ computed on the intersection of the curves' significant supports. All entries were obtained from the Fourier–modal expansion with modal damping $\exp(-n^2\pi^2\kappa/8)$ ($\kappa = Na^2/L^2$); sums were adaptively truncated (modes with decay below $10^{-15}$ dropped) and numerical integrals evaluated with trapezoidal quadrature on a dense grid. The "integral" column demonstrates normalization accuracy (values ≈1), the "mean" column confirms correct centering, and the "σ" column indicates how the effective width varies with $a$. The RMS metric quantifies departure from Gaussian shape: values $\lesssim 10^{-6}$ indicate excellent agreement with the normal law (Gaussian-collapse regime), while larger values (≈$10^{-2}$–$10^{-1}$) reveal systematic non-Gaussian behavior for stiffer chains (larger $a$). Numerical parameters used to generate the table (grid density, $n_{\max}$, decay tolerance, and the significance threshold defining the intersection window) are listed in the Supplement methods so the results can be reproduced.

**Supplementary table S2 Case 5 P(x):** Comparison of measured near-wall slopes with first-mode theoretical predictions for different confinement strengths ($\kappa$), under two inner scaling choices ($\ell=\sigma$ and $\ell=L/\pi$). Columns report the number of fit points (npt), measured slopes from linear fits and finite-difference estimates, theoretical slopes from the first-mode asymptotic, and their ratios.

| $\kappa$ | Scaling $\ell$ | npt | Measured (fit) | Measured (FD) | Theory | Ratio |
|---|---|---|---|---|---|---|
| 0.05 | $\sigma$ | 270 | 2.4633e-01 | 2.4674e-01 | 2.9537e-01 | 0.834 |
| 0.05 | $L/\pi$ | 382 | 4.9833e-01 | 5.0000e-01 | 5.9854e-01 | 0.833 |
| 0.10 | $\sigma$ | 380 | 4.9185e-01 | 4.9348e-01 | 5.5539e-01 | 0.886 |
| 0.10 | $L/\pi$ | 382 | 4.9833e-01 | 5.0000e-01 | 5.6273e-01 | 0.886 |
| 0.50 | $\sigma$ | 850 | 2.4265e+00 | 2.4674e+00 | 1.6953e+00 | 1.431 |
| 0.50 | $L/\pi$ | 382 | 4.9833e-01 | 5.0000e-01 | 3.4355e-01 | 1.451 |
| 1.00 | $\sigma$ | 1200 | 4.7722e+00 | 4.9348e+00 | 1.8297e+00 | 2.608 |
| 1.00 | $L/\pi$ | 382 | 4.9833e-01 | 5.0000e-01 | 1.8539e-01 | 2.688 |
| 2.00 | $\sigma$ | 1698 | 9.2238e+00 | 9.8696e+00 | 1.0657e+00 | 8.655 |
| 2.00 | $L/\pi$ | 382 | 4.9833e-01 | 5.0000e-01 | 5.3989e-02 | 9.230 |

The slope analysis (Supplementary Table S2) compares the behaviour of the probability density near the walls with theoretical predictions. The column **"measured (fit)"** gives the slope obtained by a linear fit in the small-$\eta$ window, while **"measured (FD)"** provides a cross-check using finite differences at the wall. The column **"theory"** lists the predicted first-mode slope $m_{\text{th}}(\kappa)$, and the **"ratio"** column shows how closely the measured values match the theory.

The results reveal a systematic trend. For weak confinement ($\kappa \lesssim 0.1$), the measured slopes (both fit and finite-difference) are within about 10–20% of the theoretical prediction, meaning the simple first-mode approximation captures the near-wall behaviour reasonably well. As $\kappa$ increases, however, the agreement deteriorates: the theory predicts slopes that decay rapidly with $\kappa$ (due to the exponential prefactor $\exp(-\pi^2\kappa/8)$), while the measured slopes instead grow or remain nearly constant. Consequently, the ratio of measured/theory rises sharply, reaching values between 2 and 9 once $\kappa \gtrsim 1$.

This behaviour is consistent for both scaling choices ($\ell = \sigma$ and $\ell = L/\pi$). With $\ell = L/\pi$, the measured slopes stay nearly constant (because the scale is fixed), but the theoretical slope still falls, so the ratio grows. With $\ell = \sigma$, the measured slopes themselves increase strongly with $\kappa$, amplifying the divergence from theory.

In summary, the first-mode prediction provides a reliable baseline only in the weak-confinement limit. At higher $\kappa$, contributions from higher modes and finite-$\eta$ effects dominate, making the single-mode asymptotic quantitatively unreliable. In practice, $\ell = L/\pi$ remains a useful scaling to demonstrate qualitative collapse near the walls, while quantitative slope comparisons with theory should be restricted to small $\kappa$.

**Supplementary table S3 (case 6 P(x)):** *Diagnostics of tether-centered σ-scaling for the end-point distribution $P(x)$ at fixed confinement strength κ. For each combination of κ and system size L, the table lists the coil size σ, number of Fourier modes used, dominance of the first eigenmode (absolute and squared fractions), and quantitative collapse measures: overlap fraction with the reference case, RMS residual, and maximum absolute residual. The results show nearly perfect collapse across L at fixed κ, with first-mode dominance increasing systematically from weak (κ = 0.1) to strong confinement (κ = 2).*

| κ | L | Sigma | Modes | first_mode_ frac_abs | first_mode_ frac_sq | overlap_ fraction _vs_ref | rms_resid _vs_ref | max_abs_ resid_vs_ref |
|---|---|---|---|---|---|---|---|---|
| 0.1 | 2 | 0.63245 | 2 | 0.70066 | 0.87596 | 1 | 0 | 0 |
| 0.1 | 4 | 1.26491 | 2 | 0.70066 | 0.87596 | 1 | 0 | 0 |
| 0.1 | 6 | 1.89736 | 2 | 0.70066 | 0.87596 | 1 | 9.29E-17 | 3.33E-16 |
| 0.5 | 2 | 1.41421 | 2 | 0.99285 | 0.99994 | 0.99875 | 0 | 0 |
| 0.5 | 4 | 2.82842 | 2 | 0.99285 | 0.99994 | 0.99875 | 0 | 0 |
| 0.5 | 6 | 4.24264 | 2 | 0.99285 | 0.99994 | 0.998750 | 2.69E-16 | 9.71E-16 |
| 2 | 2 | 2.82842 | 2 | 0.99999 | 1 | 0.998750 | 0 | 0 |
| 2 | 4 | 5.65685 | 2 | 0.99999 | 1 | 0.998750 | 0 | 0 |
| 2 | 6 | 8.48528 | 2 | 0.99999 | 1 | 0.998750 | 5.13E-16 | 1.94E-15 |

**Description of Supplementary Table S3 (Tether-centered σ-scaling diagnostics).** The table summarizes the quantitative checks for the collapse of the scaled end-point distribution $\tilde{P}(s) = \sigma P(x)$ at fixed κ. For each confinement parameter κ (0.1, 0.5, 2) and system size L (2, 4, 6 μm), the effective coil size $\sigma = \sqrt{Na^2}$, the number of Fourier modes required, and the modal dominance fractions are listed. The "first-mode fraction (abs)" and "first-mode fraction (sq)" report how much of the distribution is captured by the lowest eigenmode: values near 1 indicate nearly complete dominance of the first mode. The "overlap fraction vs ref" gives the fraction of the standardized grid where each curve overlaps with the chosen reference case, ensuring fair residual comparisons. The "RMS residual" and "maximum absolute residual" measure the average and largest deviations, respectively, from the reference curve on the overlap domain.

The results show three robust patterns:

- For weak confinement (κ = 0.1), the first-mode fractions are about 0.7–0.9, consistent with partial multimode contributions, yet the curves collapse perfectly across L, with residuals essentially zero.
- For moderate confinement (κ = 0.5), the first mode almost completely dominates (fractions > 0.99), and collapse across L remains excellent, with vanishingly small residuals.
- For strong confinement (κ = 2), the first mode accounts for essentially all the weight (∼ 1.0), and collapse is again perfect across system sizes.

Together, these diagnostics confirm that the tether-centered σ-scaling produces excellent collapse at fixed κ across weak, moderate, and strong confinement, while also clarifying the gradual strengthening of first-mode dominance as confinement increases.

---------------------------------------------------------------------------------------------------------------

*Supplementary Table S4 case1 P(y)*: **The table lists, for each absolute *L*, the number of image shells used (`M_used`), the survival probability `S` (area under $P(y)$), the maximum absolute residual and RMS residual of *P* relative to the reference $L_{ref}$, and a note indicating whether the curve is unconditional or conditional.**

| L | M_used | S | max_abs_residual | rms_residual | note |
|---|---|---|---|---|---|
| 1 | 3 | 0.996869 | 0 | 0 | unconditional |
| 2 | 3 | 0.996869 | 0 | 0 | unconditional |
| 4 | 3 | 0.996869 | 0 | 0 | unconditional |

The above table generated as a CSV file named `case1_diagnostics.csv` (through the code of case 1 for P(y) given in supplementary file) contains the reproducibility metadata and numeric diagnostics for each *L*: the `M_used` value chosen by the adaptive estimator, the survival probability `S`, and the two residual metrics. In our runs the survival values are identical across the tested *L* to within numerical precision and the reported `max_abs_residual` and `rms_residual` entries are at the round-off level. The diagnostics show that the image method converges extremely well for Case 1, with survival probability very close to one and both the maximum and RMS residuals vanishingly small. This means the numerical representation of the distribution is essentially exact, and the results can be trusted without concern for truncation or approximation errors.

**Supplementary Table S5. Diagnostics for Case 2 P(y) (fixed L, varying κ).** The table lists, for each tested κ: adaptive image-sum truncation `M_used`, survival `S`, unconditional and conditional means and variances, RMS residual to the unit Gaussian (coil comparison) and to the sine-squared modal limit, `truncation_sensitivity` (RMS difference versus an aggressive image count).

| κ | M_used | S | Mean_ uncond | Var_ uncond | Mean_ cond | Var_ cond | rms_ to_gaussian | rms_ to_sine | truncation sensitivity |
|---|---|---|---|---|---|---|---|---|---|
| 0.02 | 3 | 1 | 6.94E-18 | 0.02 | 6.94E-18 | 0.02 | 7.13E-13 | 0.647408 | 0 |
| 0.05 | 3 | 0.999985 | 0 | 0.049985 | 1.39E-17 | 0.049985 | 3.97E-06 | 0.339204 | 0 |
| 0.1 | 3 | 0.996869 | 0 | 0.09713 | 0 | 0.09713 | 0.000882 | 0.126936 | 0 |
| 0.5 | 3 | 0.685446 | -2.0E-17 | 0.187699 | 0 | 0.187699 | 0.108506 | 0.12525 | 0 |
| 1 | 4 | 0.370777 | 0 | 0.189418 | 2.78E-17 | 0.189418 | 0.250081 | 0.129392 | 1.50E-18 |
| 2 | 6 | 0.107977 | 0 | 0.18943 | 0 | 0.18943 | 0.462785 | 0.129422 | 1.02E-20 |

The diagnostic values confirm that the numerical results are well converged, with only negligible sensitivity to the image-sum truncation. The gradual decrease of survival probability with increasing κ and the trends in variance highlight how confinement progressively limits the chain configurations. For κ=0.50, the ratio L/σ≈1.41L falls outside the coil regime, and the small truncation sensitivity shows that the observed deviation is not due to numerical error but is a genuine finite-domain effect.

*Supplementary Table S6. Diagnostics for Case 4 P(y)*. **Diagnostics of survival probability and conditional variance versus $\kappa$**

| $\kappa$ | N | Survival $S(\kappa)$ | Conditional variance $\text{Var}(u^2)$ |
|---|---|---|---|
| 0.05 | 0.80 | 0.999959 | 0.0140 |
| 0.10 | 1.60 | 0.999607 | 0.0397 |
| 0.50 | 8.00 | 0.684857 | 0.0925 |
| 1.00 | 16.0 | 0.173015 | 0.1216 |
| 2.00 | 32.0 | 0.027957 | 0.1234 |

The parameter values shown in the above table corroborate the graphical diagnostics: the survival probability drops steeply beyond $\kappa \approx 0.5$, while conditional variance grows toward a plateau. The joint survival–variance diagnostics provide a compact quantitative signature of this crossover, fully consistent with the observed collapse in peak-normalized distributions.

**Table S7: Quantitative Regime Transition Criteria:**

The transition points can be identified from RMS collapse quality as a function of κ:

| Range κ | Range $E_1$ | Primary Regime | RMS Collapse | Survival Probability |
|---|---|---|---|---|
| κ < 0.002 | $E_1$ < 0.3 | Gaussian blob | ~$10^{-15}$ | ~0.95–1.0 |
| 0.002 < κ < 0.25 | 0.3 < $E_1$ < 0.8 | Transition | $10^{-7}$–$10^{-1}$ | 0.3–0.9 |
| κ > 0.25 | $E_1$ > 0.9 | Deflection | ~$10^{-9}$ | ≲0.1 |

**Table S8:** The phase diagram formalizes a hierarchy of models suited to each regime:

| Regime | κ Range | $E_1$ Range | Recommended Model | Accuracy | Computational Cost |
|---|---|---|---|---|---|
| Gaussian Blob | <0.002 | <0.1 | Full modal expansion ($n_{max}$ ~ 50–100) | RMS ≲ $10^{-15}$ | Moderate |
| Transition (lower) | 0.002–0.01 | 0.1–0.3 | Multimode ($n_{max}$ ~ 10–20) | RMS ≲ $10^{-3}$ | Low |
| Transition (upper) | 0.01–0.25 | 0.3–0.9 | Reduced multimode ($n_{max}$ ~ 5–10) or single-mode hybrid | RMS ≲ $10^{-2}$–$10^{-3}$ | Very low |
| Deflection | >0.25 | >0.9 | Single-mode $P_1(x) = c_1 \sin(\pi x/L)$ | RMS ≲ $10^{-9}$ | Minimal |

**Table S9: Modal-Theoretic Refinements Beyond Classical Theories**

| Regime | Classical Limit | Modal Prediction | Correction Factor | Physical Origin |
|---|---|---|---|---|
| Weak κ | Gaussian; $E_1$ = 0 | $E_1$ ≈ 0.06; RMS ~ $10^{-15}$ | ~1 (exact) | Many modes, each small |
| Intermediate κ | Crossover shape | $E_1$ interpolates 0.06→0.9 | O(1) | Modal competition; Flory-type scaling |
| Strong κ | Single sine; slope m_th | Measured slope ≈ 2-9 × m_th | 2-9 | Higher modes; deflection segments |
| Large a/L | Non-Gaussian; β~2 | RMS ~ 3×$10^{-2}$ at a/L = 1 | ~2 | Persistence couples to geometry |

# Supplementary file_1: Scaling strategies

## Scaling strategies for P(x)

(codes are attached separately in supplementary file)

### (i) Case 1: Scaling with $x/L$ for different system sizes

**Enforcing the collapse condition.** To test the hypothesis numerically we enforced a constant $\kappa = \kappa_\star$ (i.e. it will remain fixed) across different absolute sizes by adjusting the chain parameter $N$ for each $L$ while keeping the Kuhn length $a$ fixed:

$$N(L) = \frac{\kappa_\star L^2}{a^2}.$$

(Equivalently one may fix $N$ and vary $a$; either approach enforces $\kappa = \kappa_\star$.)

We can select a set of absolute box sizes $L_1, L_2, \ldots$ to compare (for example, 1,2,3 $\mu$m). For each $L$, we can compute the number of Kuhn segments $N$ needed to realize $\kappa^*$ using

$$N(L) = \kappa^* \frac{L^2}{a^2} \quad \text{so that} \quad \kappa = \frac{Na^2}{L^2} = \kappa^*.$$

**Using a common scaled grid.** To avoid interpolation artefacts, we evaluated all curves on a common uniform grid in $u \in [0.01, 0.99]$ and plotted $P(u)$. The curves are overlayed on one plot.

It should be mentioned that the scaling argument is dimensionless and exact in the continuum modal representation: once $\kappa$ and the dimensionless tether position $x_0/L$ are prescribed, all lengths scale with $L$ and the eigenvalues $\lambda_n = n\pi/L$ scale inversely with $L$, as a result the combination that appears in the modal weights is a function of $\kappa$ only. Consequently, provided the modal series is summed to sufficient accuracy, and the same $\kappa$ is realized, the shape of $L P(x)$ versus $x/L$ should remain invariant to the changes in absolute size.

### (ii) Case 2: Effect of tether position on $P(x)$

Two complementary numerical experiments were performed and are shown in Fig. 3:

**Left panel (tether-position sweep):** We have fixed $L = 2.0$ $\mu$m, $N = 10$, $a = 0.10$ $\mu$m (hence $\sigma = \sqrt{N} a$ and $\kappa$ fixed for this panel) and compute $P(u)$ for a family of tether ratios $\xi = \{0.1, 0.2, \ldots, 0.9\}$. Next, we evaluate the analytic modal sum with adaptive truncation and plot $L P(x)$ vs $u$ on a common grid. This panel documents how the distribution shifts and becomes skewed as the tether moves off-center.

**Right panel (overlay across $L$ at fixed $\xi$):** Here, we test the geometric similarity by fixing the dimensionless tether ratio $\xi = 0.10$ and enforcing constant $\kappa$ across different absolute sizes $L \in \{1,2,3\}$ $\mu$m. To enforce $\kappa$ we compute $N(L) = \kappa_\star L^2/a^2$ with $\kappa_\star$ (=0.025) taken from the left-panel baseline parameters; $N(L)$ is then passed to the analytic modal evaluation so that $\kappa$ is identical across the cases. Next, we evaluate $P(u)$ on the same $u$-grid and overlay. Distinct linestyles (solid, dashed, dotted) were used in addition to the colour so that multiple traces remain perceptible even when they coincide.

The formulation shows that if both $\kappa$ and $\xi$ are fixed, the scaled distribution must be invariant to absolute length scale because the eigenvalues $\lambda_n = n\pi/L$ combine with $L$ and $\sigma$ to produce dimensionless modal weights that depend only on $\kappa$ and $\xi$. Thus, the overlay test enforces the condition for geometric similarity: when $\kappa$ is held constant across different $L$ (and when $\xi$ is fixed), $P(u)$ should collapse to a common master curve. Conversely, moving $\xi$ at fixed physical parameters intentionally alters the modal weights and will therefore change $P(u)$; such changes are physical and not removable by $L$-rescaling.

### (iii) Case 3: Confinement-strength scaling of $P(x)$ ($\kappa$-dependence)

In polymer physics, the probability distribution of the DNA end point inside confinement can be expressed as a combination of simple wave-like patterns called modes. Each mode has a characteristic shape, like the standing waves of a vibrating string, and contributes with a certain weight (coefficient). The first mode (n=1) represents the broadest, smoothest variation, while higher modes add finer ripples. When confinement is weak, many modes contribute together, producing a complex distribution. As confinement becomes stronger, the higher modes are suppressed, and eventually the distribution is dominated by just the first mode. Studying which modes dominate therefore helps us to understand how confinement simplifies the behaviour of the system and whether a universal single-mode picture emerges.

Each mode contributes to the probability distribution with a weight, or coefficient $c_n$. These coefficients are not arbitrary numbers — they are calculated directly from the shape of the probability distribution P(x). Mathematically, they are obtained by projecting the distribution onto the sine functions sin(nπx/L), which form the natural building blocks inside a confined box. In practice, this means that $c_n$ measures how much of the n-th sine wave pattern is present in the overall distribution. If $c_1$ is large, the first smooth mode dominates; if higher $c_n$ values are significant, it means finer oscillations contribute strongly. Thus, the set of coefficients $c_n$ provides a compact way to describe the relative importance of each mode in shaping the distribution. The probability density can be expressed as a sum of sine modes, $P(x) = \sum_n c_n \sin(n\pi x/L)$, where each mode contributes with a coefficient $c_n$. These coefficients are obtained directly from the analytic expansion and are given by

$$c_n = \frac{\sin(n\pi x_0/L)\exp(-n^2\pi^2\kappa/8)}{L \sum_m \frac{\sin(m\pi x_0/L)}{m\pi}(1-(-1)^m)\exp(-m^2\pi^2\kappa/8)}.$$

Physically, $c_n$ measures how strongly the $n$-th standing-wave pattern contributes to the overall distribution. Studying the relative size of these coefficients allows us to assess whether the distribution is controlled mainly by the first mode or by a mixture of many modes.

In this case, our aim is twofold: (i) to reveal the entire transition from weak confinement (DNA coil much smaller than the box) to strong confinement (DNA compressed by the box), and (ii) to test how the system approaches the single-mode limit, where the distribution is dominated by just the lowest sine mode.

To achieve this, we used the following steps:

**Wide sampling of confinement ratios ($Na/L$)**: Instead of focusing on a narrow window, we explored values of $Na/L$ over several orders of magnitude (from very small, $\sim 10^{-2}$, to very large, $\sim 10^2$). This ensures that the full crossover between weak and strong confinement is visible. Representative values (0.1, 1, 5, 10, 50) were chosen for direct plotting of the function P(x) for this case.

**Modal dominance and modal fractions**: Each sine mode has a weight (coefficient $c_n$). When the system is strongly confined, the first mode ($n = 1$) dominates. To quantify this, we defined two modal fractions:

The ratio of the first coefficient to the sum of all coefficients, $|c_1|/\sum_n |c_n|$, which tells us how large the first mode is relative to the rest.

The ratio of the squared first coefficient to the sum of all squared coefficients, $c_1^2/\sum_n c_n^2$, which measures how much of the energy or intensity is carried by the first mode.

**RMS difference**: To check how close the full distribution is to the single-mode approximation, we computed the root-mean-square (RMS) difference between the exact $P(x)$ and the simplified form $P_1(x) = c_1 \sin(\pi x/L)$. The RMS difference is essentially the average deviation between the two curves, and smaller values indicate that the single-mode picture is a good approximation.

Here, the goal is not simply to display overlap of PDFs but to explain *why* overlap occurs in certain parameter regimes. The analytic damping factor provides a direct justification: as $\kappa$ increases the factor $\exp(-n^2\pi^2\kappa/8)$ suppresses high-$n$ modes exponentially in $n^2$, so for sufficiently large $\kappa$ only the $n = 1$ mode contributes appreciably. Plotting the modal fractions directly tests this. Using a logarithmic $Na/L$ axis emphasizes the rapid cross-over character and avoids misleading visual clustering of points in a narrow interval. The RMS to single-mode quantifies how close the full PDF is to the asymptotic single-mode shape in absolute units rather than visually.

### (iv) Case 4: Persistence-length (Kuhn-length) effects on $P(x)$

Practical implementation choices were selected to separate true physical trends from numerical artefacts:

**Parameter sweep.** We evaluate $a \in \{0.05, 0.10, 0.20, 0.50, 1.00\}$ $\mu$m with fixed $N = 10$ and $L = 2.0$ $\mu$m so that $Na = 1.0$ $\mu$m but $\kappa$ varies as $a^2$. This bracket covers the limit $a \ll L$ (flexible), moderate persistence, and the stiff/mesoscopic limit $a \sim L$.

**Common standardized grid.** For standardized comparisons we compute $y = (x - \langle x \rangle)/\sigma$ for each $a$ and interpolate $P_y(y)$ onto a common $y$-grid to prevent spurious tails.

**Quantitative diagnostics.** To assess whether the distributions we obtain resemble a Gaussian (normal) curve after standardization, we use a set of common statistical tools. First, the cumulative distribution function (CDF) describes how probability accumulates as we move along the horizontal axis. In simpler terms, at any position x, the CDF tells us about what fraction of the total probability lies to the left of this point. By standardizing (subtracting the mean and dividing by the standard deviation), we remove differences in overall scale and spread, so curves from different parameters can be compared fairly on the same axis. Because the extreme tails of the curves may differ or be noisy, we focus only on their common central window—the region where all distributions have reliable overlap. Within this window, comparing standardized CDFs side by side lets us see directly whether the shapes of the curves line up or diverge.

Second, a quantile–quantile (QQ) plot compares the percentiles of our data (empirical quantiles $y_p$) with those of a perfect Gaussian (reference quantiles $z_p$). If the data follow a Gaussian, the points should lie close to the diagonal line y = z. Deviations from the 45° diagonal indicate skewness or heavier/lighter tails than the normal Third, we check the probability density function (PDF) more directly by taking the ratio of the standardized density $P_y$ to the Gaussian density $\varphi(y)$. A ratio close to 1 across the central region means the shapes agree. To summarize these deviations in a single number, we also compute the root mean square (RMS) error, which measures the average vertical difference between the two curves over the region of overlap. Small RMS values mean the distributions are almost indistinguishable from the Gaussian.

Here, we present (i) standardized CDFs on the common central window (Fig. 3b left), (ii) QQ plots comparing empirical standardized quantiles $y_p$ to standard-normal quantiles $z_p$ (Fig. 3b middle), and (iii) the ratio $P_y/\varphi$ and the RMS error on the intersection window (Fig. 3b right).

### (v) Case v: Boundary-layer Inner Scaling of $P(x)$ Near Absorbing Walls

Our goal is to rigorously test whether the near-wall behavior of the polymer is universal. To do this, we designed a clear, multi-step strategy that separates genuine physical collapse from potential numerical artifacts.

**Careful Numerical Sampling Near the Wall:** The probability distribution `P(x)` changes extremely rapidly very close to the wall. We evaluated `P(x)` on a very fine grid of points that gets incredibly close to the walls (within one hundred-millionth of the box length) but never directly on them. This provides a smooth and accurate picture of the boundary layer without numerical instability.

**Testing Two Physical Hypotheses and Finding a Compromise**

We then tested our two candidate rulers or inner lengths $\ell$:

**The Polymer Ruler ($\ell = \sigma$):** Is the boundary layer structure set by the polymer's own size?
**The Box Ruler ($\ell = L/\pi$):** Is it set by the fundamental geometry of the confinement box?

For each value of the confinement strength $\kappa$, we constructed the scaled plot: $\ell P$ versus $\eta = \delta/\ell$ for both choices of $\ell$.

**Finding the Best Fit:** Furthermore, we allowed the data to tell us the best possible ruler. We defined a generalized ruler $\ell = \alpha\sigma$ and used an optimization algorithm to find the value of $\alpha$ that produced the best possible collapse of the different $\kappa$ curves onto a single master curve in the region very close to the wall ($\eta \leq 2$). This provides an empirical best compromise length scale.

**Measuring the Key Feature – The Initial Slope**

The primary signature of universality is a linear relationship between $\ell P$ and $\eta$ at very small $\eta$ (i.e., immediately next to the wall). That is why, for each scaled plot, we performed a linear fit to measure the slope $m$ in this initial linear region ($\eta \leq 0.15$). We then compared these measured slopes to the slope predicted by the theory of the first (and simplest) sine wave mode.

**Comprehensive Diagnostics for Trustworthy Results (see supplementary table S2)**

Finally, to ensure our conclusions are robust and reproducible, our code recorded a full suite of diagnostics for each run, including:

Normalization: Confirming the total probability was calculated correctly.

Survival Probability: The fraction of polymer configurations that never touch the walls.

Collapse Quality: Quantitative measures (RMS error) of how well the curves from different $\kappa$ values overlapped for each choice of $\ell$.

Slope Values: The measured and theoretical slopes for direct comparison.

### vii) Case 6: Tether-centered $\sigma$-scaling of $P(x)$

To test this hypothesis in a way that is both numerically robust and experimentally relevant we implemented the following protocol and safeguards (these are the recommendations that were encoded into the supplied scripts and used to produce the reported figures/diagnostics):

**Fixed-$\kappa$ geometric similarity.** We vary the absolute box size $L$ and set $N = \kappa L^2/a^2$ so that $\kappa$ remains constant across all cases in a collapse test. We keep the tether fraction $x_0/L$ fixed (the examples presented use central tethering $x_0/L = 0.5$ unless stated otherwise).

**Endpoint-safe evaluation grid.** We evaluate $P(x)$ on $x \in [\varepsilon L, (1-\varepsilon)L]$ with $\varepsilon \ll 1$ (we used $\varepsilon = 10^{-8}$ in production runs) to avoid singular or numerically fragile evaluations at the absorbing walls.

**Common $s$-grid with NaN padding and contiguous-segment plotting.** We interpolate each $\tilde{P}(s)$ onto a dense, symmetric common grid $s_{\text{common}}$, fill values outside a curve's native support with *NaN* (no zero-padding), and plot only contiguous finite segments to avoid artificial connectors.

**Reference residual diagnostics.** We choose one case as the reference and compute pointwise residuals $\Delta(s) = \tilde{P}_i(s) - \tilde{P}_{\text{ref}}(s)$ on the intersection of finite supports. We report per-case diagnostics: modes retained, $\int P \, dx$, $\sigma$, maximum absolute residual $\max_s|\Delta|$, RMS residual on overlap, and the fraction of the common grid used in the residual calculation.

**Modal corroboration.** For each case we compute analytic modal coefficients $c_n$ (the same analytic form as used to compute $P$) and report the first-mode fraction $c_1^2/\sum_n c_n^2$ (and an absolute fraction $|c_1|/\sum|c_n|$) to demonstrate that collapsed curves share the same spectral content.

**Negative control (sensitivity) test.** We include a contrasting run in which $\kappa$ is deliberately varied (fixed $L$, varying $\kappa$) to show that collapse fails under broken geometric similarity — this demonstrates the sensitivity and interpretive power of the test.

All of the above elements were included in the analysis code; the figure panels and the accompanying Supplementary Table S3 report the full diagnostics. The analytic modal representation makes the scaling argument transparent: for fixed $\kappa$ and fixed $x_0/L$ the modal damping factor and the geometric factors that determine the relative $c_n$ become identical across absolute system sizes, so the local superposition around the tether is identical up to the linear scale $\sigma$. Thus, the tether-centred $\sigma$-scaling is a direct consequence of geometric similarity in the modal picture.

**Valid regime.** The scaling is expected to hold (and does hold numerically) when: $\kappa$ is strictly controlled (equal across compared systems), and the tether position fraction $x_0/L$ is identical across cases, and the modal series is summed to sufficient accuracy.

**When the scaling will fail (and why).** Collapse breaks down if $\kappa$ is not the same across cases (the modal envelope changes), if $x_0/L$ differs (geometric prefactors differ), or if the local coil width $\sigma$ is so large that the scaled support overlaps boundaries in a case-dependent way (edge effects and interpolation truncation distortions). In the weak-confinement limit (very small $\kappa$) many modes contribute and practical collapse tests require significantly larger modal budgets and grid resolution; failure to increase these numerical budgets will mimic physical breakdown but is in fact a numerical artefact, reporting the mode counts and truncation tolerance guards against misinterpretation.

## Scaling strategies for P(y)

### (codes are attached separately in supplementary file)

**Case 1: Geometric-similarity collapse of $P(y)$ (scaled plot: $y/R$, $R\,P(y)$):**

From the analytic image-sum or Fourier–sine solution one sees that $P(y)$ depends only on the ratios $\sigma/R$ and $y/R$. Therefore, when $\kappa = \sigma^2/R^2$ is held fixed and $R$ is varied (adjusting $N$ so $\sigma/R$ remains constant), the scaled density $\mathcal{P}(u)$ must be identical for all $R$. The survival $S$ is likewise invariant under variation of absolute $R$ at fixed $\kappa$.

$P(y)$ is computed by the method-of-images:

$$P(y) = \frac{1}{\sqrt{2\pi}\sigma} \sum_{m=-M}^{M} (-1)^m \exp\left(-\frac{(y-2mR)^2}{2\sigma^2}\right),$$

with an adaptive choice of $M$. It may be noted that the limit of m is not set to infinity. In practice the infinite image sum is truncated to $-M \leq m \leq M$. The cutoff M is selected adaptively so that the neglected Gaussian tail terms fall below a set tolerance (typically $10^{-12}$), ensuring numerical convergence. The developed code (see supplementary files) returns `M_used` for each run. The driver accepts `conditional=True` to produce $P_{\text{cond}}$ and facilitates direct comparison of shapes independent of survival amplitude. For each $R$ the code computes pointwise residuals relative to a chosen reference $R_{\text{ref}} = 1$:

$$\Delta \mathcal{P}(u) = \mathcal{P}(u; R) - \mathcal{P}(u; R_{ref})$$

From these residuals the code reports two residual metrics such as maximum absolute residual and RMS residual, respectively in CSV format which are the primary numerical evidence for collapse (Supplementary table S4). The figures and CSV were generated with the defaults used in the code example: $\kappa = 0.1$, $R \in \{1,2,4\}$.

**Case 2: Varying polymer flexibility $\kappa$ at fixed confinement width $L$**

Here, we'll hold $R$ fixed, compute $P(y)$ for a grid of $\kappa$ values. We are going to plot unconditional scaled densities $\mathcal{P}(u) = R\,P(y)$ versus $u = y/R$ for a set of $\kappa$ values from the free-coil regime to the strongly confined regime. We'll show conditional densities $\mathcal{P}_{\text{cond}}(u) = R\,P(y)$ as a function of $u = y/R$. We'll show $\tilde{P}(s) \equiv \sigma\,P_{cond}(y)$ plotted as a function of $s = y/\sigma$. The numerical implementation (see supplementary code) follows a robust protocol and includes the following features:

$P(y)$ is computed by the method of images,

$$P(y) = \frac{1}{\sqrt{2\pi}\sigma} \sum_{m=-M}^{M} (-1)^m e^{-(y-2mR)^2/(2\sigma^2)},$$

with an adaptive estimator for $M$ chosen from a tail tolerance *eps* (~$10^{-12}$) The code (see supplementary file) records `M_used` for each $\kappa$. To investigate the robustness of the method adopted (See Supplementary table S5**)**, the code writes a diagnostics row containing the parameters: `kappa`, `M_used`, `S`, unconditional and conditional moments (`mu`, `var`), `rms_to_gaussian` (coil comparison), `rms_to_sine` (modal comparison), `truncation_sensitivity` (RMS diff versus a much larger $M$), `in_coil_regime` (operational flag), and `L_over_sigma` = $R/\sigma$ for each tested $\kappa$. The content of the generated output CSV file is shown in Supplementary Table S5.

To test whether the coil→confinement crossover is governed by the dimensionless ratio $R/\sigma$ (and not $\kappa$ alone) we performed an explicit $R$-sweep at fixed $\kappa$. For each $\kappa$ we computed coil-scaled conditional densities $\tilde{P}(s) = \sigma P_{\text{cond}}(y)$ on a common, fixed $s$-grid (so that sampling and numerical error are identical across points) and varied $R/\sigma$ over a wide range. The coil-scaled traces collapse when $R/\sigma$ is held fixed, confirming the expected invariance under the simultaneous scaling $y \to \sigma s$, $R \to \sigma(R/\sigma)$; by contrast, the same data plotted in geometry units $u = y/R$ do not collapse, revealing the $\kappa$-dependent absolute coil-size effects. We quantified deviations from the unit Gaussian using a normalized RMS metric (computed on the accessible domain $|s| \leq R/\sigma$) and complementary diagnostics (central-window RMS, KS-like CDF difference, $L_\infty$ max-abs difference), and verified numerical convergence by varying the number of image terms m $_{\text{max}}$. The resulting heatmap of normalized RMS in the $(R/\sigma, \kappa)$ plane shows vertical iso-contours (i.e. little $\kappa$-dependence) and supports the operational cutoff $R/\sigma \gtrsim 3$ as the region where Gaussian (coil) statistics are reliable. Full numerical diagnostics and convergence checks are provided in the Supplementary Information and in the CSV file `case2_Rsweep_R_as_radius_diagnostics.csv`.

**Case 3: Image-Method Distributions Across Confinement Regimes**

We'll build a common scaled grid $s$ that covers all $\lambda$ used. Then we'll compute unconditional $\tilde{P}(s)$ for each $\lambda$ (image-sum with adaptive truncation clipping tiny negative noise). Then conditional PDF $\tilde{P}_{\text{cond}}(s) = \tilde{P}(s)/S$ will be formed for shape comparison. It should be mentioned that scaling by $\sigma$ removes explicit dependence on chain length and Kuhn length; $\lambda$ then controls the confinement: $\lambda \gg 1$ (weak confinement) → bulk-like Gaussian center; $\lambda \lesssim 1$ (strong confinement) → many image terms and strong survivor bias. The $s$-representation is numerically stable for comparing shapes across widely different $\kappa$ because it keeps the free-coil kernel at unit width.

Before presenting the results, we again recall the distinction between the *unconditional* and *conditional* distributions.

The unconditional distribution is obtained directly from the image method,

$$\tilde{P}(s) = \sigma P(y),$$

where $s = y/\sigma$. Its integral gives survival probability,

$$\int_{-\lambda}^{\lambda} \tilde{P}(s)\, ds = S(\kappa) \leq 1,$$

with $\lambda = R/\sigma$. This form therefore encodes both the shape of the distribution and the probability mass lost to absorption at the walls.

The conditional distribution is obtained by renormalizing the unconditional curve by the survival probability,

$$\tilde{P}_{\text{cond}}(s) = \frac{\tilde{P}(s)}{S(\kappa)},$$

such that

$$\int_{-\lambda}^{\lambda} \tilde{P}_{\text{cond}}(s)\, ds = 1.$$

This conditional form describes the statistics of the surviving ensemble of configurations, i.e., chains that remain within the confining geometry.

**Case 4: Fixed $R$, varying $\kappa$: unconditional, conditional, and peak-normalized comparisons:**

With $u = y/R$, and $\mathcal{P}(u) = RP(Ru)$, the image sum becomes

$$\mathcal{P}(u) = \frac{R}{\sqrt{2\pi}\,\sigma} \sum_m (-1)^m \exp\left[-\frac{(Ru - 2mR)^2}{2\sigma^2}\right].$$

When plotted vs $u$ the domain and axis are identical for all κ, facilitating direct visual comparison of amplitude, survivor-fraction, and shape. Holding $R$ fixed focuses the test on how chain size (σ) changes the distribution relative to the fixed geometry. Using $u = y/R$ collapses the domain and makes amplitude changes (survival) immediately visible. Conditional and peak-normalized forms isolate shape differences from amplitude loss.

To further investigate confinement effects, we employ the image–method formulation for absorbing boundaries and analyze the endpoint distributions under varying confinement parameter $\kappa = Na^2/R^2$, with fixed Kuhn length $a$ and half-width $R$. The number of segments is determined as $N = \kappa R^2/a^2$, such that increasing κ corresponds to longer chains probing confinement more strongly.

**Case 5: Image–method collapse with α and β scalings:**

To test this formulation, we compare two approaches:

**Optimized exponents**: $\alpha = 0.5, \beta = 0.5$, chosen to minimize the spread of the collapsed curves across κ.

**Reference exponents**: $\alpha = 3.0, \beta = 1.0$, corresponding to the naïve wall–distance or free Gaussian scaling.

The idea is to evaluate whether the optimized exponents yield a near–universal curve (small residuals) while the reference exponents fail to achieve collapse.

# Supplementary file_2: additional details

**Modal (eigenfunction) coefficient analysis of the tethered end-point distribution**

**Introduction:**

This supplementary section presents a modal (eigenfunction) coefficient analysis of the tethered end-point distribution. While the main manuscript focuses on direct scaling strategies for P(x) and P(y), the modal perspective provides an additional, more fundamental view of why such scaling collapses occur. In particular, the Fourier–sine expansion expresses P(x) as a weighted sum of eigenfunctions, with coefficients $c_n$ that encode the contribution of each mode. Tracking the behavior of these coefficients clarifies when the first eigenmode dominates and when higher modes contribute significantly.

The aim of including this analysis in the Supplementary Information is to provide theoretical and numerical corroboration for the scaling results reported in the main text. By explicitly calculating and comparing the coefficients, we demonstrate how the suppression of higher modes underlies the observed PDF collapses. This analysis therefore complements the scaling tests presented in Cases 1–6 of P(x) and offers additional confidence that the reported scalings are not plotting artefacts but genuine physical simplifications rooted in modal dominance.

**(a) Geometry and Scaling Framework**

The modal (eigenfunction) analysis is carried out in the same physical setup as in the earlier cases: a polymer chain tethered at a fixed point inside a confining box of length $L$. For clarity, we illustrate the central-tether case ($x_0/L = 0.5$), though the methods apply to any tether position (Fig. 1 in manuscript).

In this setting, the natural mathematical building blocks are the sine eigenfunctions

$$\sin(n\pi x/L), \quad n = 1, 2, 3, \ldots$$

Each eigenfunction represents a mode of fluctuation, and the distribution $P(x)$ can be expressed as a weighted sum of these modes.

$$P(x) = \sum_{n \geq 1} c_n \sin\left(\frac{n\pi x}{L}\right),$$

The corresponding coefficients, $c_n$, are the modal amplitudes. They tell us how much each mode contributes to the overall distribution. These coefficients carry dimensions of inverse length and depend on the confinement strength and the tether location.

To interpret these coefficients, we introduce a few useful diagnostics:

- **Absolute amplitude $|c_n|$:** the direct size of each mode.
- **Modal energy fraction $E_n$:** how much of the total energy (it refers to the squared-amplitude contribution of each eigenmode, not to a physical energy) or variance is contained in a single mode.

$$E_n = \frac{c_n^2}{\sum_m c_m^2},$$

- **Cumulative energy** $\mathcal{E}_m$: the fraction of total energy captured when including the first $m$ modes.

$$C_M = \sum_{n \leq M} E_n$$

- **Reconstruction error** $\epsilon_m$: how much accuracy is lost if only the first $m$ modes are used to approximate $P(x)$.

$$\varepsilon_M = \sqrt{\frac{\int (P(x) - P_M(x))^2 dx}{\int P(x)^2 dx}}, \qquad P_M(x) = \sum_{n=1}^{M} c_n \sin\left(\frac{n\pi x}{L}\right).$$

These measures give us an objective way to decide when a modal expansion can be truncated (i.e., how many modes are enough to describe the distribution).

Starting from the Fourier–sine modal form used to evaluate $P(x)$ (the same modal weights that appear in Cases 1–5), the coefficients $c_n$ are obtained analytically (or numerically with analytic normalization) from

$$c_n \propto \sin\left(\frac{n\pi x_0}{L}\right) \exp\left(-\frac{n^2 \pi^2 \kappa}{8}\right),$$

up to the global normalization factor that enforces $\int_0^L P(x)\, dx = 1$. This expression makes two scaling facts immediate:

- for fixed tether position $x_0/L$ the spectral envelope decays approximately like $\exp(-\text{const} \times n^2 \kappa)$, so larger $\kappa$ produces much faster modal suppression and hence spectral concentration at low $n$;

- the factor $\sin(n\pi x_0/L)$ imposes geometric selection rules (zeros and sign changes), for example, for $x_0/L = 1/2$ all even $n$ vanish exactly, explaining the strong parity structure seen in the top-left panel of the figure.

Thus, modal coefficients encode both the confinement physics (through $\kappa$) and the tether geometry (through $\sin(n\pi\xi)$).

The analytic form of the coefficients reveals two important scaling effects:

- **Confinement dependence:** For stronger confinement ($\kappa$ large), the coefficients decay much faster with $n$. This means that only the lowest modes matter, and the distribution is spectrally concentrated.
- **Tether geometry:** The tether position introduces simple selection rules. For example, in the central-tether case, all even modes vanish exactly, so only odd modes contribute. This explains the strong parity structure observed in the spectra.

In short, the modal coefficients simultaneously encode both the confinement physics (through $\kappa$) and the tether geometry (through $x_0/L$). This makes them a powerful diagnostic tool. They allow us to see directly how confinement strength and tether placement shape the probability distribution.

**(b) Scaling strategy**

The goals of the modal study were (i) to characterize how spectral content depends on $\kappa$, (ii) to quantify how many modes are needed for faithful reconstruction as a function of $\kappa$, and (iii) to provide practical truncation/diagnostic rules for simulation and data analysis. The adopted computational strategy was:

- **Coefficient extraction consistent with analytic normalization.** We compute $c_n$ using the same analytic denominator used for $P(x)$ so that reconstructed $P_M$ matches the original analytic curve when $M$ includes all kept modes. This avoids small inconsistencies between coefficient and PDF evaluation.

- **Range and display.** Coefficients were computed for $n$ up to several hundred (display limited to first 60 odd modes in the figure because even modes vanish for the centered tether). Absolute amplitudes $|c_n|$ are plotted on a log scale to visualise the exponential envelope and the effect of $\kappa$.

- **Modal-energy diagnostics.** We compute $E_n$ and cumulative energy $C_M$. Typical truncation thresholds used in diagnostics are $C_M \geq 0.99$ (99% energy captured) or an $L^2$ error tolerance $\varepsilon_M$ below a user-prescribed value.

- **Low-mode amplitude vs $\kappa$.** Selected low-$n$ amplitudes (e.g. $n = 1,3,5,7,9$) are tracked as functions of $\kappa$ to show how confinement shifts relative modal importance.

- **Reconstruction error curves.** The $L^2$ error $\varepsilon_M$ is plotted versus $M$ to show the practical number of modes needed for given accuracy at different $\kappa$.

- **Robust numerical choices.** Modal sums used an adaptive truncation tolerance consistent with the damping factor. Care was taken to compute energy sums with stable double precision arithmetic (use of cumulative sums from small to large index where appropriate).

Modal decomposition is the natural tool to translate statements about shape and collapse (Cases 1–5) into a compact spectral language. In particular:

- If the first mode carries most of the modal energy (large $E_1$), then $P(u)$ will look like a single-sine profile and different absolute sizes will collapse when the corresponding $\kappa$ is held fixed (explaining Case 1 and Case 3 single-mode saturation).

- If several low modes have comparable weight, the shape can deviate from the single-sine form and exhibit richer structure (as in persistence-driven deviations in Case 4).

- Tether-position selection rules (zeros of $\sin(n\pi x_0/L)$) are immediately visible in the coefficient sign pattern and explain parity effects (even-mode suppression for central tether).

Quantities such as cumulative energy and $\varepsilon_M$ provide objective, model-independent criteria for truncation: they indicate how many eigenfunctions are necessary to resolve features at a given $\kappa$ and therefore how many degrees of freedom a reduced model must retain to be quantitatively accurate.

### (c) Results

The four panels in the figure 1 collectively summarize the spectral behavior:

- **Top-left: $|c_n|$ vs $n$ (log scale, analytic markers overlay).**

  - For small confinement $\kappa$ (e.g. $\kappa = 0.05$, blue markers) the modal envelope decays slowly and many modes carry appreciable amplitude; the $|c_n|$ sequence extends to high $n$.
  - As $\kappa$ increases the envelope steepens dramatically (green → brown → cyan), producing exponential-like suppression of high-$n$ modes. This behaviour is the spectral manifestation of the factor $\exp(-n^2\pi^2\kappa/8)$ in the modal weights.
  - The parity rule for the centered tether is visible: even $n$ coefficients vanish exactly (only odd indices shown/used).

- **Top-right: cumulative modal energy $C_M$ vs $M$.**

  - For large $\kappa$ nearly all energy is contained in the lowest few modes: the cumulative energy curve jumps to near unity for small $M$ (first-mode saturation). The inset zoom near small $n$ highlights this.
  - For small $\kappa$, energy accumulates slowly, and many modes are required to reach a target fraction (e.g. 0.99), reflecting the broad multimode character of the PDF.

- **Bottom-left: low-mode amplitudes $|c_{1,3,5,9}|$ vs $\kappa$.**

  - The $n = 1$ amplitude dominates for large $\kappa$ and decreases monotonically as $\kappa$ is reduced; higher odd modes (3,5,9) are progressively more important at small $\kappa$.
  - This panel provides a compact view of how the shape transitions from multimode to first-mode dominated with increasing confinement strength.

- **Bottom-right: $L^2$ reconstruction error $\varepsilon_M$ vs number of modes $M$.**

  - For large $\kappa$ the error falls to a small plateau with only a few modes; for example, $\varepsilon_M$ may be $\mathcal{O}(10^{-1})$ or smaller already at $M \sim 5$ and reaches numerical precision rapidly.
  - For small $\kappa$ the error decreases much more slowly with $M$ and plateaus at a higher floor if $M$ is limited, indicating the need for many modes for faithful reconstruction.
  - This panel gives explicit guidance: for a given $\kappa$ choose the smallest $M$ such that $C_M \geq 0.99$ or $\varepsilon_M$ is below the desired tolerance.

Additional quantitative points reported by the runs:

- **Mode counts and truncation:** The adaptive truncation criterion based on modal decay provides a sensible upper limit for practical evaluations: for large $\kappa$ only the first few odd modes survive the tolerance; for small $\kappa$ many dozens to hundreds of modes may be needed.

- **Reconstruction floors:** The reconstruction error curves sometimes display a small residual floor (numerical or model-related) that vanishes when the modal budget and quadrature resolution are increased; this was verified by raising the modal cap and refining the grid in test runs.

The modal-coefficient analysis provides a direct and quantitative bridge between the spectral content of $P(x)$ and the scaling phenomena documented in Cases 1–5:

- **First-mode dominance** at large $\kappa$ explains why scaled PDFs collapse to a single master curve and why single-mode reconstructions suffice (strong confinement / small coil limit). In that regime a compact reduced model with only the first one or two modes is both accurate and efficient.

- **Multimode character** at small $\kappa$ explains broad, Gaussian-like or complex shapes that require many modes to represent; this is the regime where standardization and modal diagnostics (as in Case 4 and Case 5) must be applied with care.

- **Tether-geometry selection rules** (zeros and sign changes in $\sin(n\pi x_0/L)$) are immediately visible in the spectrum and should be used as a diagnostic of tether placement in experimental inversion: e.g. absence of even-mode content is a strong indicator of central tethering.

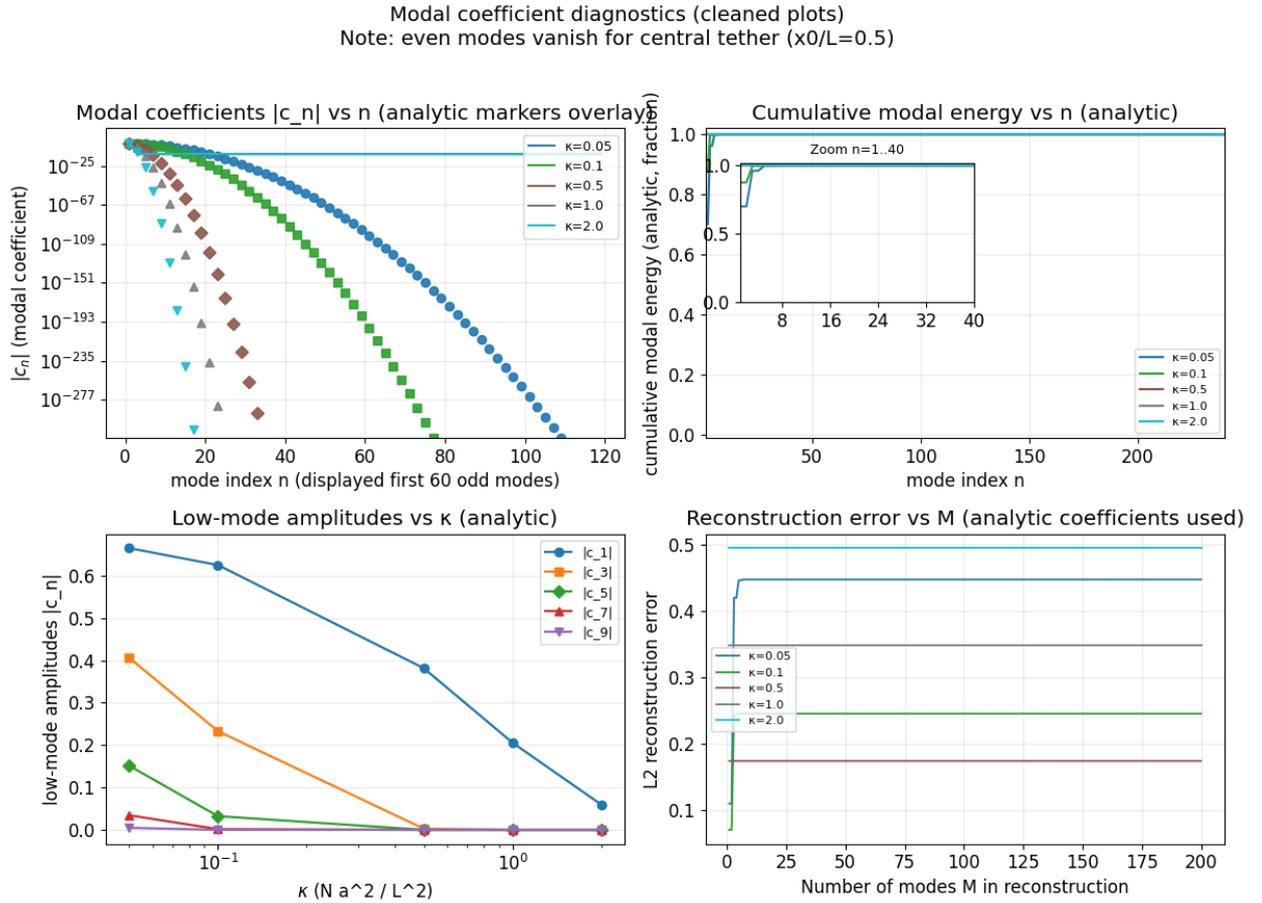

**Figure 1:** Modal decomposition diagnostics for tethered end distribution. **Top left:** absolute modal amplitudes $|a_n|$ vs mode index $n$ (semilog) for five confinement strengths $\kappa = \{0.05, 0.1, 0.5, 1.0, 2.0\}$. Even modes vanish for the central tether ($x_0/L = 0.5$); only odd modes are shown. Filled markers show analytic coefficients for the first modes; solid lines show numerical projections. **Top right:** cumulative modal energy (analytic) versus $n$; inset zoom shows the early rise for $n \leq 40$. Bottom left: selected low-mode amplitudes ($|a_1|, |a_3|, ...$) plotted versus $\kappa$ (log scale) illustrating single-mode dominance at large $\kappa$. Bottom right: $L^2$ reconstruction error of $P(x)$ when using the first $M$ analytic modes, this quantifies how many modes are required for an accurate reconstruction as a function of $\kappa$.

In summary, the modal-coefficient analysis presented here provides a quantitative spectral foundation for the scaling results discussed in the main text. By resolving how probability mass is redistributed among eigenmodes as confinement and stiffness are varied, it translates qualitative observations about distribution shape into explicit, measurable criteria. This analysis clarifies which modes dominate in Gaussian, transition, and deflection regimes, and supplies practical guidance for controlled modal truncation and error assessment. As such, it does not introduce new scaling claims, but rather substantiates and completes the unified scaling framework of Cases 1–5 by making the spectral origin of the observed collapses transparent and reproducible.

# Supplementary file_3: Conventional scaling

## Case 1 of P(x): Scaling with $x/L$ for different system sizes

### (a) Scaling variables and geometry

Here, we consider a single spatial axis along longitudinal direction $x \in [0, L]$ (box length $L$, walls at 0 and $L$, respectively), the position of the tether is at $x_0$ (here $x_0 = L/2$). The Kuhn length is $a$ and the number of Kuhn segments is $N$, giving free-coil RMS $\sigma = \sqrt{N}\, a$. We introduce a control parameter $\kappa = \kappa_\star$ across different absolute sizes (L) by adjusting the chain parameter $N$ for each $L$ while keeping the Kuhn length $a$ fixed. Thus, the dimensionless confinement parameter: $\kappa = \frac{\sigma^2}{L^2} = \frac{Na^2}{L^2}$, where $N(L) = \frac{\kappa_\star L^2}{a^2}$. Here, N(L) and L being varied while other parameters are fixed.

The scaled coordinate and density used to test collapse are $u = \frac{x}{L}$, and $P(u) = L\, P(x)$, respectively. Starting from the analytic modal representation for the tethered end-point density (Fourier–sine expansion; equivalent to Eq. 1),

$$P(x) = \frac{\sum_{n\geq 1} \sin\left(\frac{n\pi x_0}{L}\right) \sin\left(\frac{n\pi x}{L}\right) \exp\left(-\frac{n^2\pi^2\kappa}{8}\right)}{L \sum_{n\geq 1} \frac{\sin\left(\frac{n\pi x_0}{L}\right)}{n\pi} (1-(-1)^n) \exp\left(-\frac{n^2\pi^2\kappa}{8}\right)},$$

It is clearly observed that, for fixed $\kappa$ and fixed dimensionless tether position $x_0/L$, the numerator and denominator depend on $x$ only through the ratio $x/L$. Hence, the scaled density $P(u) = L\, P(x)$ is a function of $u$ alone, i.e. $P(u) = F(u; \kappa, x_0/L)$, and should therefore be invariant under changes of the absolute length scale $L$ (geometric similarity).

Thus, our goal is to test whether the longitudinal end–point distribution of a tethered polymer becomes *size-independent* after the obvious geometric rescaling i.e. whether plotting the scaled density $L\, P(x)$ against the dimensionless coordinate $u = x/L$ collapses data from different absolute box sizes $L$ onto a single master curve (see Fig. 2), when the degree of confinement is held fixed. It should be mentioned that if two systems differ only by an overall length scale (one is a uniformly larger copy of the other) and the polymer's relative size inside the box is the same, then all dimensionless observables should coincide. In other words: when the polymer's natural size and the box size keep the same ratio, absolute size should not matter only the ratio does. Testing the collapse with $u = x/L$ and $L\, P(x)$ checks this geometric similarity directly.

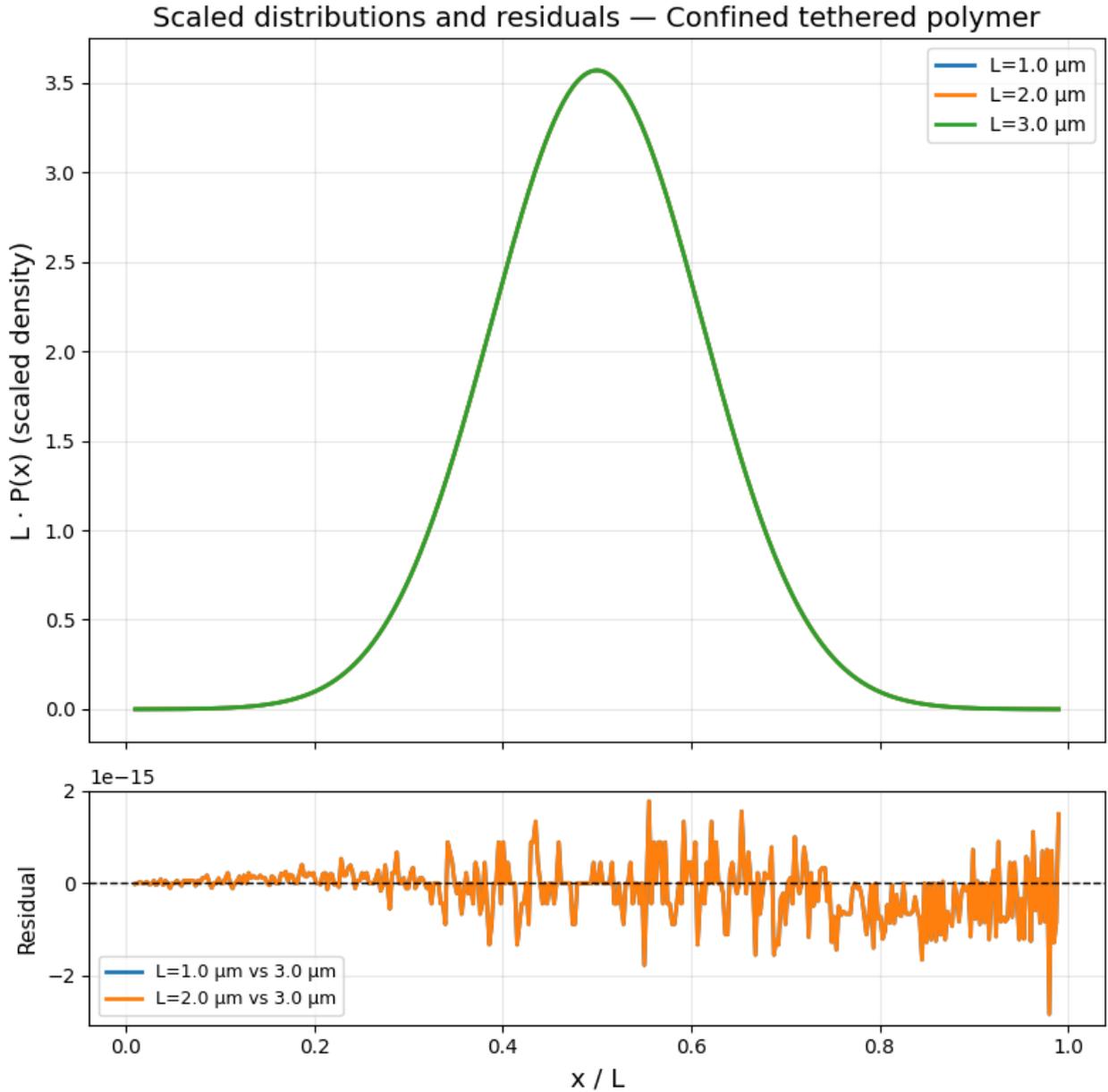

**Figure 1.** The demonstration of geometric similarity: collapse of the scaled longitudinal end-point distribution and residual diagnostics. **Top:** scaled probability density $L P(x)$ plotted against the dimensionless coordinate $u = x/L$ for three system sizes $L = 1.0, 2.0,$ and $3.0\ \mu m$, respectively (blue, orange, and green curves). For each $L$ the dimensionless confinement $\kappa = \sigma^2/L^2$ was held fixed at $\kappa_\star = 0.05$ by adjusting the chain length $N(L) = \kappa_\star L^2/a^2$ with fixed Kuhn length $a = 0.10\ \mu m$ (so $N = 5, 20, 45$ for $L = 1,2,3\ \mu m$, respectively). All three traces are visually indistinguishable, indicating collapse of the master curve $\tilde{P}(u) = L P(x)$. **Bottom:** pointwise residuals for $L = 1.0$ and $2.0\ \mu m$ are shown (orange/blue) and are everywhere $\mathcal{O}(10^{-15})$, i.e. at the level of double-precision round-off. The dashed horizontal line indicates zero. Together the main panel and residuals quantitatively confirm that, at fixed $\kappa$ and fixed dimensionless tether position ($x_0 = L/2$), the longitudinal end-point distribution is invariant to absolute system size (geometric similarity).

**(b) Results**

Fig. 1 (top) shows $P(u) = L\, P(x)$ for $L = 1, 2, 3\ \mu m$ with $\kappa_\star = 0.05$. The three traces are visually indistinguishable i.e. the curves lie on top of one another within plotting resolution. The residuals $\Delta \mathcal{P}(u) = \mathcal{P}(u; L) - \mathcal{P}(u; L_{ref})$ (differences relative to the $L = 3\ \mu m$ reference) are on the order of $10^{-15}$ in absolute value across the domain. It has been verified that when $\kappa$ was not held fixed, the curves did not coincide perfectly: smaller $L$ produced broader, flatter scaled profiles. This behavior is correct and expected, because failure to hold $\kappa$ fixed breaks geometric similarity and therefore forbids the collapse.

Thus, the combination of analytic modal evaluation and careful numerical controls confirms that the longitudinal end-point distribution of a tethered Gaussian chain exhibits strict geometric similarity. For fixed dimensionless confinement $\kappa$ and fixed dimensionless tether position $x_0/L$, the scaled density $P(u) = L\, P(x)$ is independent of absolute size $L$.

**(c) Physical interpretation:**

From a theoretical perspective, this scaling strategy demonstrates the concept of universality, a key idea in statistical mechanics. Universality refers to the idea that, under certain conditions, systems with different microscopic details exhibit the same macroscopic behaviour. By showing that the distribution depends only on the scaled coordinate u, the results reinforce the idea that the polymer's statistics are controlled by a single length scale, L, rather than the system's absolute size.

The scaling results are particularly relevant in biological systems where confined biopolymers, like chromatin, interact within bounded cellular spaces. These results suggest that the conformational behaviour of these polymers can be predicted by considering only the relative size of the confinement and not the exact dimensions of the system. This may simplify the modelling of biopolymer behaviour in complex, confined environments like the nucleus, where the size of the confinement can vary.

**(ii) Case 2 of P(x): Effect of tether position on $P(x)$**

**(a) Scaling variables and geometry**

Here, we consider one-dimensional interval $x \in [0, L]$ with absorbing walls at $x = 0$ and $x = L$. The tether is located at $x_0$ and we use the dimensionless parameter to investigate the tether position effect is $\xi \equiv x_0/L$. Chain parameters are the Kuhn length $a$, the number of Kuhn segments $N$, and the free-coil RMS $\sigma = \sqrt{N}\, a$. The confinement parameter is $\kappa = \frac{\sigma^2}{L^2} = \frac{Na^2}{L^2}$. Scaled coordinate and density used for comparison are $u = \frac{x}{L}$, $P(u) = L\, P(x)$. Using the Fourier–sine modal representation from Eq. 1, the tether position enters the modal weights through the term $\sin(n\pi\xi)$. The explicit modal form becomes

$$P(x) = \frac{\sum_{n \geq 1} \sin(n\pi\xi) \sin\left(n\pi \frac{x}{L}\right) e^{-\frac{n^2\pi^2\kappa}{8}}}{L \sum_{n \geq 1} \frac{\sin(n\pi\xi)}{n\pi} (1 - (-1)^n) e^{-\frac{n^2\pi^2\kappa}{8}}},$$

So, for fixed $\kappa$ and fixed $\xi$ the scaled density $P(u) = L\, P(x)$ is essentially a function of $u$ alone, $P(u) = F(u; \kappa, \xi)$. Changing $\xi$ changes the signs and magnitudes of the $\sin(n\pi\xi)$ factors and therefore modifies the modal superposition and the shape of $P(u)$.

**(b) Results:**

Fig. 2 demonstrates the effect of tether position on the scaled curves. Left panel shows the family of curves for $L = 2.0\ \mu m$ and $N = 10, a = 0.10\ \mu m$. The figure clearly shows the expected systematic shift of the scaled density $P(u)$ with tether ratio $\xi \equiv x_0/L$. As $\xi$ increases from 0.1 toward 0.9, the peak of $P(u)$ shifts monotonically from near the left wall toward the right wall and the shape becomes asymmetrically skewed (see Supplementary table S1). For the central tether ($\xi = 0.5$) the profile is symmetric and even modes vanish; for $\xi$ close to the wall the profile is sharply localized close to the tethered side. These changes are quantitatively verified by the first moment $\langle u \rangle$ and skewness computed from each curve (see the code). The left and right end curves of left panel are different for a physical and not numerical reason: those two curves correspond to tethers placed very close to an absorbing wall ($x_0/L = 0.1$ and $0.9$). When the tether is near a wall, the ensemble of allowed configurations is one-sided (the chain cannot explore beyond the wall), so the end-point distribution is compressed toward the interior. Because the curves are normalized, this reduced width forces the peak to be higher. In modal language, the coefficients $c_n \propto \sin(n\pi x_0/L)\, e^{-Cn^2\kappa}$ are then dominated by the lowest odd modes; the spectrum is effectively simpler, giving a narrower, taller, more skewed profile. For mid-box tethers (e.g., $x_0/L = 0.4 - 0.6$), more modes contribute, and the density is broader, so the peak height is lower. The two edge curves are mirrors of each other (left vs. right wall), so they look alike and higher than the others.

The Right panel shows the overlay of the scaled curves when $\kappa$ is kept identical across $L = 1,2,3\ \mu m$ by adjusting $N(L)$, the three overlaid curves are visually indistinguishable, and they coincide to the plotting precision. Because perfect visual overlap can conceal the presence of the multiple traces, we used distinct line styles in addition to colour; this reveals that the traces are coincident. Additionally, numerical diagnostics (see Supplementary table S1) shows identical normalization and consistent means across $L$. In addition, it has been verified that the residual values relative to a reference curve (L=3) provides quantitative confirmation that any differences are at most numerical noise (see Supplementary table S1). These steps together remove ambiguity and objectively document collapse. The enforced-$\kappa$ overlay therefore validates the scaling prediction that $P(u)$ is invariant to absolute size when $\kappa$ and $\xi$ are fixed.

The tether position $\xi = x_0/L$ is a primary geometric control parameter for the end-point distribution: moving the tether alters the modal phase factors $\sin(n\pi\xi)$ and thereby reshapes $P(u)$ in a predictable, symmetry-breaking manner. This physical effect is clearly visible in the tether-sweep panel at fixed $L$. By contrast, when the dimensionless confinement $\kappa$ and the dimensionless tether position $\xi$ are both held fixed, the scaled density $P(u)$ is invariant to absolute system size; the enforced-$\kappa$ overlay validates this geometric similarity.

**(c) Physical Interpretation:**

The observation of symmetry-breaking in the distribution is significant because it illustrates the geometric effects on the polymer behaviour. In systems like polymers or flexible chains confined in a box, the tether position is an essential parameter that determines how the polymer's configurational entropy is distributed. This result emphasizes that the geometry of confinement (i.e., the tethering position) cannot be neglected in understanding polymer dynamics, especially when the tether is off-centre.

This result is highly relevant for understanding the behaviour of tethered biopolymers in cellular contexts. For example, in the case of chromatin loci within the nucleus, the positions of the tethers (i.e., the anchored chromatin fibers) play a crucial role in determining the conformational state of chromatin. These findings suggest that the positioning of chromatin in the nuclear environment may introduce asymmetries in the spatial distribution of DNA, influencing processes like gene expression, DNA repair, and replication. Therefore, the geometry of tethered biopolymers has direct implications for understanding nuclear organization and function.

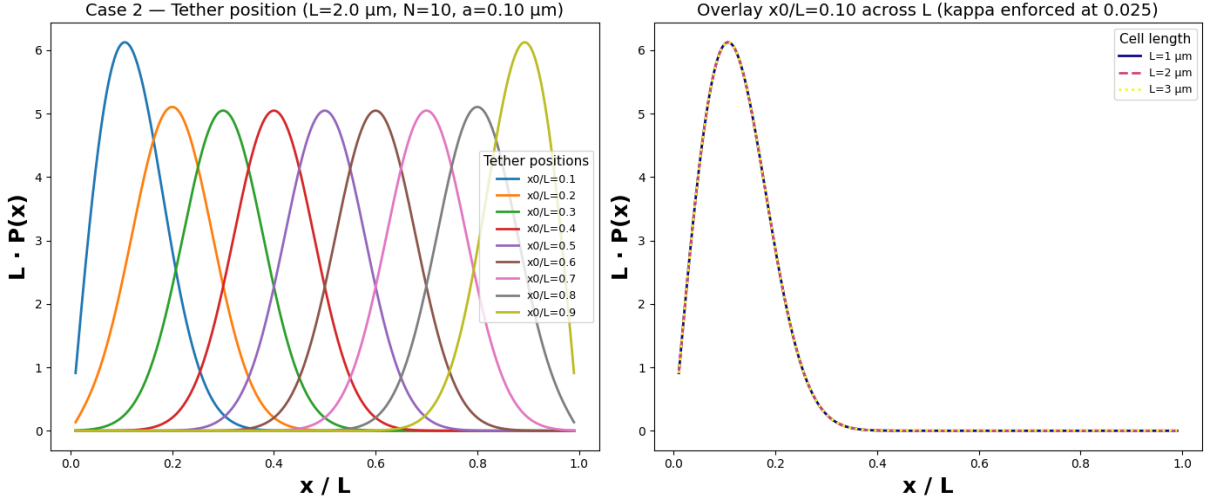

**Figure 2.** Effect of tether position on the longitudinal end-point distribution $P(x)$. **Left panel**: scaled densities $L \cdot P(x)$ versus the dimensionless coordinate $u = x/L$ for a family of tether positions $\xi = x_0/L = \{0.1, 0.2, ..., 0.9\}$ at fixed physical parameters $L = 2.0 \ \mu m$, $N = 10$, and $a = 0.10 \ \mu m$ (hence $\kappa = (Na^2)/L^2 = 0.025$). Moving the tether off center produces a clear, monotonic shift of the peak and a progressive skewing of the profile toward the nearer wall; the centered tether ($x_0/L = 0.5$) yields a symmetric profile (even modes vanish). **Right panel**: overlay of $L \cdot P(x)$ versus $u$ at fixed tether ratio $x_0/L = 0.10$ for three absolute sizes $L = \{1,2,3\} \ \mu m$ with $\kappa$ enforced equal to the left-panel baseline ($\kappa = 0.025$) by adjusting $N(L) = \kappa L^2/a^2$ (giving $N \approx 2.5, 10, 22.5$ for $L = 1,2,3 \ \mu m$, respectively). Distinct linestyles (solid, dashed, dotted) were used for three curves. The near-perfect overlap confirms geometric similarity: when both the dimensionless confinement $\kappa$ and the dimensionless tether position $x_0/L$ are held fixed, the scaled density $P(u) = L P(x)$ is invariant to absolute system size.

## Case 1 of P(y): Geometric-similarity collapse of $P(y)$ (scaled plot: $y/R, \ R P(y)$)

Here, we will show that for a tethered Gaussian chain between absorbing walls at $y = \pm R$ and at fixed dimensionless confinement $\kappa = \sigma^2/R^2$, the transverse end-point distribution $P(y)$ obeys exact geometric similarity i.e. the scaled density $P(u) = R P(y)$ with $u = y/R$ is independent of the absolute box size $R$.

### (a) Scaling variables and geometry

We consider a tethered Gaussian chain with one end fixed at the origin and the other free end with coordinate $y$ confined in the dimension $[-R, R]$ by perfectly absorbing walls. The free-coil root-mean-square is $\sigma = \sqrt{N} \ a$. The single dimensionless confinement parameter is

$$\kappa = \frac{\sigma^2}{R^2} = \frac{Na^2}{R^2}.$$

To test geometric similarity, we use the scaled coordinate and density

Scaled coordinate:

$$u = \frac{y}{R}.$$

Scaled PDFs:

$$\mathcal{P}(u) = R\, P(uR; \sigma, R),$$

$$\mathcal{P}_{\text{cond}}(u) = \frac{\mathcal{P}(u)}{S}, \qquad S = \int_{-1}^{1} \mathcal{P}(u)\, du.$$

**Specific definitions — *unconditional* vs *conditional***

It is important to report two related but distinct objects which will be used for all the cases:

Unconditional density $P(y)$ is the raw endpoint probability density obtained from the absorbing-image (or modal) solution. Because the walls are absorbing the total probability mass inside $[-R, R]$ is reduced relative to an unconstrained Gaussian; its integral

$$S = \int_{-R}^{R} P(y)\, dy$$

is the survival probability (the fraction of configurations that have not been absorbed). The unconditional scaled curve $\mathcal{P}(u) = R\, P(y)$ therefore has area $S$ on $[-1,1]$.

Conditional density $P_{\text{cond}}(y)$ is the normalized density of survivors,

$$P_{\text{cond}}(y) = \frac{P(y)}{S}, \qquad \mathcal{P}_{\text{cond}}(u) = \frac{\mathcal{P}(u)}{S},$$

so that $\int_{-1}^{1} \mathcal{P}_{\text{cond}}(u)\, du = 1$. The conditional form isolates shape information of surviving configurations and removes amplitude loss due to absorption.

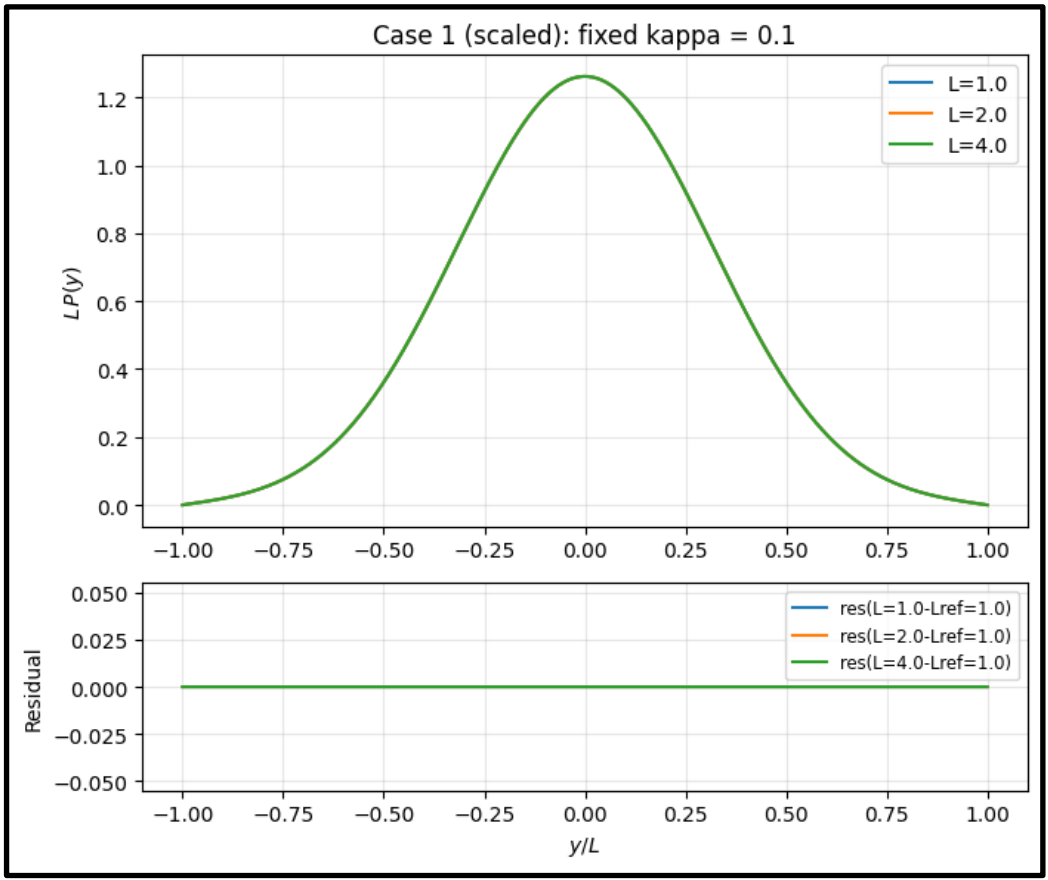

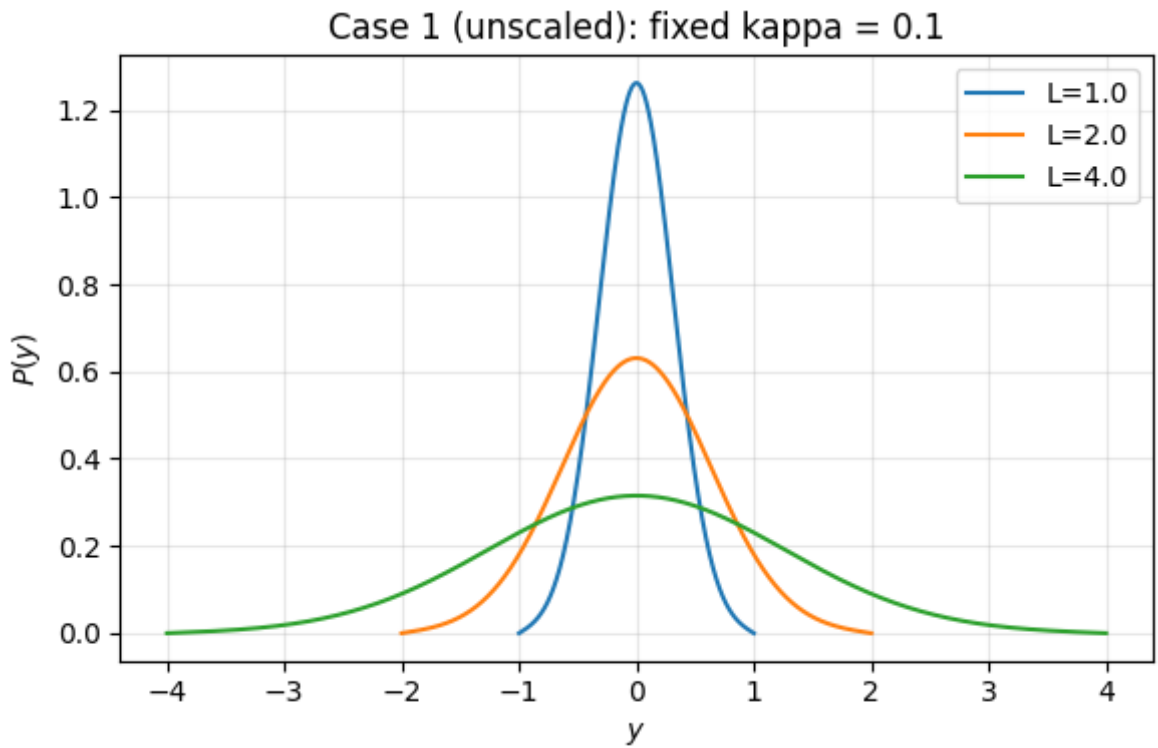

**Figure 3: (transverse collapse). Top:** scaled transverse end-point distributions $P(u) = R\,P(y)$ plotted against $u = y/R$ for three absolute box sizes $R$ with the dimensionless confinement $\kappa = \sigma^2/R^2$ held fixed ($\kappa = 0.1$). All curves collapse onto a single master curve. The pointwise residuals with $R_{ref} = 1$ is shown below. Residual metrics (maximum absolute residual and RMS residual) and the survival probabilities used to generate the curves are reported in Supplementary Table S4. **Bottom:** unscaled densities $P(y)$ vs $y$ for the same runs, demonstrating that collapse is only visible in the scaled coordinates.

### (b) Results

The top panel of Figure 3 (overlapped scaled densities along with residuals below) shows the scaled densities $\mathcal{P}(u) = R\,P(y)$ for $R = 1,2,4$ with $\kappa = 0.1$. All three scaled traces lie on top of one another within plotting resolution. The residual subplot (same figure) displays

$$\Delta\mathcal{P}(u) = \mathcal{P}(u;R) - \mathcal{P}(u;R)$$

for each $R$ relative to the reference $R_{ref} = 1$; residuals are vanishingly small (at the level of numerical round-off), confirming collapse to machine precision. The bottom panel shows the unscaled densities $P(y)$ vs $y$. As expected, different $R$ produce different amplitudes and widths in the unscaled representation while the scaled representation is invariant.

For Case 1, the unconditional distribution $P_{uncond}(y)$ represents the raw endpoint probability inside $[-R, R]$, with its total integral equal to the survival probability $S \leq 1$. The conditional distribution $P_{cond}(y) = P_{uncond}(y)/S$ is obtained by renormalizing to unit area. The conditional and unconditional PDFs are identical in shape, since the conditional case simply divides the unconditional distribution by the survival probability $S$, a constant factor independent of y. Thus, both curves coincide after scaling, and only one case is shown ($P_{uncond}(y)$). We nevertheless discuss about both cases for clarity and consistency with later sections (Cases 2–5), where conditional and unconditional distributions do differ in appearance and in physical interpretation.

### (c) Physical Interpretation:

The scaling collapse observed in Case 1 reflects a fundamental principle of statistical mechanics. Systems with a length scale (in this case, the polymer size and confinement width) can exhibit universal behaviours when rescaled by their natural length scales. The key finding is that, for fixed κ, the distribution becomes independent of the confinement width R when appropriately rescaled by R. This means that the polymer behaves similarly across different confinement widths when viewed through the lens of this scaling approach.

The implications of this result are significant for biological processes involving confined polymers, such as DNA within the cell nucleus or proteins within small cellular compartments. In such cases, confinement plays a critical role in altering the conformational dynamics of the polymer. By establishing universal scaling for polymer distributions in confined spaces, this scaling provides a framework for understanding how confined biological molecules might adapt their configurations when exposed to varying physical environments. For example, DNA packaging in the nucleus.

### Case 4 of P(y): Fixed $R$, varying $\kappa$; unconditional, conditional, and peak-normalized comparisons

Here, we extend the image–method analysis to study how the endpoint distributions change when the number of segments N is varied at fixed Kuhn length a, and cylinder width R. Since

κ=Na²/R², this allows us to probe confinement effects from the perspective of chain length. Both unconditional and conditional distributions are evaluated, along with diagnostics such as survival probability and conditional variance, to test the robustness of the scaling.

In Case 2, we varied the polymer flexibility directly by changing the stiffness parameter $\kappa$ while keeping the cylinder half-width $R$ fixed. That test isolates the effect of local bending stiffness on the endpoint statistics. In Case 4, we instead hold the Kuhn length $a$ and the cylinder half-width $R$ fixed and change the chain length $N$. Because $\kappa$ depends on $N$ for fixed $a$ and $R$, this procedure modifies the effective flexibility indirectly while also changing the modal content and finite-$N$ spectral weight of the chain.

**a) Scaling variables and geometry:**

To test geometric similarity at fixed physical box size, we hold the half-width $R$ constant and vary the confinement parameter

$$\kappa = \frac{Na^2}{R^2} \quad \text{(equivalently, } N = \kappa R^2/a^2\text{).}$$

We plot the dimensionless scaled distribution

$$\mathcal{P}(u;\kappa) = R\, P(uR;\sigma(\kappa),R), \qquad u \equiv \frac{y}{R} \in [-1,1], \quad \sigma(\kappa) = \sqrt{N}\, a.$$

Conditional version:

$$\mathcal{P}_{\text{cond}}(u;\kappa) = \frac{\mathcal{P}(u;\kappa)}{S(\kappa)}.$$

so that the horizontal axis runs over the canonical domain $[-1,1]$ for all curves. When $R$ is held fixed, variation in $\kappa$ systematically probes how the polymer's free-coil size (via $N$) changes relative to the confinement width. Therefore, three representations are used: unconditional $\mathcal{P}(u)$, conditional

$$\mathcal{P}_{\text{cond}}(u) = \frac{\mathcal{P}}{S}.$$

and peak-normalized

$$\mathcal{P}_{\text{peak}}(u;\kappa) = \mathcal{P}_{\text{cond}}(u;\kappa)/\max\mathcal{P}_{\text{cond}}(u;\kappa).$$

Three panels were used to show above three representations. Three parameters such as the survival S, $N$ and conditional variance for each κ were printed in a small diagnostics table.

**b) Results:**

**Unconditional distributions:** Figure 4 (top left) shows the unconditional scaled PDFs $\mathcal{P}(u) = RP(y)$ plotted against the scaled coordinate $u = y/R$. At small $\kappa$ (0.05,0.1), the distributions are sharply peaked near the channel center, reflecting weak confinement relative to chain size. With increasing $\kappa$, the peak progressively flattens, and the distributions broaden, consistent with stronger wall interactions.

**Conditional (survivor) distributions:** The normalized conditional PDFs $\mathcal{P}_{\text{cond}}(u)$ (Fig. 4, top right) show that after removing absorbed trajectories, the survivors converge toward nearly universal, flattened shapes for moderate-to-large $\kappa$. The inset highlights that near the central region, curves for large $\kappa$ collapse closely, reflecting the dominance of confinement. In contrast, the small–$\kappa$ cases retain strong central peaks, indicating survival-biased sampling of nearly free chains.

**Peak–normalized comparison:** The bottom right panel (Fig. 4) presents the peak–normalized distributions. Here, the differences in shape across $\kappa$ are most evident: the smallest–$\kappa$ cases (0.05,0.1) retain distinctly sharper peaks and faster tail decay, while all larger $\kappa$ values ($\geq 0.5$) collapse to nearly identical bell-like profiles. This indicates a clear crossover around $\kappa \approx 0.5$, beyond which confinement strongly dictates the conditional distribution's shape.

**Diagnostics:** The log–log diagnostic panel (bottom left, Fig. 4) quantifies survival probability $S(\kappa)$ and conditional variance $\text{Var}(u^2)$. Survival remains nearly close to unity for $\kappa \leq 0.1$, then decreases significantly with stronger confinement: $S = 0.685$ at $\kappa = 0.5$, $S = 0.173$ at $\kappa = 1.0$, and as low as $S = 0.028$ at $\kappa = 2.0$. Complementarily, the conditional variance increases monotonically from $\text{Var}(u^2) \approx 0.014$ at $\kappa = 0.05$ to $\text{Var}(u^2) \approx 0.123$ at $\kappa = 2.0$. The survival suppression and variance growth thus act as dual markers of confinement. The survival decays due to absorption at the walls, while survivors spread further in scaled coordinates.

In Case 4, the image–method formulation confirms that varying $\kappa$ with fixed Kuhn length produces a prominent crossover. For small $\kappa$, chains behave nearly freely with sharp, peaked conditional PDFs. Whereas for large $\kappa$, the confinement dominates, producing broadened survivor distributions with suppressed survival. The joint survival–variance diagnostics provide a compact quantitative signature of this crossover, fully consistent with the observed collapse in peak-normalized distributions.

It should be mentioned that like Cases 2 and 3, Case 4 demonstrates survival suppression and broadening of conditional distributions with increasing κ. However, while Cases 2 and 3 vary stiffness or confinement length directly, Case 4 achieves the same crossover by increasing chain length N, highlighting that all three routes lead to analogous suppression trends but with distinct asymptotic scaling behaviours.

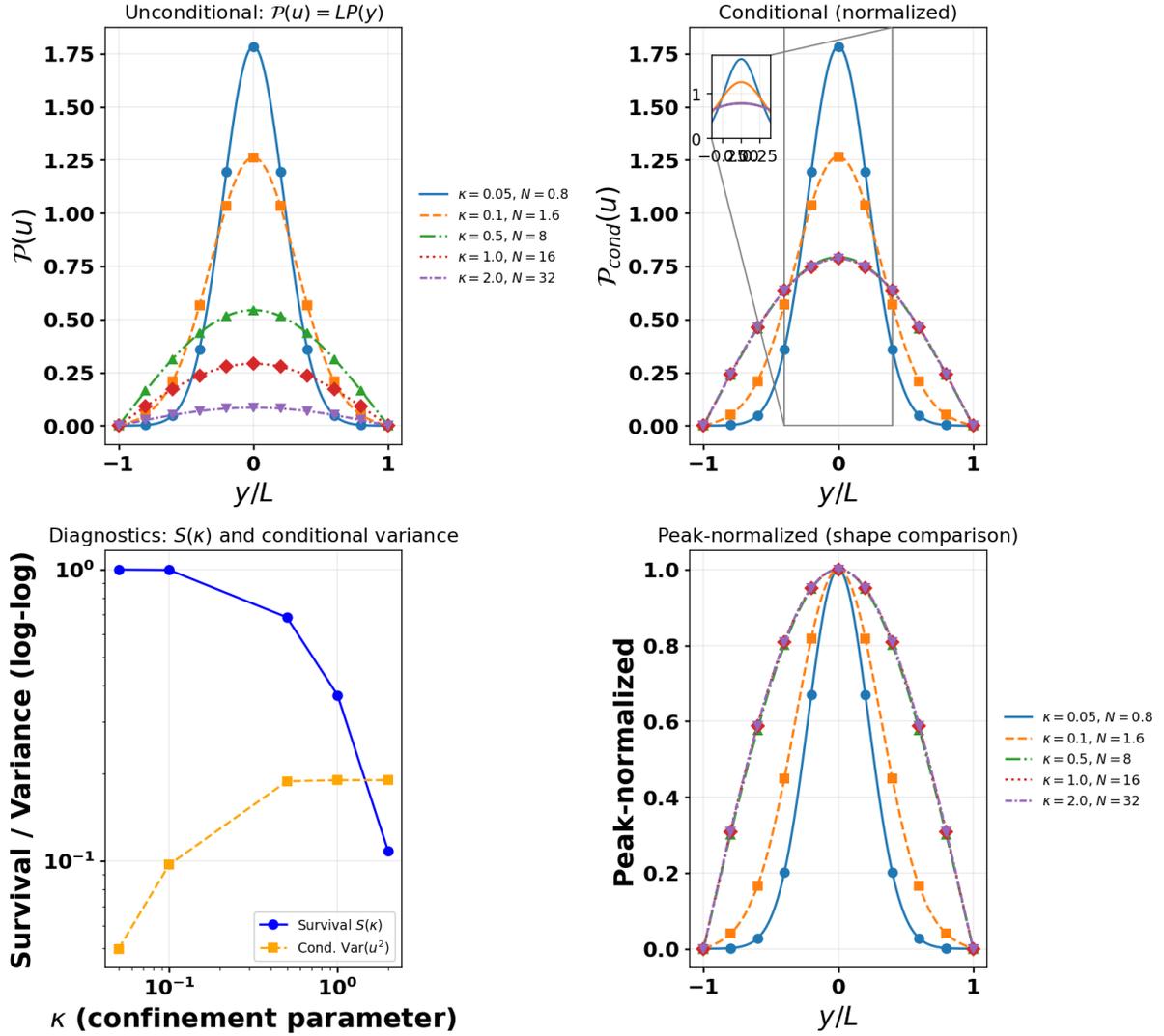

**Figure 4.** Image-method collapses with varying chain length N at fixed slit width R. **Top left:** unconditional scaled distributions $\mathcal{P}(u) = RP(y)$, with areas equal to survival probabilities $S(\kappa)$. **Top right:** conditional scaled distributions $\mathcal{P}_{\text{cond}}(u)$, normalized over survivors (inset shows zoom near origin). **Bottom left:** diagnostics of survival probability and conditional variance versus $\kappa$ (log–log). **Bottom right:** peak-normalized conditional distributions highlighting shape collapse at large $\kappa$.

**(c) Physical Interpretation:**

This case illustrates the clear impact of the confinement strength on both the distribution amplitude (via survival probability) and its shape. As κ increases, the survival probability decreases due to the polymer chain's increasing localization near the centre of the confinement region. The peak-normalized distribution reveals that the polymer's shape sharpens as confinement increases, with the distribution becoming more concentrated near the central region of confinement.

The ability of a polymer to transition from a diffuse state to a highly localized one is critical for biological processes. For example, proteins and DNA often need to undergo structural rearrangements to perform specific functions, such as in the processes of gene expression, protein folding, and enzyme activity. Our results suggest that these biomolecules could become highly localized or compartmentalized within confined regions, which might influence their functional roles. This has close relevance to the study of protein localization in cellular compartments, such as the nucleus or mitochondria, and to the mechanics of DNA condensation during cellular processes like mitosis.